\documentclass[aps,pra,twocolumn,floatfix,superscriptaddress,longbibliography,nofootinbib]{revtex4-2}

\PassOptionsToPackage{dvipsnames,usenames}{xcolor}

\usepackage{amsmath,amsfonts,amssymb,amsthm,amscd,amstext,mathtools}
\usepackage{ifthen} 
\usepackage{bm,bbm,dsfont}

\usepackage{graphicx,epsfig,epstopdf,psfrag}
\usepackage{tikz}
\usetikzlibrary{quantikz}
\usepackage{multirow}

\newcommand{\makecell}[2][c]{\begin{tabular}[c]{@{}#1@{}}#2\end{tabular}}
\newcommand{\Xhline}[1]{\specialrule{#1}{0pt}{0pt}}
\usepackage{dblfloatfix}

\usepackage{times}
\usepackage{comment}
\usepackage{soul}
\usepackage[normalem]{ulem}
\usepackage{enumerate}
\usepackage{enumitem}
\usepackage{xstring}
\usepackage{cancel}
\usepackage{footnote}
\usepackage[titletoc]{appendix}
\usepackage{import}
\usepackage{subfiles}
\usepackage{subfig}
\usepackage{ragged2e}

\usepackage{xcolor} 
\definecolor{mycolor}{RGB}{0, 118, 112}

\usepackage{hyperref}
\hypersetup{
    colorlinks=true,
    citecolor=mycolor,
    linkcolor=mycolor,
    urlcolor=mycolor,
    breaklinks=true 
}

\AtBeginDocument{
    \hypersetup{
        citecolor=mycolor,
        linkcolor=mycolor,
        urlcolor=mycolor
    }
}

\usepackage{mathtools}

\newcommand{\iden}[1]{
    \ifthenelse{\equal{1}{\string #1}}
  {
   \mathbbm{1}
  }
  {
   \mathbbm{1}^{\otimes#1}}
  }
\newcommand{\ketzero}[1]{
    \ifthenelse{\equal{1}{\string #1}}
  {
   \ket{0}
  }
  {
   \ket{0}^{\otimes#1}}
  }
\newcommand{\brazero}[1]{
    \ifthenelse{\equal{1}{\string #1}}
  {
   \bra{0}
  }
  {
   \bra{0}^{\otimes#1}}
  }
\newcommand{\ketone}[1]{
      \ifthenelse{\equal{1}{\string #1}}
    {
     \ket{1}
    }
    {
     \ket{1}^{\otimes#1}}
    }
  \newcommand{\braone}[1]{
      \ifthenelse{\equal{1}{\string #1}}
    {
     \bra{1}
    }
    {
     \bra{1}^{\otimes#1}}
    }

\renewcommand{\arraystretch}{1.2} 

\usepackage{todonotes}

\definecolor{MSgreen}{RGB}{34,139,34}

\usepackage{dsfont}
\usepackage{amssymb}
\usepackage{booktabs}
\usepackage{placeins}

\begin{document}
\title{Qutrit-Based Neural Quantum Kernels for Classification Tasks}

\author{Camila Cristiano-Romero}
\email[Corresponding author: ]{\qquad ccristiano@bcamath.org}
\affiliation{BCAM - Basque Center for Applied Mathematics, Mazarredo, 14 E48009 Bilbao, Basque Country – Spain}
\affiliation{Department of Physical Chemistry, University of the Basque Country UPV/EHU, Apartado 644, 48080 Bilbao, Spain}

\author{Pablo Rodriguez-Grasa}
\affiliation{Department of Physical Chemistry, University of the Basque Country UPV/EHU, Apartado 644, 48080 Bilbao, Spain}
\affiliation{EHU Quantum Center, University of the Basque Country UPV/EHU, Apartado 644, 48080 Bilbao, Spain}
\affiliation{TECNALIA, Basque Research and Technology Alliance (BRTA), 48160 Derio, Spain}

\author{Mikel Sanz}
\affiliation{BCAM - Basque Center for Applied Mathematics, Mazarredo, 14 E48009 Bilbao, Basque Country – Spain}
\affiliation{Department of Physical Chemistry, University of the Basque Country UPV/EHU, Apartado 644, 48080 Bilbao, Spain}
\affiliation{EHU Quantum Center, University of the Basque Country UPV/EHU, Apartado 644, 48080 Bilbao, Spain}

\begin{abstract}
Neural quantum kernels (NQKs) construct quantum kernels by pretraining a quantum neural network (QNN) and subsequently reusing the trained circuit as a task-adapted embedding. Extending this framework to qudits, with local unitaries in $\mathrm{SU}(d)$, provides a natural route to richer data embeddings through the increased local degrees of freedom and a direct interface for multiclass classification via intrinsically multi-level quantum systems. In this work, focusing on qutrits ($d=3$), we extend NQKs to the qudit setting and perform a systematic study of key design choices, including the number of encoded features, the number of qutrits, the kernel construction (1-to-$n$ and $n$-to-$n$), and the parameterization of $\mathrm{SU}(3)$ unitaries. Across binary and three-class tasks on four benchmark datasets, qutrit NQKs improve over the corresponding QNN baselines in nearly all settings considered and can benefit from scaling both the feature budget and the system size, although the magnitude of these gains may saturate, is dataset-dependent, and depends on the chosen parameterization. In particular, an ablation over $\mathrm{SU}(3)$ parameterizations shows that the unitary representation can substantially impact both optimization behaviour and classifier performance. These findings highlight the potential of qudit-based quantum models not only as a straightforward generalization of qubit-based architectures, but also as a promising means to better exploit complex data structures in quantum machine learning.
\end{abstract}

\maketitle

\section{Introduction}
\label{sec:intro}

Quantum computing has emerged as a promising computational paradigm for addressing problems that are classically intractable. One of its most compelling applications lies in the field of quantum machine learning (QML), which seeks to harness quantum resources to improve learning performance and model expressivity \cite{biamonte2017quantum, carleo2019machine, dunjko2018machine}. Among various quantum learning architectures, parameterized quantum circuits (PQCs) have become the standard approach due to their flexibility and compatibility with classical optimization techniques \cite{benedetti2019parameterized, farhi2018classification, skolik2022quantum, schuld2020circuit}.

Although there are indications of quantum advantage in carefully tailored problems \cite{gyurik2022establishing, sweke2021quantum, jerbi2021quantum, liu2021rigorous, pirnay2023superpolynomial}, demonstrating a practical advantage over classical methods remains an open question \cite{jadhav2023quantum}. A key challenge concerns the trainability and expressivity of PQCs \cite{mcclean2018barren, cerezo2021cost, holmes2101connecting, fontana2024adjoint, thanasilp2024exponential, pesah2021absence, ragone2024lie}. In particular, barren plateaus, i.e., regions of the optimization landscape where gradients vanish exponentially, pose a significant obstacle to scaling variational quantum algorithms and continue to be an active area of research.

Within this broader landscape of PQCs, data re‑uploading models represent a particularly expressive class of architectures \cite{perez2020data}. These models are characterized by the alternating application of data‑encoding unitaries and trainable gates, yielding highly expressive circuits that have seen extensive use in QML. Importantly, it has been shown that data re‑uploading models are universal \cite{perezsalinas2025universalapproximationcontinuousfunctions}, meaning they can approximate any continuous function on a compact domain, in close analogy with the universal approximation theorem for classical neural networks. This universality highlights their theoretical significance and motivates further investigation into their practical capabilities \cite{schuld-2021-effect-of, casas2023multidimensional, barthe2023gradients}.

Most existing studies in QML focus primarily on qubits, yet extending these models to qudits, i.e., $d$-level quantum systems, may offer significant advantages. The richer algebra of $\mathrm{SU}(d)$ enables more diverse embeddings and feature maps in enlarged Hilbert spaces, while also aligning naturally with multi‑valued tasks such as $d$-class classification \cite{farias2025short}. Although the experimental control of qudit operations is more demanding than in the qubit case, their structural properties and higher‑dimensional embeddings motivate the investigation of qudits as an alternative framework for QML in the NISQ era \cite{gokhale2019asymptotic, pavlidis2021quantum, gedik2015computational, deller-2023, roca-jerat-2024, anand-2026, wang-2020}. High-quality qudit control has moreover been demonstrated experimentally on trapped-ion and superconducting platforms \cite{ringbauer2022universal, blok2021quantum}.

Some recent studies have begun to explore the use of qudits in classification tasks and related machine learning applications \cite{mandilara2024classification, wach2023data, valtinos2024gell, acar2025unlocking, souza-2025, useche-2021, laino-2025}. For example, in \cite{acar2025unlocking} a comparative analysis of variational quantum neural networks (VQNNs) implemented with qubits and qutrits is performed. The results show that qutrit‑based VQNNs can achieve greater parameter efficiency and, in some cases, superior performance compared to their qubit counterparts, although the extent of these benefits depends strongly on the choice of encoding strategy and the available computational resources. In a related line of work, \cite{wach2023data} investigates data re‑uploading circuits built from single‑qudit operations. By interpreting a qudit as a spin system and employing angular momentum operators as elementary gates, the authors demonstrate that such models are naturally suited for multi‑class classification tasks. Their benchmarks further show that single‑qudit re‑uploading architectures can learn highly non‑linear decision boundaries and reach performance levels comparable to classical machine learning methods, while also revealing intrinsic biases due to operator choices and label alignment, as well as trade‑offs between circuit depth and gate sets.

At the same time, the training challenges associated with PQCs motivate hybrid strategies that reduce reliance on end‑to‑end variational optimization while still leveraging quantum feature spaces. One such strategy is provided by quantum kernel methods \cite{havlicek-2019, schuld-2019-hilbert-spaces, bartkiewicz-2020, huang-2021, peters-2021, kusumoto-2021, schuld-2021, liu-2021, kyriienko-2022, wu-2023, tomasi-2025}, in which learning is carried out by comparing quantum feature states rather than by optimizing a fixed measurement rule.

Neural quantum kernels (NQKs), introduced in Ref. \cite{nqk_pablo} and subsequently applied to real-world classification problems \cite{nqk_satellite, nqk_neutrinos}, address the challenge of engineering a suitable embedding a priori by learning the embedding once via supervised training of a quantum neural network (QNN), and subsequently freezing the learned parameters to define a problem-adapted kernel. A key practical advantage is that the kernel matrix need only be constructed once, avoiding the repeated kernel-matrix reconstruction inherent in direct kernel-training schemes. Moreover, Ref. \cite{nqk_pablo} proposes a scalable training strategy for data re-uploading QNNs that progressively increases the number of qubits, warm-starting each larger model from the optimum of the previous step. Within this framework, two complementary constructions are identified: the $1$-to-$n$ approach, in which a single-qubit QNN is trained and its learned unitary is replicated across an $n$-qubit register with fixed entanglement, and the $n$-to-$n$ approach, in which an $n$-qubit QNN is trained and directly reused as the embedding that induces the kernel. Although NQKs were originally formulated in the qubit setting, the underlying principle is not intrinsically restricted to two-level systems.

In this Article, we extend neural quantum kernels to qudits of local dimension $d>2$, focusing on qutrits ($d=3$) as the smallest nontrivial case. Moving from qubits to qudits introduces new modelling choices that are absent or much more restricted in the qubit setting, most notably the local dimension $d$ and the parameterization of single-qudit $\mathrm{SU}(d)$ operations. At the same time, the larger number of local degrees of freedom available in $\mathrm{SU}(d)$ makes the encoded feature budget $p$ a particularly relevant design parameter. These qudit-related choices interact with architectural and task-dependent ingredients already present in NQK models, such as the number of subsystems $n$, the kernel construction, and the choice of class labels, readout rule, and cost function. Table \ref{tab:intro_ingredients} provides a compact summary of the model choices considered in this work. This motivates a design-oriented perspective in which we study how selected components affect performance while keeping the remaining architectural choices and training hyperparameters fixed, including the circuit depth and entangling layout.

Our qudit extension is motivated by two complementary considerations. First, single-qubit embeddings are built from $\mathrm{SU}(2)$ operations with only three independent parameters, which limits the number of features that can be injected per subsystem through a single encoding block. In contrast, single-qudit operations belong to $\mathrm{SU}(d)$ and admit $d^2-1$ generators, enabling richer local embeddings and allowing more features to be incorporated per wire. Second, qudits naturally support multiclass learning, since their computational basis provides $d$ mutually orthogonal label states that can be directly associated with class hypotheses. Building on the NQK methodology, we generalize the data re-uploading architecture, the kernel constructions ($1$-to-$n$ and $n$-to-$n$), and the cost functions to qutrit-based binary and three-class classification. We empirically show that qutrit NQKs improve over the corresponding QNN baselines across the considered benchmarks, and we characterize how performance depends on feature budget, system size, and the parameterization of $\mathrm{SU}(3)$ unitaries.
\begin{table}[t]
    \centering
    \caption{\justifying
    Design choices in qudit Neural Quantum Kernels considered in this work.}
    \label{tab:intro_ingredients}
    \small
    \setlength{\tabcolsep}{6pt}
    \renewcommand{\arraystretch}{1.15}

    \begin{tabular}{p{0.34\columnwidth} p{0.58\columnwidth}}
    \Xhline{1.2pt}
    \textbf{Design choice} & \textbf{Description} \\
    \Xhline{1.2pt}

    Local dimension $d$ &
    Sets the local Hilbert-space dimension and the degrees of freedom of $\mathrm{SU}(d)$; enables richer local embeddings and provides a natural interface for multiclass labels. \\
    \Xhline{0.2pt}

    System size $n$ &
    Number of subsystems; enlarges the joint Hilbert space and enables richer correlations through entanglement. \\
    \Xhline{0.2pt}

    Feature budget $p$ &
    Number of encoded features; determines how much classical information is injected into the embedding per data re-uploading block. \\
    \Xhline{0.2pt}

    Single-qudit parameterization &
    Choice of coordinates / decomposition for $\mathrm{SU}(d)$ blocks. \\
    \Xhline{0.2pt}

    Kernel construction &
    How the embedding is transferred to a kernel model. \\
    \Xhline{0.2pt}

    Task definition and cost function &
    Choice of class labels, readout rule, and cost function.  \\
    \Xhline{1.2pt}
    \end{tabular}
\end{table}

\section{Methods}
\label{sec:methods}

\subsection{Neural Quantum Kernels}
\label{sec:nqk}

Kernel methods are a class of machine learning models that rely on similarity measures between data points \cite{scholkopf2002learning}. Given training inputs $\{\bm{x}_i\}_{i=1}^M$, a kernel method constructs a function 
\begin{equation}
    f_{\bm{\alpha},X}(\bm{x}) = \sum_{i=1}^{M} \alpha_i \, k(\bm{x}, \bm{x}_i),
    \label{eqn:kernel_definition}
\end{equation}
 where $k(\bm{x}, \bm{x}_i)$ is the kernel function. In standard kernel methods, the kernel function $k$ is chosen to be symmetric and positive semi-definite, so that it can be represented as an inner product in a Hilbert space $\mathcal{H}$, namely
 \begin{equation}
     k(\bm{x}_i, \bm{x}_j) = \langle \phi(\bm{x}_i), \phi(\bm{x}_j) \rangle_{\mathcal{H}},
     \label{eqn:kernel_innerproduct}
 \end{equation}  
with $\phi(\bm{x})$ denoting the associated feature map. The coefficients $\boldsymbol{\alpha} = \{\alpha_i\}$ are obtained by solving a convex optimization problem, which is carried out on a classical computer once the kernel matrix $K$ with entries $K_{ij} = k(\bm{x}_i, \bm{x}_j)$ has been constructed.

In the quantum setting, the feature map is realized by a quantum embedding that maps data points $\bm{x}$ into a quantum state $\rho(\bm{x})$ \cite{havlicek-2019, schuld-2019-hilbert-spaces, schuld-2021, schuld2021quantum}. A common way to construct a quantum kernel from such states is to compare them via their Hilbert--Schmidt inner product, yielding an embedding quantum kernel (EQK) of the form
\begin{equation}
    k(\bm{x}_i, \bm{x}_j) = \mathrm{tr}\bigl( \rho(\bm{x}_i)\rho(\bm{x}_j) \bigr).  
    \label{eqn:eqk_definition}
\end{equation}
When the quantum embedding is a unitary transformation $S(\bm{x})$, the feature vector is a pure state $\rho(\bm{x}) = S(\bm{x})|0\rangle \langle 0|S(\bm{x})^\dagger$ and (\ref{eqn:eqk_definition}) takes the form
\begin{equation}
    k(\bm{x}_i, \bm{x}_j) = |\langle 0| S(\bm{x}_i)^\dagger S(\bm{x}_j) |0\rangle|^2,
\end{equation}
which corresponds to the squared overlap between quantum feature states. This overlap can be estimated on a quantum computer by preparing the state $S(\bm{x}_i)^\dagger S(\bm{x}_j)|0\rangle$ and measuring the probability of obtaining $|0\rangle$ in all subsystems \cite{buhrman-2001, fanizza-2020, cincio-2018}. EQKs of this type have been shown to be universal, making them a widely studied approach in quantum machine learning \cite{gil2024expressivity}.

A central limitation of quantum kernel methods is that the performance is largely dictated by the choice of the quantum embedding (feature map). In practice, this means that one must engineer a quantum embedding $S(\bm{x})$ whose induced geometry in Hilbert space matches the structure of the learning task. This requirement is problem-dependent, and it is closely related to the broader principle that meaningful quantum machine learning models must incorporate inductive biases and problem-inspired structure, rather than relying on generic, problem-agnostic embeddings~\cite{lloyd-2020, blank-2020, kubler-2021, salmenpera-2024, shirai-2024}.

When such prior structure (e.g., symmetries or known invariances) is not available, a natural alternative is to learn a task-adapted embedding by optimizing a parameterized feature map $S_{\gamma}(\cdot)$, which induces a trainable kernel $k_{\gamma}(\bm{x}_i,\bm{x}_j)$ \cite{vedaie-2020, hubregtsen-2022, ghukasyan-2023}. Existing approaches along these lines typically rely on iteratively updating $\gamma$ while repeatedly estimating the full kernel matrix during optimization. This can be prohibitively expensive in practice, since kernel construction already scales as $\mathcal{O}(M^2)$ in the number of training samples, and it can be further exacerbated in parametrized-kernel training since objectives such as kernel target alignment may exhibit exponentially flat landscapes under conditions analogous to barren plateaus, hindering convergence \cite{thanasilp2024exponential, thanasilp2023subtleties, shaydulin-2022, kairon2025equivalenceexponentialconcentrationquantum}.

Neural Quantum Kernels \cite{nqk_pablo, nqk_satellite, nqk_neutrinos} address these limitations by shifting the learning effort from repeatedly training the kernel to learning a task-adapted embedding once. The key idea is to pretrain a QNN on the supervised task and then freeze its trained parameters, reinterpreting the resulting circuit as the embedding $S(\bm{x})$ used to define the quantum feature states. After this pretraining stage, the kernel matrix is computed only once and the final predictor is obtained via a classical convex optimization procedure (e.g., an SVM) on the fixed kernel matrix, avoiding the repeated kernel-matrix reconstruction required by direct kernel-training schemes.

This two-stage construction can be viewed as replacing the QNN's fixed measurement-based decision rule with an optimal linear readout in the induced feature space. In practice, this often improves over the corresponding QNN classifier, since the kernel model can adjust the separating hyperplane even when the QNN was not trained to optimality.

\subsubsection{Pretraining a data re-uploading QNN as a scalable quantum embedding}
\label{sec:nqk_qnn}

In Neural Quantum Kernels, the quantum embedding is realized by a parameterized quantum circuit trained in a supervised manner. The architecture considered in \cite{nqk_pablo} follows a data re-uploading structure \cite{perez2020data}, in which the classical features are encoded repeatedly throughout the circuit and interleaved with trainable variational blocks. At a schematic level, a single-subsystem block can be written as
\begin{equation}
    S_{\bm{\theta}}(\bm{x})
    \;=\;
    \prod_{\ell=1}^{L} U(\bm{\theta}_\ell)\,U(\bm{x}),
    \label{eqn:datareuploading_circuit}
\end{equation}
where $U(\bm{x})$ denotes an angle-encoding map and $U(\bm{\theta}_\ell)$ denotes a trainable single-subsystem unitary at re-uploading step $\ell$.

For an $n$-qubit register, the multiqubit QNN is obtained by applying the single-qubit blocks in parallel and introducing parametrized nearest-neighbour entangling operations between subsystems $E(\bm{\varphi})$. The resulting $n$-qubit embedding circuit can be written as
\begin{equation}
    S^{(n)}_{\bm{\theta},\bm{\varphi}}(\bm{x})
    \;=\;
    \prod_{\ell=1}^{L}
    \left[
        \prod_{s=1}^{n-1} E^{s}_{s+1}(\bm{\varphi}^{(s)}_{\ell})\,
        \Bigl(\bigotimes_{r=1}^{n} U(\bm{\theta}^{(r)}_\ell)\Bigr)\,
        U(\bm{x})^{\otimes n}
    \right].
    \label{eq:n_qubit_qnn_general_entangler}
\end{equation}
The circuit prepares a data-dependent quantum feature state
\begin{equation}
    \rho_{\bm{\theta},\bm{\varphi}}^{(n)}(\bm{x})
    \;=\;
    S^{(n)}_{\bm{\theta},\bm{\varphi}}(\bm{x})\,\rho_0\,S^{(n)}_{\bm{\theta},\bm{\varphi}}(\bm{x})^{\dagger},
\end{equation}
where $\rho_0$ is a fixed reference state (typically $|0\rangle\!\langle 0|$).  The parameters $(\bm{\theta},\bm{\varphi})$ are obtained by minimizing a cost function $f_{\mathrm{cost}}$ defined from measurement outcomes, yielding optimal parameters $(\bm{\theta}^\star,\bm{\varphi}^\star)$.

A central ingredient enabling scalable pretraining is a progressive growth strategy in the number of qubits \cite{nqk_pablo}. We first train a single-qubit QNN to obtain optimal parameters $\bm{\theta}^{\star}_{(1)}$. We then enlarge the model by appending a second qubit and introducing trainable nearest-neighbour entangling operations between adjacent qubits, denoted by $E^{s}_{s+1}(\bm{\varphi})$. The two-qubit QNN is initialized by copying the previously learned parameters on the first qubit, while setting all newly introduced parameters to zero, namely the parameters of the second qubit as well as the entangling parameters $\bm{\varphi}$. This initialization yields an effectively decoupled starting point for the added subsystem. The enlarged model is then trained to obtain the corresponding optimal parameters for both qubits together with the entangling parameters.

This procedure is repeated iteratively. At growth step $n$, the $(n-1)$-qubit parameters are initialized with the optimum from the previous step, whereas the parameters of the newly added $n$-th qubit and the new nearest-neighbour entangling block coupling qubits $(n-1,n)$ are initialized at zero. Training then proceeds to produce $(\bm{\theta}^{\star}_{(n)},\bm{\varphi}^{\star}_{(n)})$. In this way, model capacity is increased gradually while warm-starting optimization from the previously learned solution.

\begin{figure*}[t]
    \centering
    \includegraphics[width=0.78\linewidth]{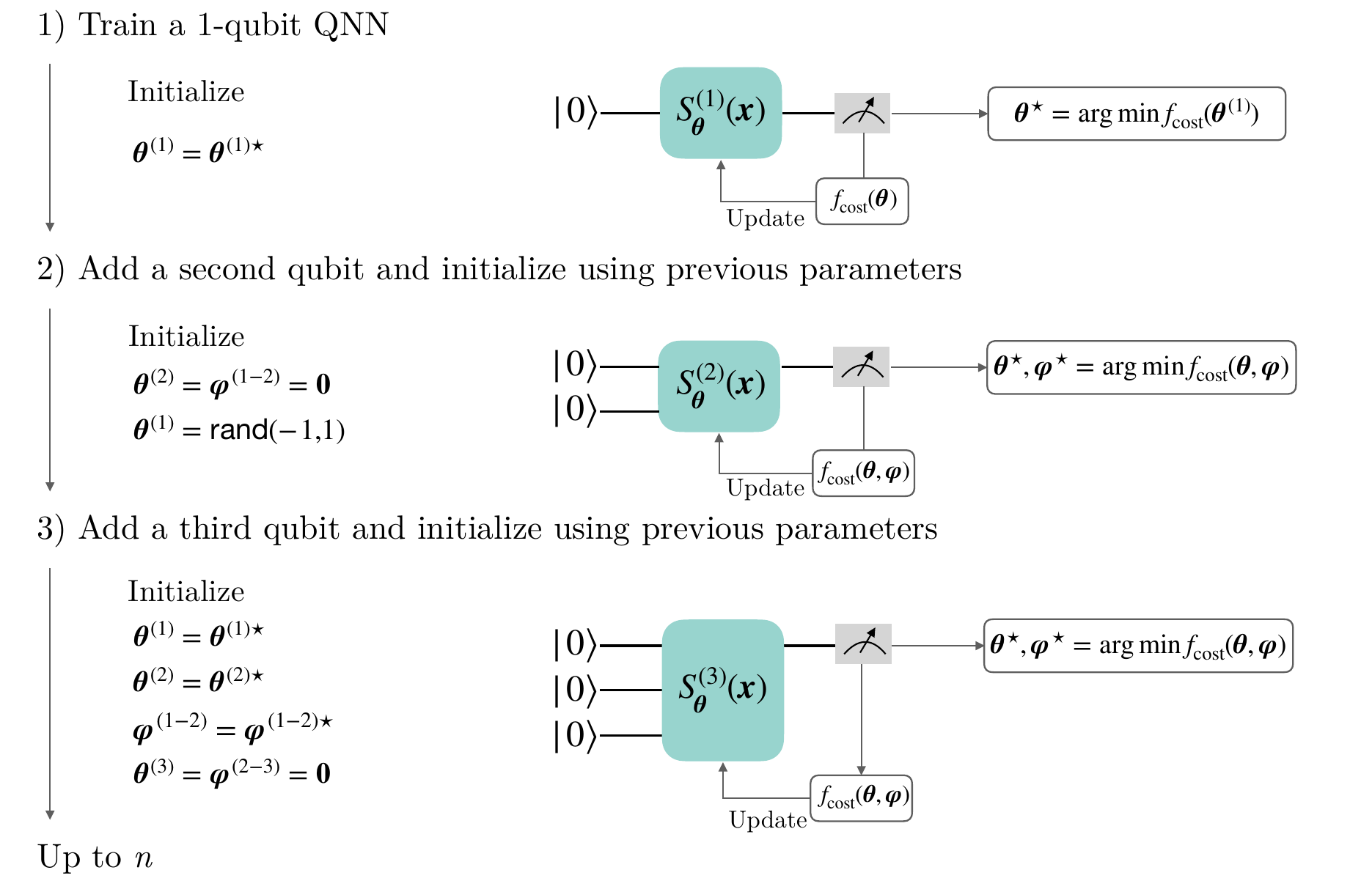}
    \caption{\justifying Progressive-growth training strategy used to scale the data re-uploading QNN embedding. Training starts from a single subsystem and iteratively appends additional subsystems, initializing new parameters (including entangling parameters) at zero while reusing the previously learned optimum for the existing blocks. Inspired by the procedure introduced in Ref. \cite{nqk_pablo}.}
    \label{fig:qnn_progressive_training}
\end{figure*}

The progressive growth procedure is illustrated in Fig. \ref{fig:qnn_progressive_training}. This training scheme can be viewed as a warm-start approach, in which the solution obtained for $(n-1)$ subsystems is reused to initialize the optimization at size $n$, while the parameters associated with the newly appended subsystem and its couplings are initialized so that the added degrees of freedom are initially weakly activated. Warm-start strategies for scaling variational quantum models and mitigating optimization pathologies have been recently gaining attention and studied in detail in \cite{egger-2021, mhiri-2025, puig-2025}.

\subsubsection{Kernel construction}
\label{sec:nqk_kcons}

\begin{figure*}[t] 
  \centering
  \includegraphics[width=0.78\linewidth]{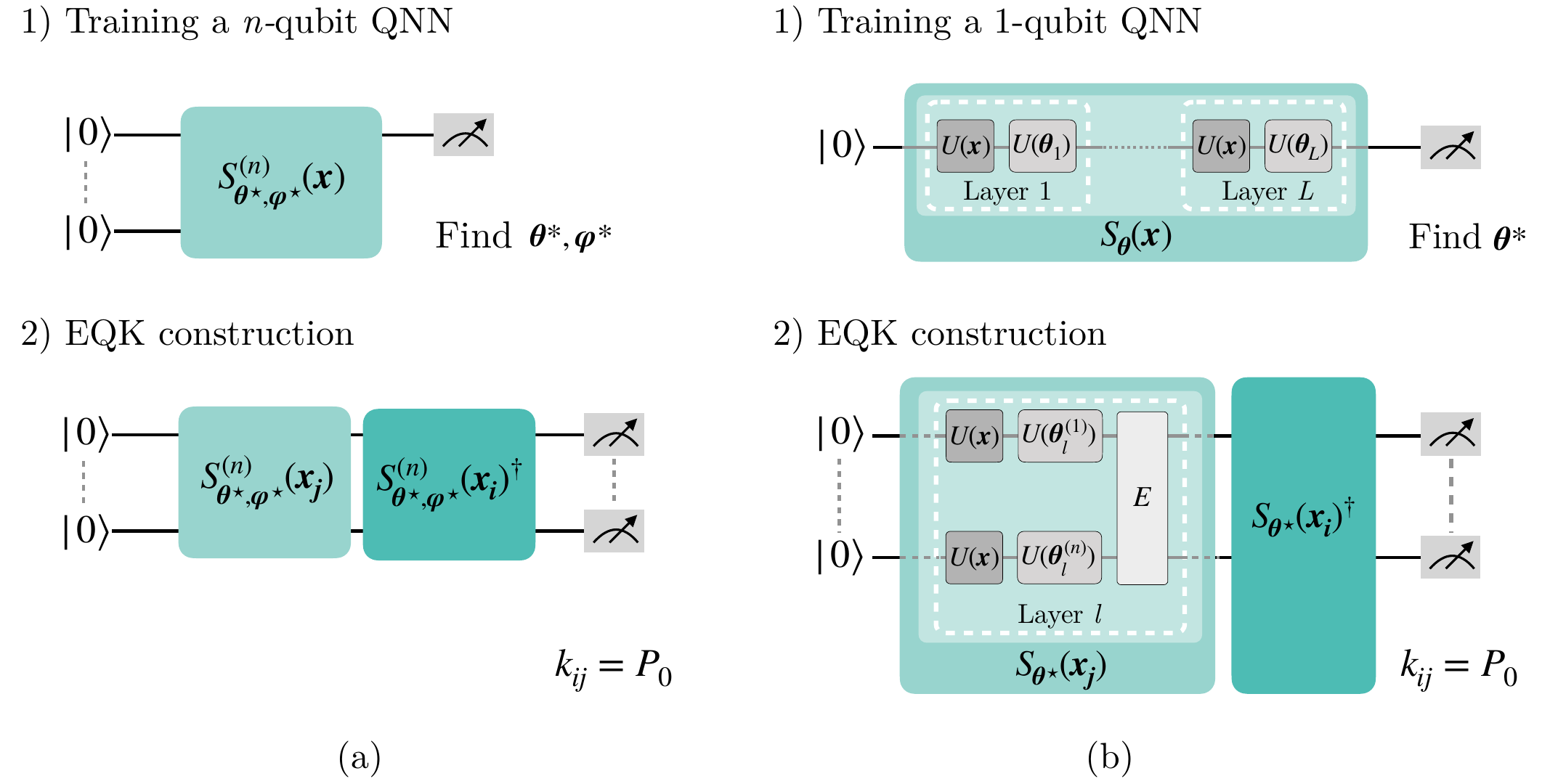}
  \caption{\justifying Schematic description of the two neural quantum kernel constructions considered in this work. (a) $n$-to-$n$: an $n$-qubit QNN is trained to obtain optimized embedding parameters $(\boldsymbol{\theta}^\star, \boldsymbol{\varphi}^\star)$, which are then frozen and used to construct the EQK by estimating $k_{ij}=P_0$, where $P_0$ denotes the probability of measuring the all-zero outcome after preparing $S^{(n)}_{\bm{\theta}^\star,\bm{\varphi}^\star}(\bm{x}_i)^{\dagger}\,S^{(n)}_{\bm{\theta}^\star,\bm{\varphi}^\star}(\bm{x}_j)\lvert 0\rangle$. (b) $1$-to-$n$: a single-qubit QNN is trained to obtain $\boldsymbol{\theta}^\star$, replicated across an $n$-qubit register, and interleaved at each re-uploading layer with a fixed entangling operation $E$; the resulting embedding is used to construct the EQK via the same overlap-estimation procedure.}
  \label{fig:nqk_constructions}
\end{figure*}

We consider two neural quantum kernel constructions: the $n\text{-to-}n$ and $1\text{-to-}n$ approaches, which are summarized in Fig. \ref{fig:nqk_constructions}. Both constructions define an EQK by (i) learning a task-adapted quantum feature map via QNN pretraining and (ii) freezing the resulting embedding to construct the EQK according to Eq. (\ref{eqn:eqk_definition}).

\paragraph{$n\text{-to-}n$ NQK.}
Following \cite{nqk_pablo}, the $n\text{-to-}n$ construction trains an $n$-qubit data re-uploading QNN using the progressive growth strategy described in Sec. \ref{sec:nqk_qnn}. After training saturates, the optimized embedding parameters $(\bm{\theta}^\star,\bm{\varphi}^\star)$ are fixed to define the feature map $S_{\bm{\theta}^\star,\bm{\varphi}^\star}(\bm{x})$, from which the EQK is constructed via Eq.~(\ref{eqn:eqk_definition}).

\paragraph{$1\text{-to-}n$ NQK.}
Also following \cite{nqk_pablo}, the $1\text{-to-}n$ construction first trains a single-qubit data re-uploading QNN to obtain optimal embedding parameters $\bm{\theta}^\star$. The learned single-subsystem unitary is then replicated across an $n$-subsystem register and interleaved with a fixed entangling operation, yielding the feature map
\begin{equation}
    S(\bm{x})
    \;=\;
    \prod_{\ell=1}^{L}
     E \; U(\bm{\theta}^\star)^{\otimes n}\;U(\bm{x})^{\otimes n},
    \label{eq:1toN_feature_map}
\end{equation}
where $U(\bm{x})$ denotes the data-encoding unitary applied to each qubit, $U(\bm{\theta}^\star)$ is the trained single-qubit unitary replicated across the register, and $E$ is a fixed entangling operation acting on the $n$-qubit Hilbert space. The corresponding EQK is then obtained from the feature states induced by $S(\bm{x})$.

While the constructions above are presented in the qubit setting, the NQK framework is not intrinsically restricted to two-level systems. In the remainder of this work, we extend neural quantum kernels to qudits of local dimension $d>2$ by generalizing the data re-uploading embedding and entangling architecture, thereby constructing task-adapted quantum kernels on $(\mathbb{C}^d)^{\otimes n}$.

\subsection{Extension of NQK to Qudits}
\label{sec:nqk_qudits}

We extend NQK from qubits to qudits motivated by two complementary considerations. First, qubit embeddings are built from local $\mathrm{SU}(2)$ operations, which have only three independent parameters; consequently, each local data-encoding block can incorporate at most three independent features per qubit. This imposes a structural bottleneck on feature encoding in qubit-based NQKs. In contrast, single-qudit operations belong to $\mathrm{SU}(d)$ and admit $d^2-1$ independent parameters, enabling richer local embeddings and allowing a larger number of features to be incorporated within each subsystem. 

Second, qudits provide a natural framework for multiclass learning: a $d$-level system offers $d$ mutually orthogonal basis states that can be directly associated with class labels. In particular, qutrits ($d=3$) enable a direct three-class prediction by measuring in the computational basis, aligning measurement outcomes with class hypotheses in a conceptually transparent manner.

Beyond these motivations, moving to qudits introduces additional design considerations, in particular the local dimension $d$, the parameterization of single-qudit $\mathrm{SU}(d)$ operations, the number of features that can be encoded locally, and the readout/loss structure used for multiclass tasks. These choices interact with architectural ingredients already present in NQK models, such as the number of subsystems, the entangling strategy, and the kernel construction. Understanding the role of these components is key for making informed design choices in qudit-based NQKs.

\subsubsection{Qudit preliminaries}
\label{sec:qudit_pre}

A \emph{qudit} is a generalization of the qubit to a $d$-dimensional Hilbert space $\mathcal{H}_d$. The computational basis is given by the set of orthonormal states \cite{balantekin-2024, nikolaeva-2024, wang-2020}
\begin{equation*}
    |0\rangle, |1\rangle, \dots, |d-1\rangle.
\end{equation*}

An arbitrary pure state can be expressed as
\begin{equation*}
    |\psi\rangle = \sum_{k=0}^{d-1} c_k |k\rangle,
\end{equation*}
where $c_k \in \mathbb{C}$ and $\sum_{k=0}^{d-1} |c_k|^2 = 1$.

Single-qudit unitary evolutions are described by operators $U \in \mathrm{SU}(d)$. Any such unitary can be written in exponential form as
\begin{equation*}
    U = \exp\;\left( i \sum_{j=1}^{d^2-1} \alpha_j \Lambda_j \right),
\end{equation*}
where $\{\Lambda_j\}_{j=1}^{d^2-1}$ are generators of the Lie algebra $\mathfrak{su}(d)$. In general, $\mathrm{SU}(d)$ has $d^2-1$ independent generators.

Although the framework is formulated for general $d$, in the experimental section we focus on qutrits ($d=3$). In this case we take $\{\lambda_k\}_{k=1}^{8}$ as a basis of $\mathfrak{su}(3)$ (e.g., the Gell--Mann matrices) and use the one-parameter generator rotations
\begin{equation}
    R_k(\varphi)\;:=\;\exp\!\left(i\,\varphi\,\lambda_k\right),
    \qquad k\in\{1,\ldots,8\}.
    \label{eq:Rk_def}
\end{equation}
We will also consider alternative parameterizations of $\mathrm{SU}(3)$ unitaries for constructing data embeddings and variational layers.

\subsubsection{$\mathrm{SU}(3)$ parameterizations}
\label{sec:qudit_parameterizations}

In the qutrit setting ($d=3$), we consider three alternative parameterizations of single-qutrit unitaries in $\mathrm{SU}(3)$, which we use to define both data-encoding maps and variational blocks. 

\paragraph{Geometric (Lie-algebra exponential).}
A direct parameterization is obtained from the exponential map of $\mathfrak{su}(3)$,
\begin{equation}
    U(\bm{\theta}) \;=\; \exp\!\left(i\sum_{j=1}^{8}\theta_j \lambda_j\right),
    \label{eq:su3_geometric}
\end{equation}
where $\{\lambda_j\}_{j=1}^{8}$ are the $\mathfrak{su}(3)$ generators \cite{mandilara2024classification}. This is the default parameterization used in most of our experiments.

\paragraph{Euler-angles decomposition.}
Alternatively, a general $\mathrm{SU}(3)$ unitary can be written as an ordered product of one-parameter generator rotations $R_k(\cdot)$ defined in Eq. (\ref{eq:Rk_def}) \cite{tilma-2002},
\begin{eqnarray}
    U(\bm{\theta}) \;&=&\; 
    R_3(\theta_8)\,R_2(\theta_7)\,R_3(\theta_6)\,R_5(\theta_5)\, \times \nonumber\\
    &\times& \, R_3(\theta_4)\,R_2(\theta_3)\,R_3(\theta_2)\,R_8(\theta_1),
    \label{eq:su3_euler}
\end{eqnarray}

\paragraph{Givens-rotations form.}
A third parameterization expresses a general $\mathrm{SU}(3)$ as a product of two-level (Givens) rotations acting on selected mode pairs, followed by a diagonal phase gate,
\begin{equation}
    U(\bm{\theta}) \;=\; G_{01}(\theta_7,\theta_8)\,G_{12}(\theta_5,\theta_6)\,G_{01}(\theta_3,\theta_4)\,D(\theta_1,\theta_2),
    \label{eq:su3_givens}
\end{equation}
where each $G_{ab}$ acts non-trivially only on the subspace spanned by $\{|a\rangle,|b\rangle\}$ and $D$ is diagonal (with unit determinant) \cite{vitanov-2012}.

In all cases, the resulting unitary serves as a single-qutrit building block inside the data re-uploading embedding and the variational layers. While these parameterizations are designed to capture general $\mathrm{SU}(3)$ unitaries, they induce different parameter landscapes, which can translate into different optimization behaviour, sensitivity to initialization, and numerical performance.

\subsubsection{Qutrit entangling primitives}
\label{sec:qutrit_entanglers}

To construct multiqutrit embeddings we combine local $\mathrm{SU}(3)$ blocks with two-qutrit entangling operations. In this work we consider two families of entangling primitives, which are used in different kernel constructions.

\paragraph{SUM (generalized C\!X) gate.}
A standard generalization of the qubit controlled-$X$ to qudits is the SUM gate, defined on the computational basis as \cite{pudda-2024}
\begin{equation}
    \mathrm{SUM}_d\,|c\rangle|t\rangle \;=\; |c\rangle\,|t \oplus c\rangle,
    \label{eq:sum_gate}
\end{equation}
where $\oplus$ denotes addition modulo $d$. In the qutrit case we write $\mathrm{SUM}_3$ and use it as a fixed entangling operation within the $1\text{-to-}n$ construction.

\paragraph{Controlled unitaries.}
For the $n\text{-to-}n$ construction we employ parametrized controlled unitaries acting between neighbouring qutrits. Given the single-qutrit rotation $R_k(\varphi)\in \mathrm{SU}(3)$ defined in Eq.~(\ref{eq:Rk_def}), its controlled version with control $c$ and target $t$ is \cite{de-souza-farias-2025}
\begin{equation}
    CR^{c\to t}_{k}(\varphi)
    \;=\;
    \sum_{m=0}^{d-1} |m\rangle\!\langle m|_{c} \otimes R_k(m\varphi)_{t},
    \label{eq:controlled_unitary_general}
\end{equation}
which reduces to the usual qubit controlled rotations when $d=2$ (with Pauli generators). In our implementation, these controlled rotations are arranged according to a nearest-neighbour connectivity pattern. For each neighbouring pair, the entangling block is instantiated as a fixed sequence of three controlled $X$ rotations generated by the Gell--Mann matrices $\lambda_1$, $\lambda_4$, and $\lambda_6$, i.e.,
$CR_{1}(\cdot)\,CR_{4}(\cdot)\,CR_{6}(\cdot)$ with trainable angles.

\subsubsection{From qubits to qutrits}
\label{sec:qudit_extension}

The extension from qubit-based NQKs to qutrits is conceptually straightforward since the overall data re-uploading architecture, the progressive growth strategy, and the subsequent kernel construction remain unchanged, while the circuit building blocks are promoted from local $\mathrm{SU}(2)$ operations to local $\mathrm{SU}(3)$ unitaries. Concretely, for an $n$-qutrit register we prepare feature states
\begin{equation}
    \rho_{\theta}^{(n)}(\bm{x})
    \;=\;
    S_{\theta}^{(n)}(\bm{x})\,\rho_0\,S_{\theta}^{(n)}(\bm{x})^{\dagger},
\end{equation}
where $S_{\theta}^{(n)}(\bm{x})$ is constructed from local $\mathrm{SU}(3)$ blocks and qutrit entangling gates. The encoded feature budget $p$ is implemented at the level of the single-qutrit data-encoding unitary as follows. Each input vector is written as $\bm{x}=(x_1,\ldots,x_p)$ after preprocessing, with $p\leq 8$ in the qutrit case. For the \textit{Geometric} parameterization, features are assigned sequentially to the $\mathrm{SU}(3)$ generators,
\begin{equation}
    U_p^{\mathrm{geom}}(\bm{x})
    =
    \exp\!\left(
    i\sum_{j=1}^{p} x_j \lambda_j
    \right),
    \label{eq:qutrit_feature_encoding_geom}
\end{equation}
where the remaining generator coefficients are set to zero when $p<8$. For the \textit{Euler-angles} and \textit{Givens-rotations} parameterizations, we use the same sequential convention at the level of the parameterization coordinates: the input features are assigned to the first $p$ parameters of Eqs. \eqref{eq:su3_euler} and \eqref{eq:su3_givens}, respectively, while the remaining coordinates are set to zero. Unlike the \textit{Geometric} parameterization, where all encoded features enter symmetrically through a single exponential of $\mathrm{SU}(3)$ generators, the \textit{Euler-angles} and \textit{Givens-rotations} decompositions assign different structural roles to different coordinates, such as phases or two-level rotations. As a result, the sequential feature assignment may introduce a parameterization-dependent inductive bias, an effect that we examine empirically in Sec. \ref{sec:results_parameterization}. In all parameterizations, the same data-encoding map $U_p(\bm{x})$ is applied identically at each re-uploading layer and on each qutrit of the register.

The embedding quantum kernel is then defined as in Eq. (\ref{eqn:eqk_definition}), with $\rho(\bm{x})$ acting on $(\mathbb{C}^3)^{\otimes n}$.

In this work we consider two classification settings: (i) binary and (ii) three-class, and employ distinct cost functions and readout rules for each.

\paragraph{Binary classification}
For binary tasks, we read out the first qutrit using the spin-1 operator
\begin{equation}
    S_z \;=\; \mathrm{diag}(1,\,0,\,-1),
    \label{eq:sz_qutrit}
\end{equation}
and define the model output as its expectation value
\begin{equation}
    \langle S_z \rangle_{\bm{x}}
    \;:=\;
    \mathrm{tr}\!\left[\left(S_z \otimes \mathds{1}^{\otimes(n-1)}\right)\rho_{\theta}^{(n)}(\bm{x})\right].
    \label{eq:sz_expectation_global}
\end{equation}
Given labels $y_j\in\{-1,+1\}$, the parameters are trained by minimizing a mean-squared error loss,
\begin{equation}
    f_{\mathrm{cost}}(\theta)
    \;=\;
    \frac{1}{M}\sum_{j=1}^{M}
    \left(
    \langle S_z \rangle_{\bm{x}_j} - y_j
    \right)^2.
    \label{eq:binary_mse}
\end{equation}
At inference time, predictions are obtained via a threshold at zero,
\begin{equation}
    \hat{y}(\bm{x})
    \;=\;
    \begin{cases}
    +1, & \text{if } \langle S_z \rangle_{\bm{x}} \ge 0,\\[2pt]
    -1, & \text{if } \langle S_z \rangle_{\bm{x}} < 0.
    \end{cases}
    \label{eq:binary_threshold}
\end{equation}

\paragraph{Three-class classification}
For three-class tasks with labels $y_j\in\{0,1,2\}$, we measure the first qutrit in the computational basis and define class probabilities from the corresponding outcome projectors. Specifically, for $k\in\{0,1,2\}$ we set
\begin{equation}
    p_k(\bm{x})
    \;:=\;
    \mathrm{tr}\!\left[
    \left(|k\rangle\!\langle k| \otimes \mathds{1}^{\otimes(n-1)}\right)
    \rho_{\theta}^{(n)}(\bm{x})
    \right],
    \label{eq:qutrit_probs}
\end{equation}
which satisfies $\sum_{k=0}^{2} p_k(\bm{x})=1$. The QNN parameters are trained by minimizing the multiclass cross-entropy loss
\begin{equation}
    f_{\mathrm{cost}}(\theta)
    \;=\;
    -\frac{1}{M}\sum_{j=1}^{M}\log p_{y_j}(\bm{x}_j).
    \label{eq:qutrit_cross_entropy}
\end{equation}
While we use cross-entropy for the main three-class experiments, we also explore alternative cost functions in Appendix \ref{sec:appendix_cost_functions}; in our experiments, the corresponding kernel construction appears less sensitive to this choice than the standalone 1-qutrit QNN.

\section{Results}
\label{sec:results}

This section reports our main results and provides a systematic study of qutrit-based Neural Quantum Kernels under ideal simulation. Guided by the model choices summarized in Table \ref{tab:intro_ingredients}, we vary selected components of the construction, namely the number of subsystems $n$, the encoded feature budget $p$, the kernel construction, the single-qutrit $\mathrm{SU}(3)$ parameterization, and the task-dependent readout and cost function. Unless stated otherwise, we focus on qutrits ($d=3$), use a data re-uploading architecture with $L=6$ layers, and keep the entangling ansatz and connectivity fixed as specified in Sec. \ref{sec:nqk_qudits}.

We begin by describing the datasets and the preprocessing/evaluation protocol. We then study feature scaling in the $1$-qutrit QNN by varying the number of encoded features. Next, we analyze the $1$-to-$n$ construction by first fixing the feature budget and varying the number of qutrits, and then fixing the system size and varying the number of encoded features. We repeat the same two-step scaling analysis for the $n$-to-$n$ construction. Finally, we present an ablation study on the impact of different $\mathrm{SU}(3)$ parameterizations.

\subsection{Datasets}
\label{sec:datasets}

We evaluate our models on four widely used benchmarks spanning heterogeneous data modalities. In the main text, we focus on Fashion-MNIST, a grayscale image dataset commonly used for lightweight vision benchmarks \cite{fmnist, fmnist_github}. To avoid redundancy, we report only these results in the main body, as they are representative of the behaviour observed across all datasets. Additional experiments on HAR, MAGIC Gamma Telescope, and Covertype are provided in Appendix \ref{sec:appendix_datasets_results}, and dataset access details are summarized in Appendix \ref{sec:appendix_datasets_description}.

\subsection{Preprocessing, training, and evaluation protocol}
\label{sec:preprocessing_protocol}

All datasets are preprocessed in a unified manner to produce feature vectors compatible with the data-encoding unitary of the QNN embedding (Sec. \ref{sec:nqk_qnn}). Each sample is first represented as a one-dimensional real-valued feature vector (by flattening the pixel array for image data). The features are then standardized using \texttt{StandardScaler}, PCA is applied to control the number of encoded features, and the resulting components are rescaled to the interval $[-1,1]$ prior to embedding.

Model evaluation is performed using stratified $K$-fold cross-validation with $K=5$. For each fold, the \texttt{StandardScaler} and PCA are fitted exclusively on the training split and subsequently applied to the corresponding test split, ensuring that no information from the test data leaks into preprocessing. 

To ensure balanced class distributions and comparable computational budgets across benchmarks, we subsample each task to a fixed number of examples per class. For binary classification tasks we use $2000$ samples in total ($1000$ per class), and for three-class tasks we use $3000$ samples in total ($1000$ per class). All results are obtained from ideal circuit simulations. Unless otherwise stated, performance is reported in terms of classification accuracy, averaged across the five folds.

For reproducibility, we also specify the main training and classifier hyperparameters. The QNN pretraining was implemented in PyTorch using the Adam optimizer with weight decay. For the single-qutrit models, we used learning rate $\mathrm{LR}=0.006$, weight decay $\mathrm{WD}=10^{-5}$, batch size $64$, and trained for $70$ epochs. Since the optimization is sensitive to parameter initialization, for each fold and model configuration we performed $7000$ random initializations and retained the run achieving the lowest training cost.

For the $n>1$ QNNs used in the $n$-to-$n$ construction, training was initialized through the progressive growth procedure described in Sec. \ref{sec:nqk_qnn},, starting from the optimized solution at the previous system size. Since each growth step corresponds to a fine-tuning stage around an already trained model, we used a smaller learning rate $\mathrm{LR}=0.002$ and trained for $20$ epochs.

The final NQK classifier was implemented using \texttt{sklearn.svm.SVC} with a precomputed kernel matrix, i.e., \texttt{SVC(kernel="precomputed")}. The regularization parameter $C$ was selected by grid search over $C\in\{0.01,0.1,1,10,100\}$. For three-class tasks, we used the default one-vs-one multiclass strategy implemented by \texttt{sklearn.svm.SVC}. As a classical baseline, we also trained an RBF-kernel SVM using the same preprocessing and data splits. For this baseline, we performed a grid search over $C\in\{0.01,0.1,1,10,100\}$ and $\gamma\in\{\texttt{scale},0.001,0.01,0.1,1\}$.

\subsection{Feature scaling in a $1$-qutrit QNN}
\label{sec:results_features_1qutrit}

We first investigate feature scaling in a pretrained $1$-qutrit QNN on Fashion-MNIST by varying the number of encoded features $p$ while keeping the remaining architectural choices fixed. In this experiment we use the \textit{Geometric} $\mathrm{SU}(3)$ parameterization (Eq. \ref{eq:su3_geometric}). Binary models are trained using the $S_z$-based mean-squared error cost function, whereas three-class models are trained using cross-entropy over computational-basis probabilities (see Appendix \ref{sec:appendix_cost_functions} for a sensitivity study to alternative cost functions). To mitigate sensitivity to parameter initialization, for each configuration we perform multiple random restarts and retain the run achieving the lowest training loss.
\begin{figure}[t]
    \centering
    \includegraphics[width=0.94\linewidth]{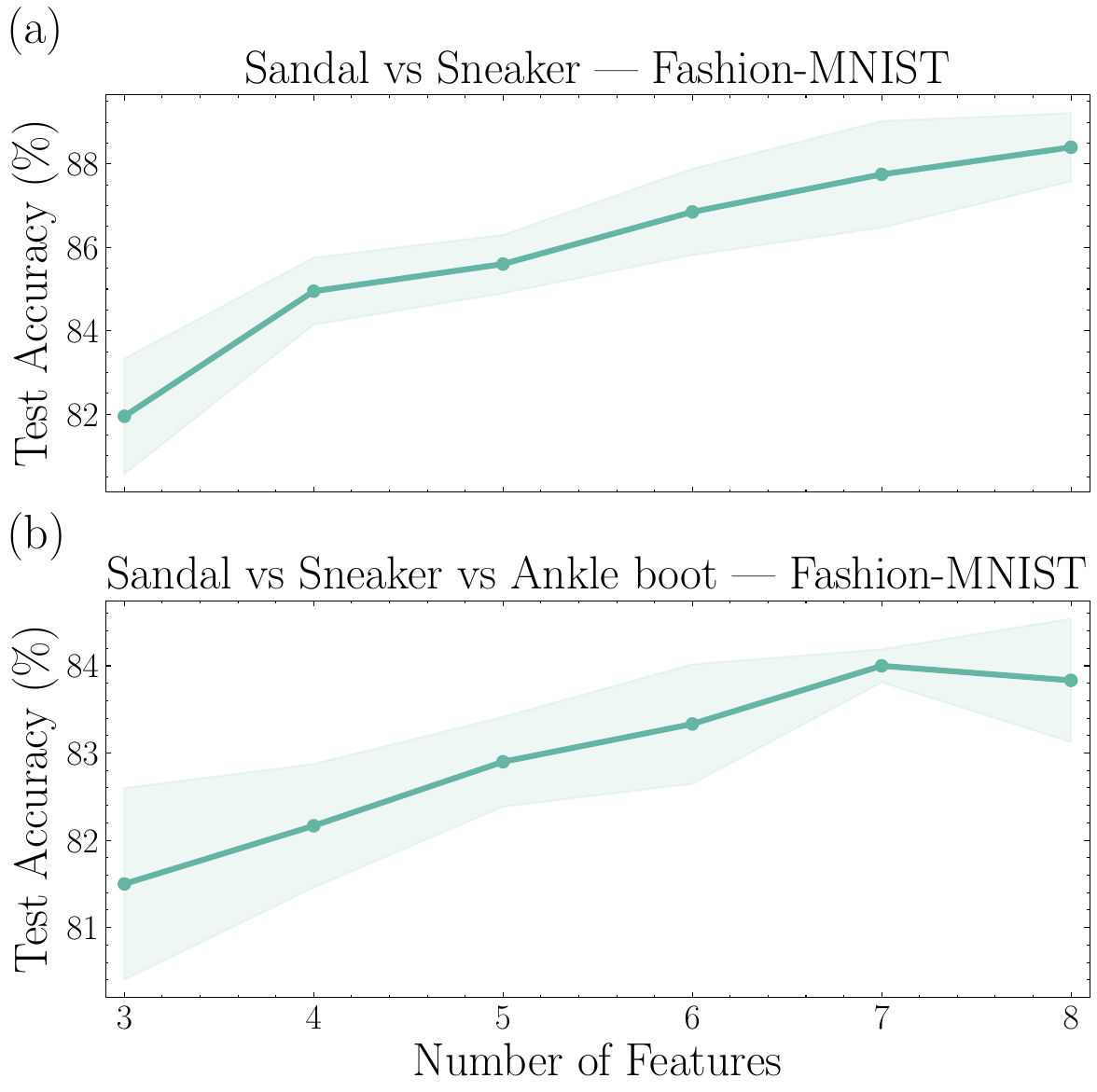}
    \caption{\justifying Test accuracy of the $1$-qutrit QNN on Fashion-MNIST as a function of the number of encoded features (mean $\pm$ standard error over 5 stratified folds). (a) Binary task (classes \emph{Sandal} vs. \emph{Sneaker}). (b) Three-class task (classes \emph{Sandal} vs. \emph{Sneaker} vs. \emph{Ankle boot}).}
    \label{fig:fmnist_acc_vs_features}
\end{figure}

Figure \ref{fig:fmnist_acc_vs_features} shows how test accuracy varies with the number of encoded features under this fixed configuration. Accuracy generally improves as $p$ increases, with gains that tend to saturate at larger feature counts. This behaviour is consistent with the motivation for qutrit embeddings, namely that the larger local parameter space of $\mathrm{SU}(3)$ enables richer local transformations and allows additional encoded features to be incorporated within the same data re-uploading architecture. In additional benchmarks (Appendix \ref{sec:appendix_datasets_results}), saturation often occurs already around $p\simeq 5$--$6$, for example in the HAR and MAGIC datasets, whereas for Covertype the dependence is non-monotonic, with accuracy peaking at intermediate feature counts, indicating that the effective feature budget is dataset-dependent. Importantly, the strength and monotonicity of feature scaling depend on the dataset and on the choice of $\mathrm{SU}(3)$ parameterization, as further illustrated by the parameterization ablation in Sec. \ref{sec:results_parameterization}.

\subsection{1-to-n NQK}
\label{sec:results_1ton_scaling}

We next evaluate the $1$-to-$n$ qutrit NQK construction, which reuses a trained $1$-qutrit embedding and lifts it to larger system sizes by replication across the register together with a fixed entangling operation (Sec. \ref{sec:nqk_kcons}). In this section we study two complementary scaling behaviours. We first fix the feature budget and vary the number of qutrits to assess system-size scaling, and we then fix the system size and vary the number of encoded features to assess feature scaling within the induced kernel model.

\begin{figure}[t]
    \centering
    \includegraphics[width=0.94\linewidth]{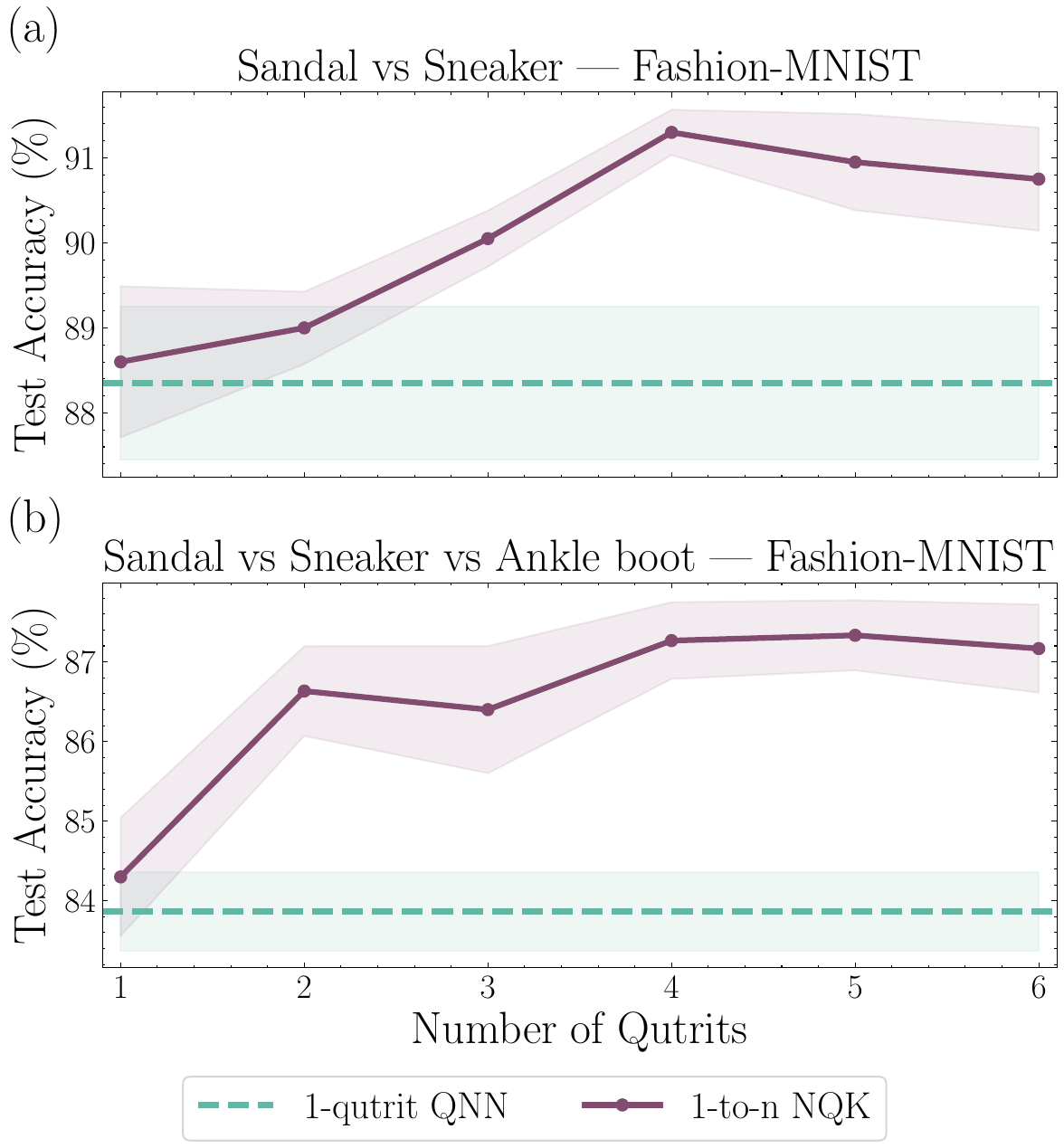}
    \caption{\justifying Test accuracy on Fashion-MNIST as a function of the number of qutrits for the 1-to-$n$ qutrit NQK (mean $\pm$ standard error over 5 stratified folds), with the number of encoded features fixed to $p=8$. The dashed horizontal line shows the corresponding $1$-qutrit QNN baseline used to construct the embedding. (a) Binary task (classes \emph{Sandal} vs.\ \emph{Sneaker}). (b) Three-class task (classes \emph{Sandal} vs.\ \emph{Sneaker} vs.\ \emph{Ankle boot}).}
    \label{fig:fmnist_1ton_acc_vs_qutrits}
\end{figure}

Figure \ref{fig:fmnist_1ton_acc_vs_qutrits} reports mean test accuracy over 5 stratified folds as a function of the number of qutrits for both the binary and three-class Fashion-MNIST tasks, while fixing the number of encoded features to $p=8$. In both settings, the $1$-to-$n$ NQK improves over the $1$-qutrit QNN baseline across the explored system sizes. Accuracy increases from small registers up to approximately $n=4$, after which the performance stabilizes, with only marginal changes for larger registers. In the binary task, the curve slightly decreases beyond $n=4$, whereas in the three-class task it remains approximately saturated. 

\begin{figure}[t]
    \centering
    \includegraphics[width=0.94\linewidth]{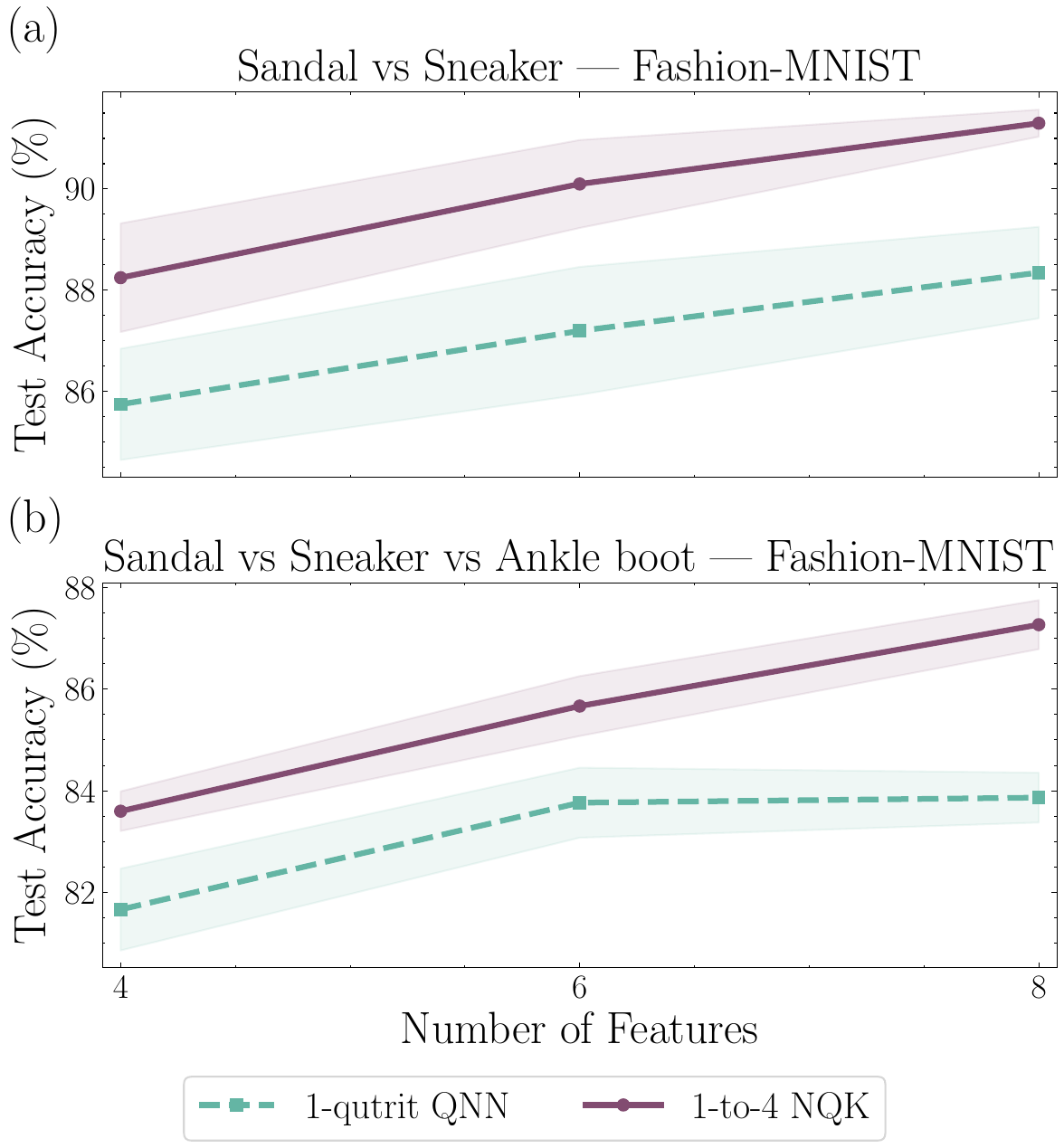}
    \caption{\justifying Test accuracy on Fashion-MNIST as a function of the number of encoded features at fixed system size $n=4$. The $1$-to-$n$ NQK (4 qutrits) is compared against the corresponding $1$-qutrit QNN baseline (mean $\pm$ standard error over 5 stratified folds). (a) Binary task (classes \emph{Sandal} vs.\ \emph{Sneaker}). (b) Three-class task (classes \emph{Sandal} vs.\ \emph{Sneaker} vs.\ \emph{Ankle boot}).}
    \label{fig:fmnist_1ton_acc_vs_features}
\end{figure}

We next fix the system size to $n=4$ qutrits and vary the number of encoded features. Figure \ref{fig:fmnist_1ton_acc_vs_features} shows that the $1$-to-$4$ kernel model consistently outperforms the underlying $1$-qutrit QNN across the explored feature budgets. In the binary task, both models improve with $p$ with a broadly similar trend, while the kernel model maintains a higher accuracy throughout. In the three-class task, the improvement with $p$ is more pronounced for the $1$-to-$4$ NQK, whereas the $1$-qutrit QNN shows a weaker increase and tends to saturate. This suggests that, when the learned $1$-qutrit embedding is used to construct a larger entangled register, the resulting kernel model can exploit additional encoded features at least as effectively as the single-qutrit variational classifier.

\subsection{n-to-n NQK}
\label{sec:results_nton_scaling}

\begin{figure}[t]
    \centering
    \includegraphics[width=0.94\linewidth]{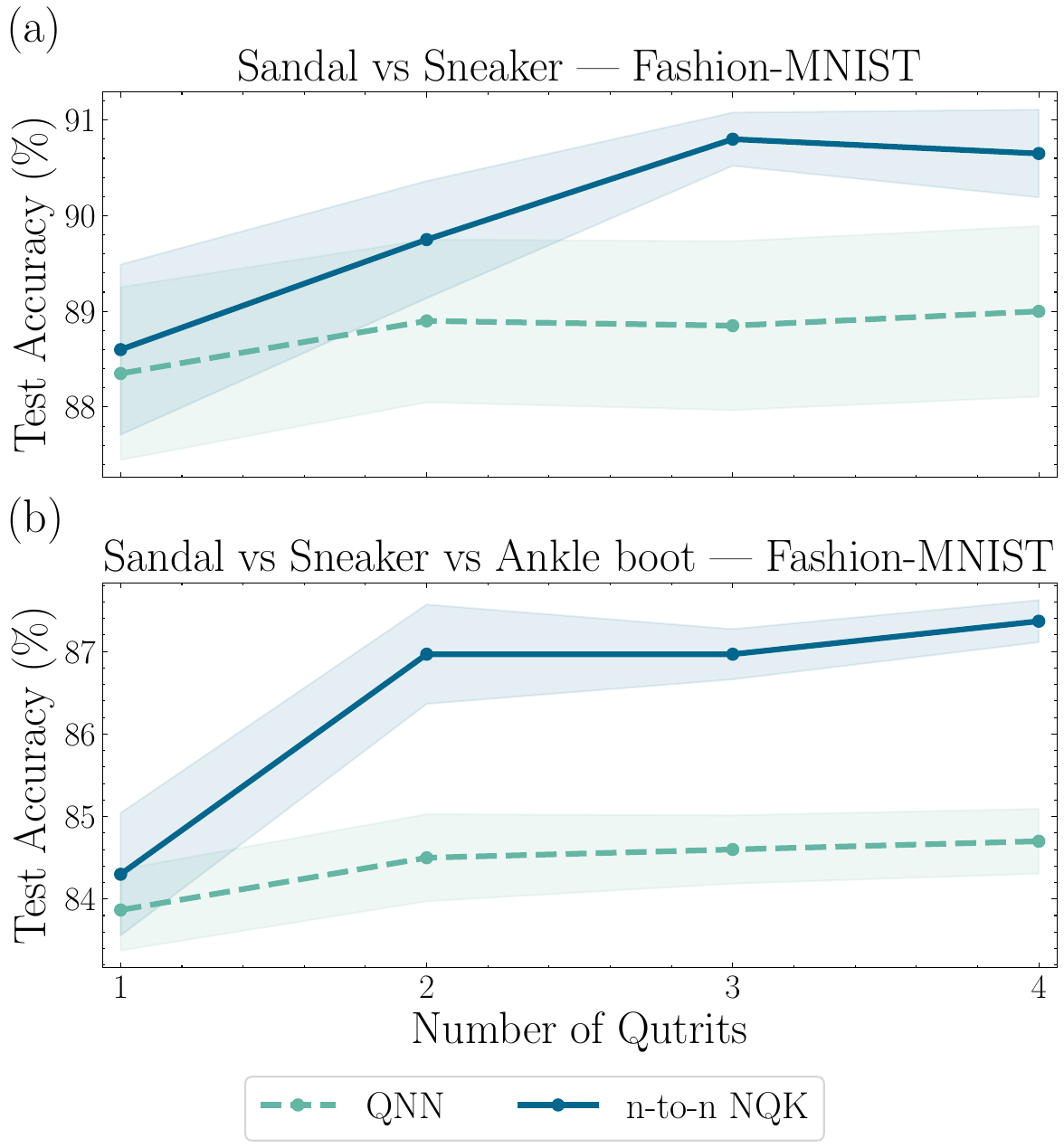}
    \caption{\justifying Test accuracy on Fashion-MNIST as a function of the number of qutrits for the $n$-to-$n$ qutrit NQK (mean $\pm$ standard error over 5 stratified folds), with the number of encoded features fixed to $p=8$. The dashed curve reports the corresponding $n$-qutrit QNN baseline trained for each system size, enabling a direct comparison between the kernel model and the variational classifier at matched $n$. (a) Binary task (classes \emph{Sandal} vs.\ \emph{Sneaker}). (b) Three-class task (classes \emph{Sandal} vs.\ \emph{Sneaker} vs.\ \emph{Ankle boot}).}
    \label{fig:fmnist_nton_acc_vs_qutrits}
\end{figure}

We next evaluate the $n$-to-$n$ qutrit NQK construction. The underlying $n$-qutrit data re-uploading QNN is trained using the progressive growth strategy described in Sec. \ref{sec:nqk_qnn}, and the resulting optimized parameters are fixed to construct the corresponding EQK (Sec. \ref{sec:nqk_kcons}). We consider two complementary scaling analyses: (i) system-size scaling, where we fix the feature budget to $p=8$ and vary the number of qutrits $n$, and (ii) feature scaling at fixed system size, where we fix $n=4$ and vary the number of encoded features $p$. Figure \ref{fig:fmnist_nton_acc_vs_qutrits} reports test accuracy (mean and standard error over 5 stratified folds) as a function of $n\in{1,2,3,4}$ for both the binary and three-class Fashion-MNIST tasks. We restrict to $n\le 4$ since, for the present architecture and feature budget, the improvement from $n=3$ to $n=4$ is already marginal.

In the Fashion-MNIST tasks considered here, accuracy improves as additional qutrits are introduced. Moreover, for these tasks, the $n$-to-$n$ NQK consistently outperforms the corresponding $n$-qutrit QNN baseline across all system sizes considered. Overall, these results indicate that, on Fashion-MNIST, jointly training the multi-qutrit embedding and then constructing the kernel yields a more effective classifier than the variational QNN at matched $n$, while improvements gradually plateau as the register grows under the fixed feature budget $p=8$.

\begin{figure}[t]
    \centering
    \includegraphics[width=0.94\linewidth]{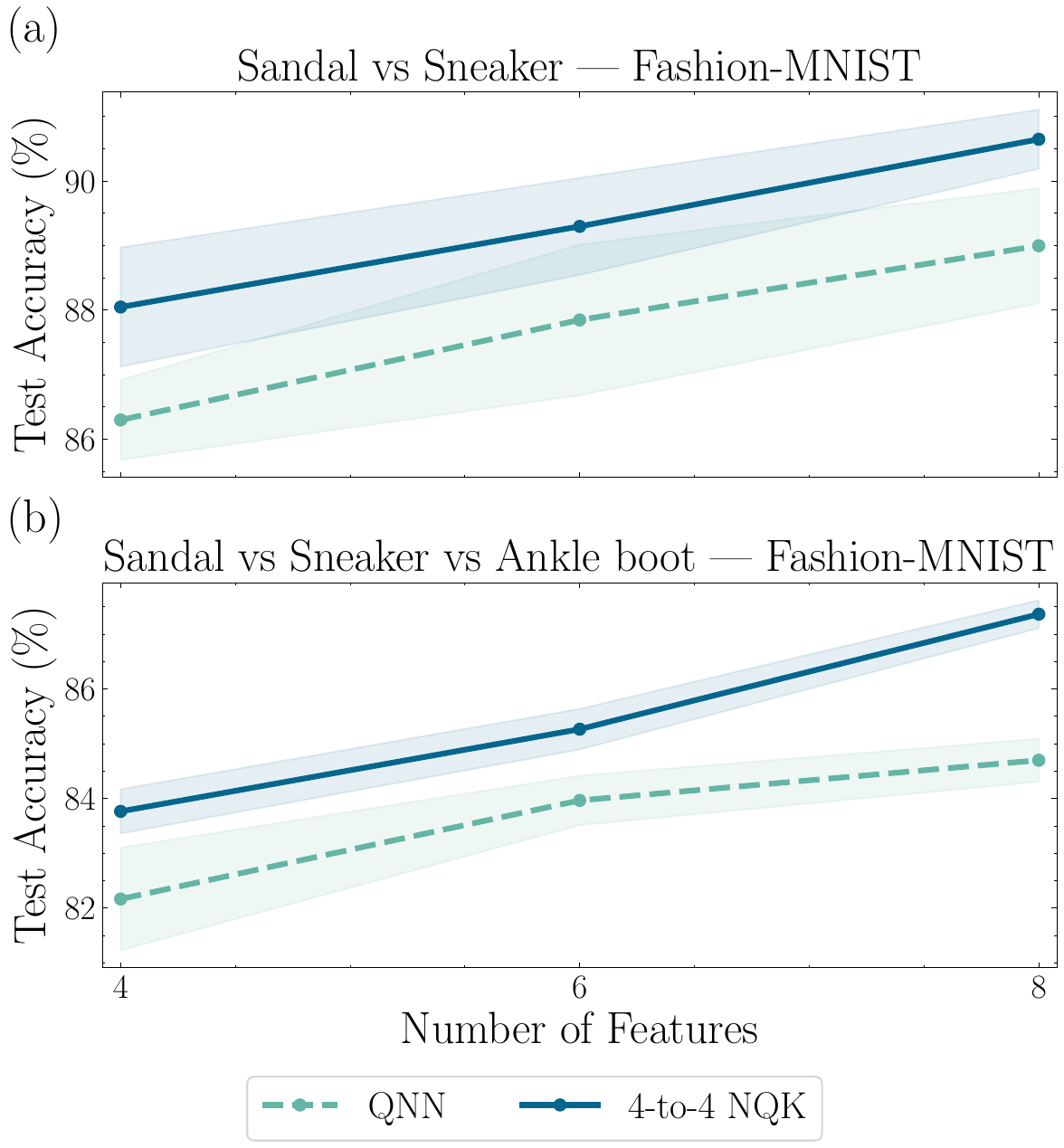}
        \caption{\justifying Test accuracy on Fashion-MNIST as a function of the number of encoded features at fixed system size $n=4$ for the $n$-to-$n$ qutrit NQK (mean $\pm$ standard error over 5 stratified folds). The dashed curve shows the matched 4-qutrit QNN baseline. (a) Binary task (classes \emph{Sandal} vs.\ \emph{Sneaker}). (b) Three-class task (classes \emph{Sandal} vs.\ \emph{Sneaker} vs.\ \emph{Ankle boot}).}
    \label{fig:fmnist_nton_acc_vs_features}
\end{figure}

We also examine feature scaling within the $n$-to-$n$ construction by fixing the system size to $n=4$ and varying the number of encoded features. Figure \ref{fig:fmnist_nton_acc_vs_features} shows that, for Fashion-MNIST at fixed $n$, increasing $p$ improves test accuracy for both the $4$-qutrit QNN and the corresponding $4$-to-$4$ NQK. In the binary task, both models display similar feature-scaling trends, with the NQK remaining consistently above the QNN baseline across the explored feature budgets. In the three-class task, the NQK also remains above the QNN baseline and exhibits a slightly stronger improvement as $p$ increases. This indicates that, once a multi-qutrit embedding is trained, the induced kernel model provides a consistently stronger classifier than the matched variational QNN, while additional encoded features benefit both models.

Table \ref{tab:fmnist_model_comparison_n4} summarizes the performance of the QNN baseline, the two NQK constructions, and a classical RBF-SVM baseline at fixed system size $n=4$ and feature budget $p=8$. In both the binary and three-class Fashion-MNIST tasks, the quantum kernel models improve over the corresponding QNN baseline. The $1$-to-$4$ and $4$-to-$4$ NQK constructions achieve very similar accuracies in both tasks, with differences that are small compared with the reported statistical uncertainty. Compared with the classical RBF-SVM, both NQK models remain competitive: the RBF-SVM has a slightly higher mean accuracy, but the differences are again within the reported uncertainty.
\begin{table}[t]
    \centering
    \caption{\justifying
    Test accuracy (\%) on Fashion-MNIST at $n=4$ qutrits and feature budget $p=8$ (mean $\pm$ standard error over 5 folds).}
    \label{tab:fmnist_model_comparison_n4}
    \small
    \setlength{\tabcolsep}{6pt}
    \renewcommand{\arraystretch}{1.2}

    \begin{tabular}{lcccc}
    \Xhline{1.2pt}
    \textbf{Task} & \textbf{QNN} & \textbf{1-to-$4$} & \textbf{$4$-to-$4$} & \textbf{RBF-SVM}\\
    \Xhline{1.2pt}
    Binary   & 89.0$\pm$0.9 & 90.7$\pm$0.6 & 90.7$\pm$0.6& 91.2$\pm$1.3 \\
    \Xhline{0.2pt}
    3-class  & 84.7$\pm$0.4 & 87.2$\pm$0.5 & 87.4$\pm$0.3 & 87.6$\pm$0.9 \\
    \Xhline{1.2pt}
    \end{tabular}
\end{table}

Notably, despite their similar performance at $n=4$, the two NQK constructions allocate trainable degrees of freedom differently. In the $4$-to-$4$ setting, the local single-qutrit unitaries, independent for each qutrit, and the nearest-neighbour entangling parameters are optimized jointly. In the $1$-to-$4$ construction, a single-qutrit feature map is replicated across the register and interleaved with a fixed nearest-neighbour entangling layer. The fact that both constructions achieve comparable performance suggests that the simpler $1$-to-$4$ strategy can remain competitive at this system size, even though the $4$-to-$4$ model has more trainable flexibility.

\subsubsection{Effect of $\mathrm{SU}(3)$ parameterization}
\label{sec:results_parameterization}

We next examine how the choice of single-qutrit $\mathrm{SU}(3)$ parameterization impacts performance. We compare the \textit{Geometric} (Lie-algebra exponential) form (Eq. \ref{eq:su3_geometric}), the \textit{Euler-angles} decomposition (Eq. \ref{eq:su3_euler}), and the \textit{Givens-rotation} form (Eq. \ref{eq:su3_givens}), keeping the rest of the experimental setup unchanged.

\begin{figure}[t]
    \centering
    \includegraphics[width=0.94\linewidth]{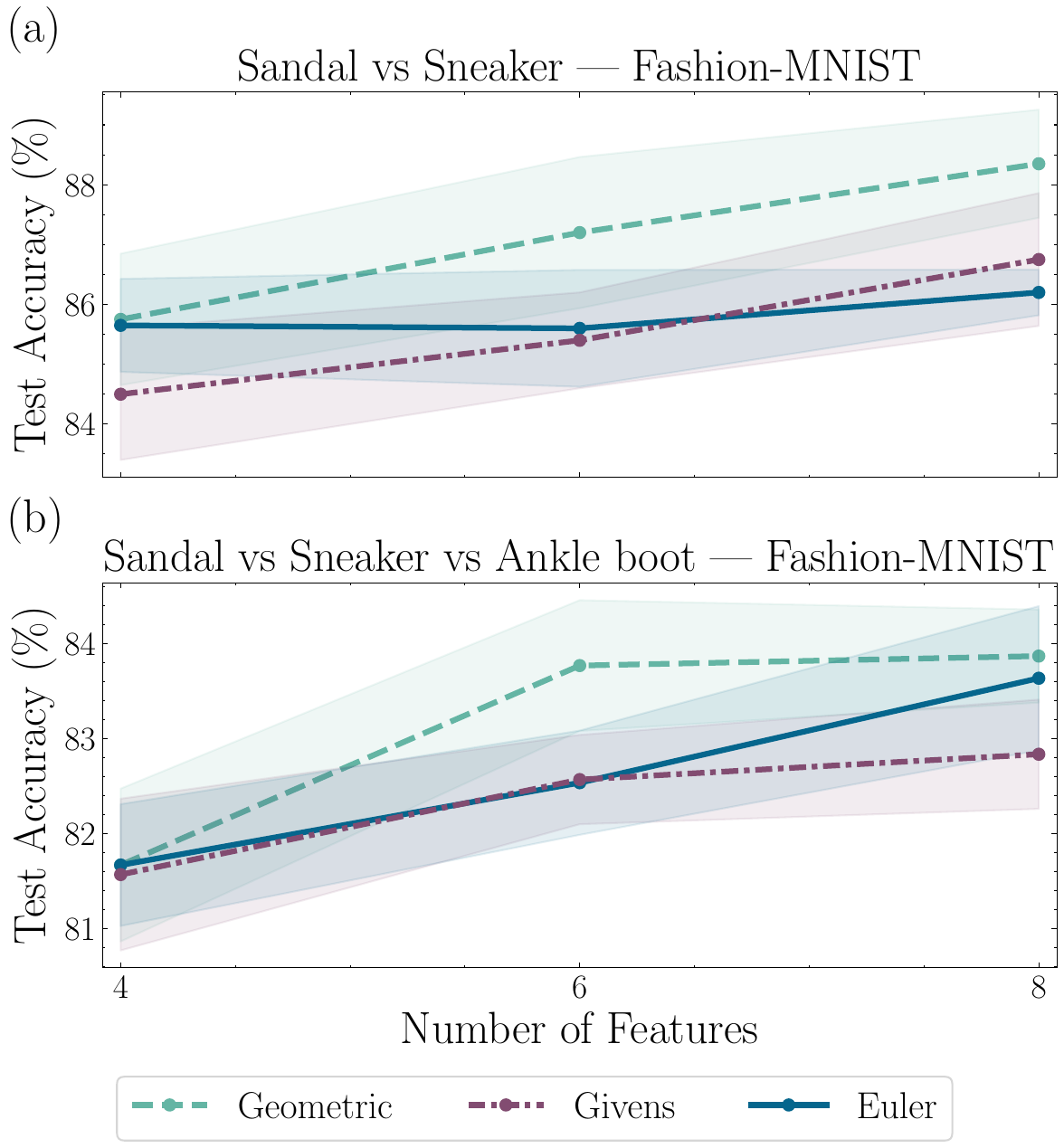}
    \caption{\justifying Effect of $\mathrm{SU}(3)$ parameterization on the $1$-qutrit QNN. Test accuracy on Fashion-MNIST as a function of the number of encoded features for the \textit{Geometric} parameterization, \textit{Givens-rotations} and \textit{Euler-angles} parameterization. Curves show the mean test accuracy over 5 stratified folds; shaded regions indicate the standard error. (a) Binary task (classes \emph{Sandal} vs. \emph{Sneaker}). (b) Three-class task (classes \emph{Sandal} vs. \emph{Sneaker} vs. \emph{Ankle boot}). }
    \label{fig:param_1qutrit_features}
\end{figure}

This choice affects performance already at the level of a $1$-qutrit QNN, as shown in Fig.~\ref{fig:param_1qutrit_features}. The figure reports test accuracy as a function of the number of encoded features. While accuracy increases with $p$ in both tasks, the improvement is not uniform across parameterizations. The \textit{Geometric} parameterization consistently achieves the best accuracy in the explored range and benefits most from increasing $p$, particularly for three-class classification, whereas \textit{Euler-angles} and \textit{Givens-rotations} exhibit smaller gains. To better understand these differences, we also tracked training diagnostics under identical data and optimizer settings. As shown in Appendix \ref{sec:appendix_param_diagnostics}, the \textit{Geometric} parameterization typically reaches lower training loss and maintains larger gradient norms than the \textit{Givens-rotations} parameterization, suggesting more favorable optimization dynamics in this setting.

\begin{figure}[t]
    \centering
    \includegraphics[width=0.94\linewidth]{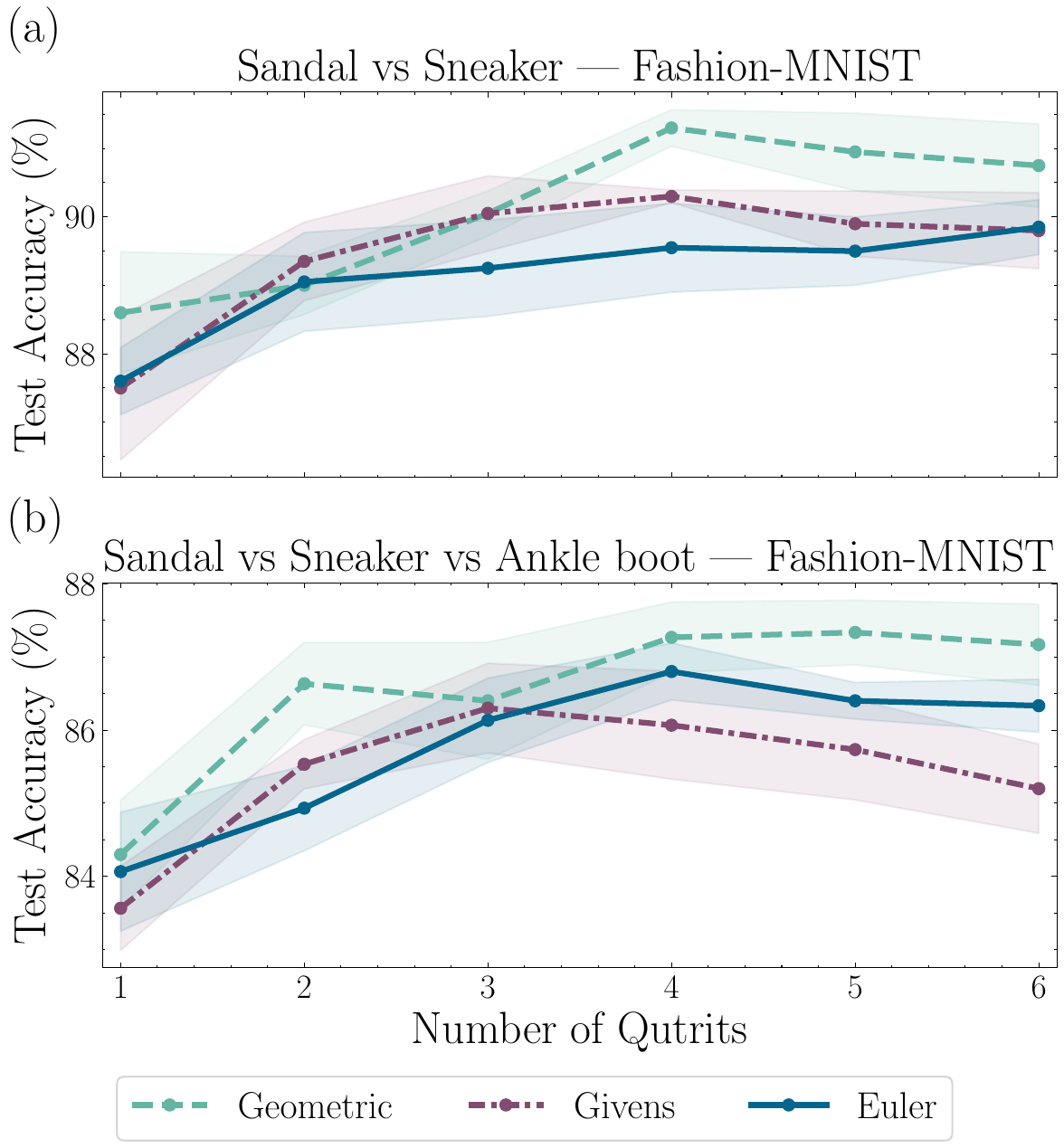}
    \caption{\justifying Effect of $\mathrm{SU}(3)$ parameterization on the $1\text{-to-}n$ NQK construction. Test accuracy on Fashion-MNIST as a function of the number of qutrits for the \textit{Geometric}, \textit{Givens-rotations}, and \textit{Euler-angles} parameterizations (feature budget fixed to $p=8$). Curves show the mean test accuracy over 5 stratified folds; shaded regions indicate the standard error. (a) Binary task (classes \emph{Sandal} vs. \emph{Sneaker}). (b) Three-class task (classes \emph{Sandal} vs. \emph{Sneaker} vs. \emph{Ankle boot}). }
    \label{fig:param_1ton_qutrits}
\end{figure}

The same dependence carries over to the kernel construction. Fixing the feature budget to $p=8$, Fig. \ref{fig:param_1ton_qutrits} reports the scaling with the number of qutrits in the $1$-to-$n$ NQK. In both the binary and three-class tasks, the three parameterizations generally improve as the system size increases, although with different rates and final performance levels. The \textit{Geometric} parameterization yields the strongest overall performance across the explored range, while \textit{Euler-angles} and \textit{Givens-rotations} remain below it. These results show that the parameterization affects not only the absolute accuracy, but also how effectively the kernel construction benefits from increasing the number of qutrits.

\begin{figure}[t]
    \centering
    \includegraphics[width=0.94\linewidth]{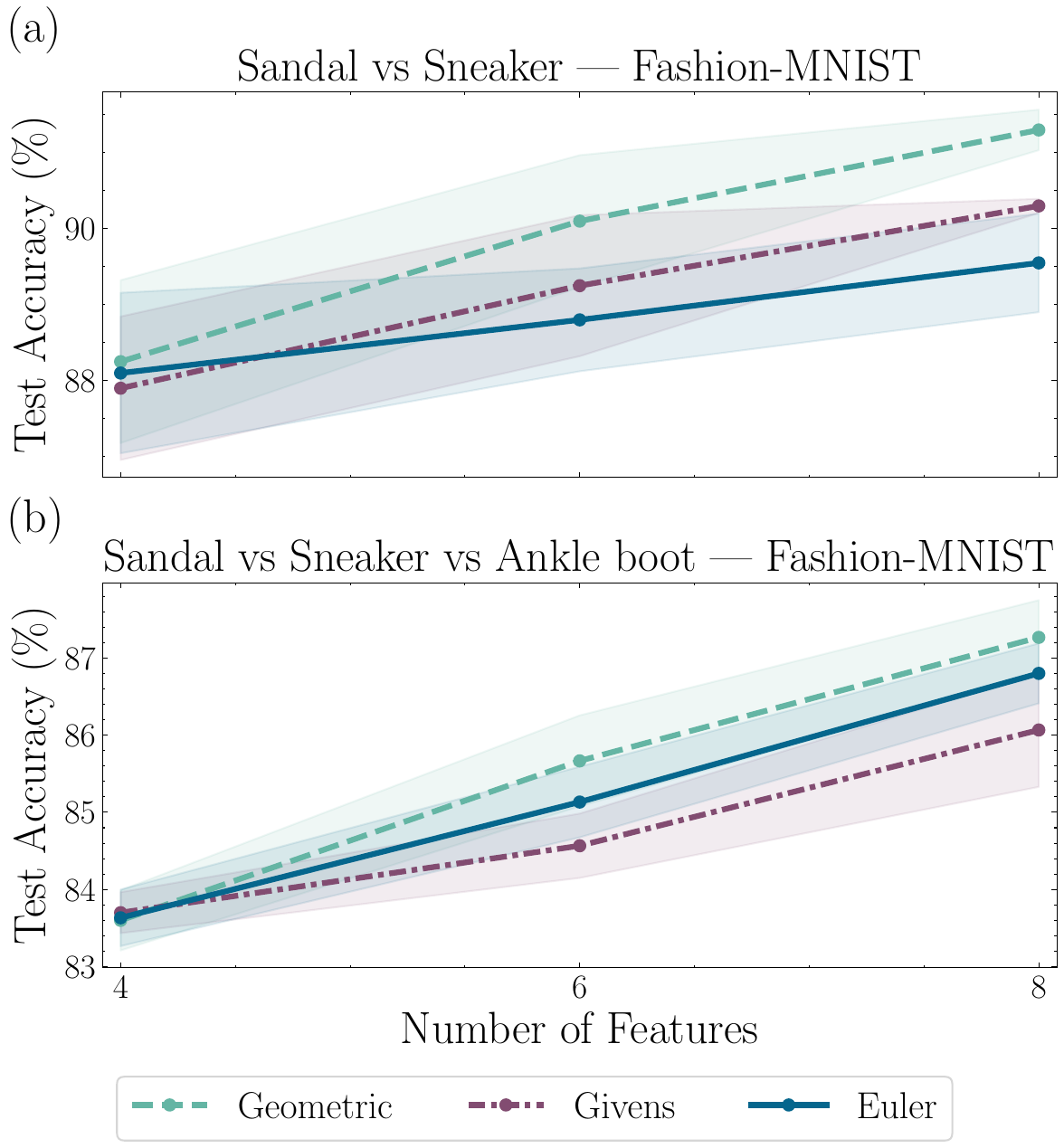}
    \caption{\justifying Effect of $\mathrm{SU}(3)$ parameterization on feature scaling in the $1$-to-$n$ NQK at fixed system size $n=4$. Test accuracy on Fashion-MNIST as a function of the number of encoded features for the \textit{Geometric}, \textit{Euler-angles}, and \textit{Givens-rotations} parameterizations (mean $\pm$ standard error over 5 stratified folds). (a) Binary task (classes \emph{Sandal} vs.\ \emph{Sneaker}). (b) Three-class task (classes \emph{Sandal} vs.\ \emph{Sneaker} vs.\ \emph{Ankle boot}).}
    \label{fig:param_1ton_features}
\end{figure}

Complementarily, fixing the system size to $n=4$, Fig. \ref{fig:param_1ton_features} shows feature scaling within the $1$-to-$n$ kernel model. In this setting, all three parameterizations exhibit a clear and essentially monotonic improvement with the number of encoded features in both tasks. The \textit{Geometric} parameterization again gives the highest accuracy throughout the explored range. For three-class classification, \textit{Euler-angles} consistently performs between \textit{Geometric} and \textit{Givens-rotations}, whereas in the binary task \textit{Givens-rotations} overtakes \textit{Euler-angles} at larger feature budgets. Compared with the $1$-qutrit QNN, the dependence on $p$ is more regular in the kernel setting, suggesting that lifting the learned single-qutrit map to a larger register makes additional encoded features translate into more consistent performance improvements.

Taken together, these results indicate that the choice of $\mathrm{SU}(3)$ parameterization plays an important role in qutrit-based models. It can affect both the optimization behaviour and how additional encoded features translate into effective variations of the model. A plausible explanation is that different parameterizations distribute degrees of freedom unevenly across parameters. In the \textit{Geometric} form, parameters enter symmetrically through a single exponential of a linear combination of generators, whereas more structured decompositions can assign parameters to qualitatively different roles (e.g., phases versus two-level rotations), so increasing the feature budget may not always activate equally informative directions. These observations motivate treating the parameterization/ansatz as a key design choice in qudit NQKs, and a more systematic characterization of its interaction with feature encoding and system-size scaling is left for future work.

\section{Discussion and conclusions}
\label{sec:discussion}

We extended Neural Quantum Kernels (NQKs) to qudits and carried out a systematic study in the qutrit setting ($d=3$) for both binary and three-class classification tasks. Moving to higher-dimensional local systems increases the space of available gate sets and parameterizations, and consequently makes architectural choices more consequential. From a practical standpoint, this motivates treating qudit NQKs as a compositional model whose performance depends on several interacting building blocks, including the encoded feature budget, system size, parameterization of single-qudit unitaries, entangling strategy, and task-dependent readout rule and cost function. Understanding how these components affect performance is essential for designing qudit-based models in an informed and efficient manner. Tables \ref{tab:summary_1} and \ref{tab:summary_2} provide cross-dataset qualitative summaries of the main trends, while Table \ref{tab:summary_ingredients} distills these results into a compact set of observations organized by design choice. We elaborate on these findings below.

Across benchmarks, we observe that increasing the number of encoded features can improve accuracy, although the magnitude of the improvement is not universal and depends on the dataset and on the chosen $\mathrm{SU}(3)$ parameterization (see Tables \ref{tab:summary_1}--\ref{tab:summary_2}). In particular, feature scaling is typically more regular in the kernel setting at fixed system size (both $1$-to-$n$ and $n$-to-$n$ at $n=4$) than for the $1$-qutrit QNN, and the \textit{Geometric} parameterization yields the most consistent gains across our experiments. This highlights that feature scaling should not be viewed as a guaranteed route to better performance, but rather as a controllable design axis whose benefit must be evaluated jointly with the parameterization.

Scaling the number of qutrits generally improves performance for both $1$-to-$n$ and $n$-to-$n$ kernel constructions within the explored range, although the magnitude and monotonicity of the improvement depend on the dataset and on the specific construction. Moreover, qutrit NQKs outperform their corresponding QNN baselines in nearly all settings considered --- with binary Covertype at small register sizes as the main exception (Figs.~\ref{fig:appendix_1ton_qutrits}(d) and \ref{fig:appendix_nton_qutrits}(d)) --- supporting the NQK principle that learning a task-adapted representation via QNN pretraining and then constructing a fixed kernel can yield a stronger classifier than the measurement-based variational model at matched resources. At $n=4$, the $1$-to-$n$ and $n$-to-$n$ constructions achieve comparable accuracy despite allocating trainable degrees of freedom differently, with the $n$-to-$n$ model slightly outperforming the $1$-to-$n$ model on Fashion-MNIST after SVM hyperparameter selection. This suggests that both strategies can be competitive at moderate system sizes under nearest-neighbour connectivity.

When compared with a classical RBF-SVM baseline using the same preprocessing and cross-validation splits, the qutrit NQKs remain competitive across the considered benchmarks. The RBF-SVM attains slightly higher accuracy in several binary and HAR settings, while the $4$-to-$4$ NQK achieves comparable performance within statistical uncertainty and slightly exceeds the classical baseline in the Covertype three-class task. This suggests that the proposed qudit kernels can approach strong classical kernel baselines, while leaving room for further improvements through more systematic choices of embeddings, parameterizations, and entangling structures.

The ablation over $\mathrm{SU}(3)$ parameterizations further indicates that the parameterization choice plays an important role in qutrit-based models. It can materially affect both the achieved accuracy and the scaling trends with respect to $p$ and $n$, and it is consistent with distinct optimization behaviour across parameterizations. Taken together, these results emphasize that careful selection of qudit building blocks is crucial for obtaining strong performance and reliable training behaviour.

Our study also has limitations. All results are obtained under ideal (noiseless) simulation, and a natural next step is to evaluate robustness under realistic noise channels and finite-shot sampling \cite{wang2021noise, thanasilp2024exponential}. In addition, due to the computational cost of multi-qutrit training, $n$-to-$n$ experiments were limited to small system sizes, and extending feature-scaling studies in the $n$-to-$n$ setting would require retraining multi-qutrit embeddings for multiple feature budgets. Finally, while we focus on qutrits as the minimal nontrivial qudit, extending the analysis to $d>3$ would clarify how the increased local degrees of freedom translate into practical gains for feature encoding and multiclass learning.

In summary, our qutrit extension shows that Neural Quantum Kernels remain effective beyond qubits and naturally accommodate multiclass learning, while their practical performance remains tightly linked to architectural choices such as parameterization, feature budget, and system size. These results support a design-driven approach to qudit NQKs, where careful selection of building blocks is as important as scaling resources, and motivate further study under realistic noise and larger local dimensions.
\onecolumngrid

{\captionsetup[table]{width=\textwidth}
\begin{table}[h]
    \centering
    \caption{\justifying
    Cross-dataset qualitative summary of the main experimental trends for the $1$-qutrit QNN and the $1$-to-$n$ NQK construction. Entries report how performance scales with the encoded feature budget $p$ and with system size $n$ (at fixed $p=8$), and are shown separately for the \textit{Geometric} (Geom), \textit{Euler-angles} (Euler), and \textit{Givens-rotations} (Givens) $\mathrm{SU}(3)$ parameterizations. Detailed plots are provided in Appendix \ref{sec:appendix_datasets_results}.}
    \label{tab:summary_1}
    \footnotesize
    \setlength{\tabcolsep}{4pt}
    \renewcommand{\arraystretch}{1.2}
    \begin{tabular}{clp{0.25\textwidth}p{0.25\textwidth}p{0.25\textwidth}}
    \Xhline{1.2pt}
    \textbf{Dataset / task} &
    \textbf{Param.} &
    \textbf{$1$-qutrit QNN: $p$-scaling} &
    \textbf{$1$-to-$n$: $n$-scaling ($p{=}8$)} &
    \textbf{$1$-to-$n$: $p$-scaling ($n{=}4$)} \\
    \Xhline{1.2pt}
    \multirow[c]{3}{*}{\parbox{0.15\textwidth}{\centering Fashion-MNIST\\(binary)}} &
    Geom &
    \makecell[l]{Increases with $p$ (best performance)} &
    \makecell[l]{Non-monotonic, saturates at $n=4$\\(best performance)} &
    \makecell[l]{Increases with $p$} \\
    &
    Euler &
    \makecell[l]{Mild increase with $p$} &
    \makecell[l]{Non-monotonic, saturates at $n=4$} &
    \makecell[l]{Increases with $p$} \\
    &
    Givens &
    \makecell[l]{Increases with $p$} &
    \makecell[l]{Increases with $n$} &
    \makecell[l]{Increases with $p$ (lowest performance)} \\
    \Xhline{0.2pt}

    \multirow[c]{3}{*}{\parbox{0.15\textwidth}{\centering Fashion-MNIST\\(three-class)}} &
    Geom &
    \makecell[l]{Increases with $p$ (best performance)} &
    \makecell[l]{Increases and saturates at $n=4$ \\ (best performance)} &
    \makecell[l]{Increases with $p$} \\
    &
    Euler &
    \makecell[l]{Mild increase with $p$} &
    \makecell[l]{Non-monotonic, saturates at $n=4$} &
    \makecell[l]{Increases with $p$} \\
    &
    Givens &
    \makecell[l]{Increases with $p$ and saturates at $p=6$} &
    \makecell[l]{Increases and saturates at $n=3$} &
    \makecell[l]{Increases with $p$ (lowest performance)} \\
    \Xhline{0.2pt}

    \multirow[c]{3}{*}{\parbox{0.15\textwidth}{\centering HAR\\(binary)}} &
    Geom &
    \makecell[l]{Increases with $p$ and saturates at $p=6$} &
    \makecell[l]{Increases with $n$ \\ (best performance)} &
    \makecell[l]{Increases with $p$ (best performance)} \\
    &
    Euler &
    \makecell[l]{Increases with $p$ and saturates at $p=6$} &
    \makecell[l]{Increases until $n=5$} &
    \makecell[l]{Increases and saturates at $p=6$} \\
    &
    Givens &
    \makecell[l]{Increases with $p$ and saturates at $p=6$ \\ (lowest performance)} &
    \makecell[l]{Increases and saturates at $n=2$} &
    \makecell[l]{Increases and saturates at $p=6$} \\
    \Xhline{0.2pt}

    \multirow[c]{3}{*}{\parbox{0.15\textwidth}{\centering HAR\\(three-class)}} &
    Geom &
    \makecell[l]{Increases with $p$ and saturates at $p=6$} &
    \makecell[l]{Increases with $n$ (best performance)} &
    \makecell[l]{Increases with $p$} \\
    &
    Euler &
    \makecell[l]{Increases with $p$ and saturates at $p=6$} &
    \makecell[l]{Increases with $n$} &
    \makecell[l]{Increases with $p$} \\
    &
    Givens &
    \makecell[l]{Increases with $p$ and saturates at $p=6$ \\ (lowest performance)} &
    \makecell[l]{Increases and saturates at $n=2$} &
    \makecell[l]{Increases and saturates at $p=6$} \\
    \Xhline{0.2pt}

    \multirow[c]{3}{*}{\parbox{0.15\textwidth}{\centering MAGIC\\(binary)}} &
    Geom &
    \makecell[l]{Increases, peaks at $p=6$ and decreases} &
    \makecell[l]{Non-monotonic, saturates at $n=3$} &
    \makecell[l]{Increases with $p=6$} \\
    &
    Euler &
    \makecell[l]{Increases, peaks at $p=6$ and decreases \\ (lowest performance)} &
    \makecell[l]{Non-monotonic, saturates at $n=4$} &
    \makecell[l]{Increases and saturates at $p=6$} \\
    &
    Givens &
    \makecell[l]{Increases, peaks at $p=6$ and decreases} &
    \makecell[l]{Saturates at $n=1$} &
    \makecell[l]{Increases and saturates at $p=6$} \\
    \Xhline{0.2pt}

    \multirow[c]{3}{*}{\parbox{0.15\textwidth}{\centering Covertype\\(binary)}} &
    Geom &
    \makecell[l]{Increases, peaks at $p=6$ and \\ decreases (best performance)} &
    \makecell[l]{Non-monotonic, saturates at $n=4$} &
    \makecell[l]{Increases with $p$ (best performance)} \\
    &
    Euler &
    \makecell[l]{Increases, peaks at $p=6$ and saturates} &
    \makecell[l]{Non-monotonic, but generally increases \\ with $n$} &
    \makecell[l]{Increases and saturates at $p=6$} \\
    &
    Givens &
    \makecell[l]{Increases, peaks at $p=6$ and \\ decreases} &
    \makecell[l]{Non-monotonic, saturates at $n=3$} &
    \makecell[l]{Peaks at $p=6$ and decreases} \\
    \Xhline{0.2pt}

    \multirow[c]{3}{*}{\parbox{0.15\textwidth}{\centering Covertype\\(three-class)}} &
    Geom &
    \makecell[l]{Non-monotonic, peaks at $p=5$ \\ (best performance)} &
    \makecell[l]{Non-monotonic, saturates at $n=4$} &
    \makecell[l]{Increases with $p$ (best performance)} \\
    &
    Euler &
    \makecell[l]{Increases, peaks at $p=6$ and decreases} &
    \makecell[l]{Non-monotonic, saturates at $n=4$} &
    \makecell[l]{Increases and saturates at $p=6$} \\
    &
    Givens &
    \makecell[l]{Increases, peaks at $p=6$ and decreases} &
    \makecell[l]{Increases and saturates at $n=4$} &
    \makecell[l]{Increases and saturates at $p=6$} \\
    \Xhline{1.2pt}
    \end{tabular}
\end{table}}

{\captionsetup[table]{width=\textwidth}
\begin{table}[h]
    \centering
    \caption{\justifying
    Cross-dataset qualitative summary of scaling trends for the $n$-qutrit QNN and the $n$-to-$n$ NQK under the \textit{Geometric} $\mathrm{SU}(3)$ parameterization (default setting). Detailed plots are provided in Appendix \ref{sec:appendix_datasets_results}.}
    \label{tab:summary_2}
    \footnotesize
    \setlength{\tabcolsep}{4pt}
    \renewcommand{\arraystretch}{1.2}
    \begin{tabular}{cp{0.26\textwidth}p{0.26\textwidth}p{0.26\textwidth}}
    \Xhline{1.2pt}
    \textbf{Dataset / task} &
    \textbf{QNN: $n$-scaling} &
    \textbf{$n$-to-$n$: $n$-scaling ($p{=}8$)} &
    \textbf{$n$-to-$n$: $p$-scaling ($n{=}4$)} \\
    \Xhline{1.2pt}

    \addlinespace[1.2pt]
    
    \parbox{0.15\textwidth}{\centering Fashion-MNIST\\(binary)} &
    \makecell[l]{Mild increase with $n$} &
    \makecell[l]{Increases with $n$, peaks at $n=3$} &
    \makecell[l]{Increases with $p$} \\

    \addlinespace[1.2pt]
    \Xhline{0.2pt}
    \addlinespace[1.2pt]
    
    \parbox{0.15\textwidth}{\centering Fashion-MNIST\\(three-class)} &
    \makecell[l]{Mild increase with $n$} &
    \makecell[l]{Increases with $n$} &
    \makecell[l]{Increases with $p$} \\

    \addlinespace[1.2pt]
    \Xhline{0.2pt}
    \addlinespace[1.2pt]

    \parbox{0.15\textwidth}{\centering HAR\\(binary)} &
    \makecell[l]{Mild increase with $n$} &
    \makecell[l]{Increases with $n$} &
    \makecell[l]{Increases with $p$} \\

    \addlinespace[1.2pt]
    \Xhline{0.2pt}
    \addlinespace[1.2pt]

    \parbox{0.15\textwidth}{\centering HAR\\(three-class)} &
    \makecell[l]{Increases with $n$} &
    \makecell[l]{Increases with $n$} &
    \makecell[l]{Increases with $p$} \\

    \addlinespace[1.2pt]
    \Xhline{0.2pt}
    \addlinespace[1.2pt]

    \parbox{0.15\textwidth}{\centering MAGIC\\(binary)} &
    \makecell[l]{Increases with $n$} &
    \makecell[l]{Non-monotonic, peaks at $n=3$} &
    \makecell[l]{Increases with $p$} \\

    \addlinespace[1.2pt]
    \Xhline{0.2pt}
    \addlinespace[1.2pt]

    \parbox{0.15\textwidth}{\centering Covertype\\(binary)} &
    \makecell[l]{Non-monotonic, peaks at $n=2$} &
    \makecell[l]{Increases with $n$} &
    \makecell[l]{Increases with $p$} \\

    \addlinespace[1.2pt]
    \Xhline{0.2pt}
    \addlinespace[1.2pt]

    \parbox{0.15\textwidth}{\centering Covertype\\(three-class)} &
    \makecell[l]{Mild increase with $n$} &
    \makecell[l]{Increases with $n$} &
    \makecell[l]{Saturates at $p=6$} \\

    \addlinespace[1.2pt]
    \Xhline{0.2pt}
    \end{tabular}
\end{table}}

\clearpage
{\captionsetup[table]{width=\textwidth}
\begin{table}[h]
    \centering
    \caption{\justifying
    Summary of the main observations by design choice, distilled from the cross-dataset results in Tables~\ref{tab:summary_1}--\ref{tab:summary_2} (see Appendix~\ref{sec:appendix_datasets_results} for plots).}
    \label{tab:summary_ingredients}
    \small
    \setlength{\tabcolsep}{6pt}
    \renewcommand{\arraystretch}{1.15}

    \begin{tabular*}{\textwidth}{@{\extracolsep{\fill}} p{0.32\textwidth} p{0.64\textwidth}}
    \Xhline{1.2pt}
    \textbf{Design choice} & \textbf{Main observation} \\
    \Xhline{1.2pt}
    Local dimension $d$ &
    We focus on qutrits ($d=3$); results support qutrit NQKs as an effective qudit extension in both binary and three-class settings. Extending to $d>3$ is left for future work. \\
    \Xhline{0.2pt}
    Feature budget $p$ &
    Increasing $p$ can improve accuracy, but the magnitude and monotonicity of the gains depend on the dataset and the chosen $\mathrm{SU}(3)$ parameterization, and improvements may saturate for larger feature counts. In the kernel setting, feature scaling tends to be more regular than in the $1$-qutrit QNN. \\
    \Xhline{0.2pt}
    System size $n$ &
    For both $1$-to-$n$ and $n$-to-$n$ constructions, accuracy improves when increasing $n$, with a gradual plateau at larger system sizes (within the explored range). \\
    \Xhline{0.2pt}
    Single-qutrit $\mathrm{SU}(3)$ parameterization &
    parameterization materially affects both performance level and scaling trends; \textit{Geometric} tends to yield stronger gains and better optimization behaviour than structured decompositions in our experiments. \\
    \Xhline{0.2pt}
    Kernel construction &
    Both $1$-to-$n$ and $n$-to-$n$ NQKs improve over their corresponding QNN baselines and reach comparable accuracy at $n=4$. Both use nearest-neighbour entanglement, but $n$-to-$n$ learns entangling parameters jointly with independent single-qutrit unitaries, whereas $1$-to-$n$ replicates a learned single-qutrit map and keeps entanglement fixed. \\
    \Xhline{0.2pt}
    Task definition and cost function &
    We consider binary and three-class classification settings, with task-dependent readout rules and cost functions. Both settings exhibit similar qualitative trends, but differ in absolute accuracy and sensitivity to modelling choices. \\
    \Xhline{1.2pt}
    \end{tabular*}
\end{table}}
\twocolumngrid

\section{Acknowledgements}
\label{sec:acknowledgements}

We acknowledge financial support from OpenSuperQ+100 (Grant No. 101113946) of the EU Flagship on Quantum Technologies, from Project Grant No. PID2024-156808NB-I00 and Spanish Ramón y Cajal Grant No. RYC-2020-030503-I funded by MICIU/AEI/10.13039/501100011033 and by “ERDF A way of making Europe” and “ERDF Invest in your Future”, from the Spanish Ministry for Digital Transformation
and of Civil Service of the Spanish Government through the QUANTUM ENIA project call-Quantum Spain, and by the EU through the Recovery, Transformation and Resilience Plan–NextGenerationEU within the framework of the Digital Spain 2026 Agenda, and from the Elkartek project KUBIBIT - kuantikaren berrikuntzarako ibilbide teknologikoak (ELKARTEK25/79). We acknowledge funding from Basque Government through the IKUR Strategy under the collaboration agreement between Ikerbasque Foundation and BCAM on behalf of the Department of Education of the Basque Government. 

\FloatBarrier

\bibliography{main}

\begin{thebibliography}{88}%
\makeatletter
\providecommand \@ifxundefined [1]{%
 \@ifx{#1\undefined}
}%
\providecommand \@ifnum [1]{%
 \ifnum #1\expandafter \@firstoftwo
 \else \expandafter \@secondoftwo
 \fi
}%
\providecommand \@ifx [1]{%
 \ifx #1\expandafter \@firstoftwo
 \else \expandafter \@secondoftwo
 \fi
}%
\providecommand \natexlab [1]{#1}%
\providecommand \enquote  [1]{``#1''}%
\providecommand \bibnamefont  [1]{#1}%
\providecommand \bibfnamefont [1]{#1}%
\providecommand \citenamefont [1]{#1}%
\providecommand \href@noop [0]{\@secondoftwo}%
\providecommand \href [0]{\begingroup \@sanitize@url \@href}%
\providecommand \@href[1]{\@@startlink{#1}\@@href}%
\providecommand \@@href[1]{\endgroup#1\@@endlink}%
\providecommand \@sanitize@url [0]{\catcode `\\12\catcode `\$12\catcode `\&12\catcode `\#12\catcode `\^12\catcode `\_12\catcode `\%12\relax}%
\providecommand \@@startlink[1]{}%
\providecommand \@@endlink[0]{}%
\providecommand \url  [0]{\begingroup\@sanitize@url \@url }%
\providecommand \@url [1]{\endgroup\@href {#1}{\urlprefix }}%
\providecommand \urlprefix  [0]{URL }%
\providecommand \Eprint [0]{\href }%
\providecommand \doibase [0]{https://doi.org/}%
\providecommand \selectlanguage [0]{\@gobble}%
\providecommand \bibinfo  [0]{\@secondoftwo}%
\providecommand \bibfield  [0]{\@secondoftwo}%
\providecommand \translation [1]{[#1]}%
\providecommand \BibitemOpen [0]{}%
\providecommand \bibitemStop [0]{}%
\providecommand \bibitemNoStop [0]{.\EOS\space}%
\providecommand \EOS [0]{\spacefactor3000\relax}%
\providecommand \BibitemShut  [1]{\csname bibitem#1\endcsname}%
\let\auto@bib@innerbib\@empty
\bibitem [{\citenamefont {Biamonte}\ \emph {et~al.}(2017)\citenamefont {Biamonte}, \citenamefont {Wittek}, \citenamefont {Pancotti}, \citenamefont {Rebentrost}, \citenamefont {Wiebe},\ and\ \citenamefont {Lloyd}}]{biamonte2017quantum}%
  \BibitemOpen
  \bibfield  {author} {\bibinfo {author} {\bibfnamefont {J.}~\bibnamefont {Biamonte}}, \bibinfo {author} {\bibfnamefont {P.}~\bibnamefont {Wittek}}, \bibinfo {author} {\bibfnamefont {N.}~\bibnamefont {Pancotti}}, \bibinfo {author} {\bibfnamefont {P.}~\bibnamefont {Rebentrost}}, \bibinfo {author} {\bibfnamefont {N.}~\bibnamefont {Wiebe}},\ and\ \bibinfo {author} {\bibfnamefont {S.}~\bibnamefont {Lloyd}},\ }\bibfield  {title} {\bibinfo {title} {{Quantum machine learning}},\ }\href {https://doi.org/10.1038/nature23474} {\bibfield  {journal} {\bibinfo  {journal} {Nature}\ }\textbf {\bibinfo {volume} {549}},\ \bibinfo {pages} {195} (\bibinfo {year} {2017})}\BibitemShut {NoStop}%
\bibitem [{\citenamefont {Carleo}\ \emph {et~al.}(2019)\citenamefont {Carleo}, \citenamefont {Cirac}, \citenamefont {Cranmer}, \citenamefont {Daudet}, \citenamefont {Schuld}, \citenamefont {Tishby}, \citenamefont {Vogt-Maranto},\ and\ \citenamefont {Zdeborov{\'a}}}]{carleo2019machine}%
  \BibitemOpen
  \bibfield  {author} {\bibinfo {author} {\bibfnamefont {G.}~\bibnamefont {Carleo}}, \bibinfo {author} {\bibfnamefont {I.}~\bibnamefont {Cirac}}, \bibinfo {author} {\bibfnamefont {K.}~\bibnamefont {Cranmer}}, \bibinfo {author} {\bibfnamefont {L.}~\bibnamefont {Daudet}}, \bibinfo {author} {\bibfnamefont {M.}~\bibnamefont {Schuld}}, \bibinfo {author} {\bibfnamefont {N.}~\bibnamefont {Tishby}}, \bibinfo {author} {\bibfnamefont {L.}~\bibnamefont {Vogt-Maranto}},\ and\ \bibinfo {author} {\bibfnamefont {L.}~\bibnamefont {Zdeborov{\'a}}},\ }\bibfield  {title} {\bibinfo {title} {{Machine learning and the physical sciences}},\ }\bibfield  {journal} {\bibinfo  {journal} {Reviews of Modern Physics}\ }\textbf {\bibinfo {volume} {91}},\ \href {https://doi.org/10.1103/revmodphys.91.045002} {10.1103/revmodphys.91.045002} (\bibinfo {year} {2019})\BibitemShut {NoStop}%
\bibitem [{\citenamefont {Dunjko}\ and\ \citenamefont {Briegel}(2018)}]{dunjko2018machine}%
  \BibitemOpen
  \bibfield  {author} {\bibinfo {author} {\bibfnamefont {V.}~\bibnamefont {Dunjko}}\ and\ \bibinfo {author} {\bibfnamefont {H.~J.}\ \bibnamefont {Briegel}},\ }\bibfield  {title} {\bibinfo {title} {{Machine learning \& artificial intelligence in the quantum domain: a review of recent progress}},\ }\href {https://doi.org/10.1088/1361-6633/aab406} {\bibfield  {journal} {\bibinfo  {journal} {Reports on Progress in Physics}\ }\textbf {\bibinfo {volume} {81}},\ \bibinfo {pages} {074001} (\bibinfo {year} {2018})}\BibitemShut {NoStop}%
\bibitem [{\citenamefont {Benedetti}\ \emph {et~al.}(2019)\citenamefont {Benedetti}, \citenamefont {Lloyd}, \citenamefont {Sack},\ and\ \citenamefont {Fiorentini}}]{benedetti2019parameterized}%
  \BibitemOpen
  \bibfield  {author} {\bibinfo {author} {\bibfnamefont {M.}~\bibnamefont {Benedetti}}, \bibinfo {author} {\bibfnamefont {E.}~\bibnamefont {Lloyd}}, \bibinfo {author} {\bibfnamefont {S.}~\bibnamefont {Sack}},\ and\ \bibinfo {author} {\bibfnamefont {M.}~\bibnamefont {Fiorentini}},\ }\bibfield  {title} {\bibinfo {title} {{Parameterized quantum circuits as machine learning models}},\ }\href {https://doi.org/10.1088/2058-9565/ab4eb5} {\bibfield  {journal} {\bibinfo  {journal} {Quantum Science and Technology}\ }\textbf {\bibinfo {volume} {4}},\ \bibinfo {pages} {043001} (\bibinfo {year} {2019})}\BibitemShut {NoStop}%
\bibitem [{\citenamefont {Farhi}\ and\ \citenamefont {Neven}(2018)}]{farhi2018classification}%
  \BibitemOpen
  \bibfield  {author} {\bibinfo {author} {\bibfnamefont {E.}~\bibnamefont {Farhi}}\ and\ \bibinfo {author} {\bibfnamefont {H.}~\bibnamefont {Neven}},\ }\href {https://arxiv.org/abs/1802.06002} {\bibinfo {title} {{Classification with Quantum Neural Networks on Near Term Processors}}} (\bibinfo {year} {2018})\BibitemShut {NoStop}%
\bibitem [{\citenamefont {Skolik}\ \emph {et~al.}(2022)\citenamefont {Skolik}, \citenamefont {Jerbi},\ and\ \citenamefont {Dunjko}}]{skolik2022quantum}%
  \BibitemOpen
  \bibfield  {author} {\bibinfo {author} {\bibfnamefont {A.}~\bibnamefont {Skolik}}, \bibinfo {author} {\bibfnamefont {S.}~\bibnamefont {Jerbi}},\ and\ \bibinfo {author} {\bibfnamefont {V.}~\bibnamefont {Dunjko}},\ }\bibfield  {title} {\bibinfo {title} {{Quantum agents in the Gym: a variational quantum algorithm for deep Q-learning}},\ }\href {https://doi.org/10.22331/q-2022-05-24-720} {\bibfield  {journal} {\bibinfo  {journal} {Quantum}\ }\textbf {\bibinfo {volume} {6}},\ \bibinfo {pages} {720} (\bibinfo {year} {2022})}\BibitemShut {NoStop}%
\bibitem [{\citenamefont {Schuld}\ \emph {et~al.}(2020)\citenamefont {Schuld}, \citenamefont {Bocharov}, \citenamefont {Svore},\ and\ \citenamefont {Wiebe}}]{schuld2020circuit}%
  \BibitemOpen
  \bibfield  {author} {\bibinfo {author} {\bibfnamefont {M.}~\bibnamefont {Schuld}}, \bibinfo {author} {\bibfnamefont {A.}~\bibnamefont {Bocharov}}, \bibinfo {author} {\bibfnamefont {K.~M.}\ \bibnamefont {Svore}},\ and\ \bibinfo {author} {\bibfnamefont {N.}~\bibnamefont {Wiebe}},\ }\bibfield  {title} {\bibinfo {title} {{Circuit-centric quantum classifiers}},\ }\bibfield  {journal} {\bibinfo  {journal} {Physical review. A/Physical review, A}\ }\textbf {\bibinfo {volume} {101}},\ \href {https://doi.org/10.1103/physreva.101.032308} {10.1103/physreva.101.032308} (\bibinfo {year} {2020})\BibitemShut {NoStop}%
\bibitem [{\citenamefont {Gyurik}\ and\ \citenamefont {Dunjko}(2022)}]{gyurik2022establishing}%
  \BibitemOpen
  \bibfield  {author} {\bibinfo {author} {\bibfnamefont {C.}~\bibnamefont {Gyurik}}\ and\ \bibinfo {author} {\bibfnamefont {V.}~\bibnamefont {Dunjko}},\ }\href {https://arxiv.org/abs/2208.06339} {\bibinfo {title} {{On establishing learning separations between classical and quantum machine learning with classical data}}} (\bibinfo {year} {2022})\BibitemShut {NoStop}%
\bibitem [{\citenamefont {Sweke}\ \emph {et~al.}(2021)\citenamefont {Sweke}, \citenamefont {Seifert}, \citenamefont {Hangleiter},\ and\ \citenamefont {Eisert}}]{sweke2021quantum}%
  \BibitemOpen
  \bibfield  {author} {\bibinfo {author} {\bibfnamefont {R.}~\bibnamefont {Sweke}}, \bibinfo {author} {\bibfnamefont {J.-P.}\ \bibnamefont {Seifert}}, \bibinfo {author} {\bibfnamefont {D.}~\bibnamefont {Hangleiter}},\ and\ \bibinfo {author} {\bibfnamefont {J.}~\bibnamefont {Eisert}},\ }\bibfield  {title} {\bibinfo {title} {{On the Quantum versus Classical Learnability of Discrete Distributions}},\ }\bibfield  {journal} {\bibinfo  {journal} {Quantum}\ }\href {https://doi.org/10.22331/q-2021-03-23-417} {10.22331/q-2021-03-23-417} (\bibinfo {year} {2021})\BibitemShut {NoStop}%
\bibitem [{\citenamefont {Jerbi}\ \emph {et~al.}(2021)\citenamefont {Jerbi}, \citenamefont {Trenkwalder}, \citenamefont {Nautrup}, \citenamefont {Briegel},\ and\ \citenamefont {Dunjko}}]{jerbi2021quantum}%
  \BibitemOpen
  \bibfield  {author} {\bibinfo {author} {\bibfnamefont {S.}~\bibnamefont {Jerbi}}, \bibinfo {author} {\bibfnamefont {L.~M.}\ \bibnamefont {Trenkwalder}}, \bibinfo {author} {\bibfnamefont {H.~P.}\ \bibnamefont {Nautrup}}, \bibinfo {author} {\bibfnamefont {H.~J.}\ \bibnamefont {Briegel}},\ and\ \bibinfo {author} {\bibfnamefont {V.}~\bibnamefont {Dunjko}},\ }\bibfield  {title} {\bibinfo {title} {{Quantum Enhancements for Deep Reinforcement Learning in Large Spaces}},\ }\bibfield  {journal} {\bibinfo  {journal} {PRX Quantum}\ }\textbf {\bibinfo {volume} {2}},\ \href {https://doi.org/10.1103/prxquantum.2.010328} {10.1103/prxquantum.2.010328} (\bibinfo {year} {2021})\BibitemShut {NoStop}%
\bibitem [{\citenamefont {Liu}\ \emph {et~al.}(2021{\natexlab{a}})\citenamefont {Liu}, \citenamefont {Arunachalam},\ and\ \citenamefont {Temme}}]{liu2021rigorous}%
  \BibitemOpen
  \bibfield  {author} {\bibinfo {author} {\bibfnamefont {Y.}~\bibnamefont {Liu}}, \bibinfo {author} {\bibfnamefont {S.}~\bibnamefont {Arunachalam}},\ and\ \bibinfo {author} {\bibfnamefont {K.}~\bibnamefont {Temme}},\ }\bibfield  {title} {\bibinfo {title} {{A rigorous and robust quantum speed-up in supervised machine learning}},\ }\href {https://doi.org/10.1038/s41567-021-01287-z} {\bibfield  {journal} {\bibinfo  {journal} {Nature Physics}\ }\textbf {\bibinfo {volume} {17}},\ \bibinfo {pages} {1013} (\bibinfo {year} {2021}{\natexlab{a}})}\BibitemShut {NoStop}%
\bibitem [{\citenamefont {Pirnay}\ \emph {et~al.}(2023)\citenamefont {Pirnay}, \citenamefont {Sweke}, \citenamefont {Eisert},\ and\ \citenamefont {Seifert}}]{pirnay2023superpolynomial}%
  \BibitemOpen
  \bibfield  {author} {\bibinfo {author} {\bibfnamefont {N.}~\bibnamefont {Pirnay}}, \bibinfo {author} {\bibfnamefont {R.}~\bibnamefont {Sweke}}, \bibinfo {author} {\bibfnamefont {J.}~\bibnamefont {Eisert}},\ and\ \bibinfo {author} {\bibfnamefont {J.-P.}\ \bibnamefont {Seifert}},\ }\bibfield  {title} {\bibinfo {title} {{Superpolynomial quantum-classical separation for density modeling}},\ }\bibfield  {journal} {\bibinfo  {journal} {Physical review. A/Physical review, A}\ }\textbf {\bibinfo {volume} {107}},\ \href {https://doi.org/10.1103/physreva.107.042416} {10.1103/physreva.107.042416} (\bibinfo {year} {2023})\BibitemShut {NoStop}%
\bibitem [{\citenamefont {Jadhav}\ \emph {et~al.}(2023)\citenamefont {Jadhav}, \citenamefont {Rasool},\ and\ \citenamefont {Gyanchandani}}]{jadhav2023quantum}%
  \BibitemOpen
  \bibfield  {author} {\bibinfo {author} {\bibfnamefont {A.}~\bibnamefont {Jadhav}}, \bibinfo {author} {\bibfnamefont {A.}~\bibnamefont {Rasool}},\ and\ \bibinfo {author} {\bibfnamefont {M.}~\bibnamefont {Gyanchandani}},\ }\bibfield  {title} {\bibinfo {title} {{Quantum Machine Learning: Scope for real-world problems}},\ }\href {https://doi.org/10.1016/j.procs.2023.01.235} {\bibfield  {journal} {\bibinfo  {journal} {Procedia Computer Science}\ }\textbf {\bibinfo {volume} {218}},\ \bibinfo {pages} {2612} (\bibinfo {year} {2023})}\BibitemShut {NoStop}%
\bibitem [{\citenamefont {McClean}\ \emph {et~al.}(2018)\citenamefont {McClean}, \citenamefont {Boixo}, \citenamefont {Smelyanskiy}, \citenamefont {Babbush},\ and\ \citenamefont {Neven}}]{mcclean2018barren}%
  \BibitemOpen
  \bibfield  {author} {\bibinfo {author} {\bibfnamefont {J.~R.}\ \bibnamefont {McClean}}, \bibinfo {author} {\bibfnamefont {S.}~\bibnamefont {Boixo}}, \bibinfo {author} {\bibfnamefont {V.~N.}\ \bibnamefont {Smelyanskiy}}, \bibinfo {author} {\bibfnamefont {R.}~\bibnamefont {Babbush}},\ and\ \bibinfo {author} {\bibfnamefont {H.}~\bibnamefont {Neven}},\ }\bibfield  {title} {\bibinfo {title} {{Barren plateaus in quantum neural network training landscapes}},\ }\href {https://doi.org/10.1038/s41467-018-07090-4} {\bibfield  {journal} {\bibinfo  {journal} {Nature Communications}\ }\textbf {\bibinfo {volume} {9}},\ \bibinfo {pages} {4812} (\bibinfo {year} {2018})}\BibitemShut {NoStop}%
\bibitem [{\citenamefont {Cerezo}\ \emph {et~al.}(2021)\citenamefont {Cerezo}, \citenamefont {Sone}, \citenamefont {Volkoff}, \citenamefont {Cincio},\ and\ \citenamefont {Coles}}]{cerezo2021cost}%
  \BibitemOpen
  \bibfield  {author} {\bibinfo {author} {\bibfnamefont {M.}~\bibnamefont {Cerezo}}, \bibinfo {author} {\bibfnamefont {A.}~\bibnamefont {Sone}}, \bibinfo {author} {\bibfnamefont {T.}~\bibnamefont {Volkoff}}, \bibinfo {author} {\bibfnamefont {L.}~\bibnamefont {Cincio}},\ and\ \bibinfo {author} {\bibfnamefont {P.~J.}\ \bibnamefont {Coles}},\ }\bibfield  {title} {\bibinfo {title} {{Cost function dependent barren plateaus in shallow parametrized quantum circuits}},\ }\href {https://doi.org/10.1038/s41467-021-21728-w} {\bibfield  {journal} {\bibinfo  {journal} {Nature Communications}\ }\textbf {\bibinfo {volume} {12}},\ \bibinfo {pages} {1791} (\bibinfo {year} {2021})}\BibitemShut {NoStop}%
\bibitem [{\citenamefont {Holmes}\ \emph {et~al.}(2022)\citenamefont {Holmes}, \citenamefont {Sharma}, \citenamefont {Cerezo},\ and\ \citenamefont {Coles}}]{holmes2101connecting}%
  \BibitemOpen
  \bibfield  {author} {\bibinfo {author} {\bibfnamefont {Z.}~\bibnamefont {Holmes}}, \bibinfo {author} {\bibfnamefont {K.}~\bibnamefont {Sharma}}, \bibinfo {author} {\bibfnamefont {M.}~\bibnamefont {Cerezo}},\ and\ \bibinfo {author} {\bibfnamefont {P.~J.}\ \bibnamefont {Coles}},\ }\bibfield  {title} {\bibinfo {title} {{Connecting Ansatz Expressibility to Gradient Magnitudes and Barren Plateaus}},\ }\bibfield  {journal} {\bibinfo  {journal} {PRX Quantum}\ }\textbf {\bibinfo {volume} {3}},\ \href {https://doi.org/10.1103/prxquantum.3.010313} {10.1103/prxquantum.3.010313} (\bibinfo {year} {2022})\BibitemShut {NoStop}%
\bibitem [{\citenamefont {Fontana}\ \emph {et~al.}(2024)\citenamefont {Fontana}, \citenamefont {Herman}, \citenamefont {Chakrabarti}, \citenamefont {Kumar}, \citenamefont {Yalovetzky}, \citenamefont {Heredge}, \citenamefont {Sureshbabu},\ and\ \citenamefont {Pistoia}}]{fontana2024adjoint}%
  \BibitemOpen
  \bibfield  {author} {\bibinfo {author} {\bibfnamefont {E.}~\bibnamefont {Fontana}}, \bibinfo {author} {\bibfnamefont {D.}~\bibnamefont {Herman}}, \bibinfo {author} {\bibfnamefont {S.}~\bibnamefont {Chakrabarti}}, \bibinfo {author} {\bibfnamefont {N.}~\bibnamefont {Kumar}}, \bibinfo {author} {\bibfnamefont {R.}~\bibnamefont {Yalovetzky}}, \bibinfo {author} {\bibfnamefont {J.}~\bibnamefont {Heredge}}, \bibinfo {author} {\bibfnamefont {S.~H.}\ \bibnamefont {Sureshbabu}},\ and\ \bibinfo {author} {\bibfnamefont {M.}~\bibnamefont {Pistoia}},\ }\bibfield  {title} {\bibinfo {title} {{Characterizing barren plateaus in quantum ans{\"a}tze with the adjoint representation}},\ }\href {https://doi.org/10.1038/s41467-024-49910-w} {\bibfield  {journal} {\bibinfo  {journal} {Nature Communications}\ }\textbf {\bibinfo {volume} {15}},\ \bibinfo {pages} {7171} (\bibinfo {year} {2024})}\BibitemShut {NoStop}%
\bibitem [{\citenamefont {Thanasilp}\ \emph {et~al.}(2024)\citenamefont {Thanasilp}, \citenamefont {Wang}, \citenamefont {Cerezo},\ and\ \citenamefont {Holmes}}]{thanasilp2024exponential}%
  \BibitemOpen
  \bibfield  {author} {\bibinfo {author} {\bibfnamefont {S.}~\bibnamefont {Thanasilp}}, \bibinfo {author} {\bibfnamefont {S.}~\bibnamefont {Wang}}, \bibinfo {author} {\bibfnamefont {M.}~\bibnamefont {Cerezo}},\ and\ \bibinfo {author} {\bibfnamefont {Z.}~\bibnamefont {Holmes}},\ }\bibfield  {title} {\bibinfo {title} {Exponential concentration in quantum kernel methods},\ }\href {https://doi.org/https://doi.org/10.1038/s41467-024-49287-w} {\bibfield  {journal} {\bibinfo  {journal} {Nature communications}\ }\textbf {\bibinfo {volume} {15}},\ \bibinfo {pages} {5200} (\bibinfo {year} {2024})}\BibitemShut {NoStop}%
\bibitem [{\citenamefont {Pesah}\ \emph {et~al.}(2021)\citenamefont {Pesah}, \citenamefont {Cerezo}, \citenamefont {Wang}, \citenamefont {Volkoff}, \citenamefont {Sornborger},\ and\ \citenamefont {Coles}}]{pesah2021absence}%
  \BibitemOpen
  \bibfield  {author} {\bibinfo {author} {\bibfnamefont {A.}~\bibnamefont {Pesah}}, \bibinfo {author} {\bibfnamefont {M.}~\bibnamefont {Cerezo}}, \bibinfo {author} {\bibfnamefont {S.}~\bibnamefont {Wang}}, \bibinfo {author} {\bibfnamefont {T.}~\bibnamefont {Volkoff}}, \bibinfo {author} {\bibfnamefont {A.~T.}\ \bibnamefont {Sornborger}},\ and\ \bibinfo {author} {\bibfnamefont {P.~J.}\ \bibnamefont {Coles}},\ }\bibfield  {title} {\bibinfo {title} {{Absence of Barren Plateaus in Quantum Convolutional Neural Networks}},\ }\bibfield  {journal} {\bibinfo  {journal} {Physical Review X}\ }\textbf {\bibinfo {volume} {11}},\ \href {https://doi.org/10.1103/physrevx.11.041011} {10.1103/physrevx.11.041011} (\bibinfo {year} {2021})\BibitemShut {NoStop}%
\bibitem [{\citenamefont {Ragone}\ \emph {et~al.}(2024)\citenamefont {Ragone}, \citenamefont {Bakalov}, \citenamefont {Sauvage}, \citenamefont {Kemper}, \citenamefont {Ortiz~Marrero}, \citenamefont {Larocca},\ and\ \citenamefont {Cerezo}}]{ragone2024lie}%
  \BibitemOpen
  \bibfield  {author} {\bibinfo {author} {\bibfnamefont {M.}~\bibnamefont {Ragone}}, \bibinfo {author} {\bibfnamefont {B.~N.}\ \bibnamefont {Bakalov}}, \bibinfo {author} {\bibfnamefont {F.}~\bibnamefont {Sauvage}}, \bibinfo {author} {\bibfnamefont {A.~F.}\ \bibnamefont {Kemper}}, \bibinfo {author} {\bibfnamefont {C.}~\bibnamefont {Ortiz~Marrero}}, \bibinfo {author} {\bibfnamefont {M.}~\bibnamefont {Larocca}},\ and\ \bibinfo {author} {\bibfnamefont {M.}~\bibnamefont {Cerezo}},\ }\bibfield  {title} {\bibinfo {title} {A lie algebraic theory of barren plateaus for deep parameterized quantum circuits},\ }\href@noop {} {\bibfield  {journal} {\bibinfo  {journal} {Nature Communications}\ }\textbf {\bibinfo {volume} {15}},\ \bibinfo {pages} {7172} (\bibinfo {year} {2024})}\BibitemShut {NoStop}%
\bibitem [{\citenamefont {P{\'e}rez-Salinas}\ \emph {et~al.}(2020)\citenamefont {P{\'e}rez-Salinas}, \citenamefont {Cervera-Lierta}, \citenamefont {Gil-Fuster},\ and\ \citenamefont {Latorre}}]{perez2020data}%
  \BibitemOpen
  \bibfield  {author} {\bibinfo {author} {\bibfnamefont {A.}~\bibnamefont {P{\'e}rez-Salinas}}, \bibinfo {author} {\bibfnamefont {A.}~\bibnamefont {Cervera-Lierta}}, \bibinfo {author} {\bibfnamefont {E.}~\bibnamefont {Gil-Fuster}},\ and\ \bibinfo {author} {\bibfnamefont {J.~I.}\ \bibnamefont {Latorre}},\ }\bibfield  {title} {\bibinfo {title} {{Data re-uploading for a universal quantum classifier}},\ }\href {https://doi.org/10.22331/q-2020-02-06-226} {\bibfield  {journal} {\bibinfo  {journal} {Quantum}\ }\textbf {\bibinfo {volume} {4}},\ \bibinfo {pages} {226} (\bibinfo {year} {2020})}\BibitemShut {NoStop}%
\bibitem [{\citenamefont {P{\'e}rez-Salinas}\ \emph {et~al.}(2025)\citenamefont {P{\'e}rez-Salinas}, \citenamefont {Rad}, \citenamefont {Barthe},\ and\ \citenamefont {Dunjko}}]{perezsalinas2025universalapproximationcontinuousfunctions}%
  \BibitemOpen
  \bibfield  {author} {\bibinfo {author} {\bibfnamefont {A.}~\bibnamefont {P{\'e}rez-Salinas}}, \bibinfo {author} {\bibfnamefont {M.~Y.}\ \bibnamefont {Rad}}, \bibinfo {author} {\bibfnamefont {A.}~\bibnamefont {Barthe}},\ and\ \bibinfo {author} {\bibfnamefont {V.}~\bibnamefont {Dunjko}},\ }\bibfield  {title} {\bibinfo {title} {{Universal approximation of continuous functions with minimal quantum circuits}},\ }\bibfield  {journal} {\bibinfo  {journal} {Physical Review Research}\ }\textbf {\bibinfo {volume} {7}},\ \href {https://doi.org/10.1103/t49h-jmty} {10.1103/t49h-jmty} (\bibinfo {year} {2025})\BibitemShut {NoStop}%
\bibitem [{\citenamefont {Schuld}\ \emph {et~al.}(2021)\citenamefont {Schuld}, \citenamefont {Sweke},\ and\ \citenamefont {Meyer}}]{schuld-2021-effect-of}%
  \BibitemOpen
  \bibfield  {author} {\bibinfo {author} {\bibfnamefont {M.}~\bibnamefont {Schuld}}, \bibinfo {author} {\bibfnamefont {R.}~\bibnamefont {Sweke}},\ and\ \bibinfo {author} {\bibfnamefont {J.~J.}\ \bibnamefont {Meyer}},\ }\bibfield  {title} {\bibinfo {title} {{Effect of data encoding on the expressive power of variational quantum-machine-learning models}},\ }\bibfield  {journal} {\bibinfo  {journal} {Physical review. A/Physical review, A}\ }\textbf {\bibinfo {volume} {103}},\ \href {https://doi.org/10.1103/physreva.103.032430} {10.1103/physreva.103.032430} (\bibinfo {year} {2021})\BibitemShut {NoStop}%
\bibitem [{\citenamefont {Casas}\ and\ \citenamefont {Cervera-Lierta}(2023)}]{casas2023multidimensional}%
  \BibitemOpen
  \bibfield  {author} {\bibinfo {author} {\bibfnamefont {B.}~\bibnamefont {Casas}}\ and\ \bibinfo {author} {\bibfnamefont {A.}~\bibnamefont {Cervera-Lierta}},\ }\bibfield  {title} {\bibinfo {title} {Multidimensional fourier series with quantum circuits},\ }\href@noop {} {\bibfield  {journal} {\bibinfo  {journal} {Physical Review A}\ }\textbf {\bibinfo {volume} {107}},\ \bibinfo {pages} {062612} (\bibinfo {year} {2023})}\BibitemShut {NoStop}%
\bibitem [{\citenamefont {Barthe}\ and\ \citenamefont {P{\'e}rez-Salinas}(2023)}]{barthe2023gradients}%
  \BibitemOpen
  \bibfield  {author} {\bibinfo {author} {\bibfnamefont {A.}~\bibnamefont {Barthe}}\ and\ \bibinfo {author} {\bibfnamefont {A.}~\bibnamefont {P{\'e}rez-Salinas}},\ }\href@noop {} {\bibinfo {title} {Gradients and frequency profiles of quantum re-uploading models}} (\bibinfo {year} {2023}),\ \Eprint {https://arxiv.org/abs/2311.10822} {arXiv:2311.10822 [quant-ph]} \BibitemShut {NoStop}%
\bibitem [{\citenamefont {De~Souza~Farias}\ \emph {et~al.}(2025{\natexlab{a}})\citenamefont {De~Souza~Farias}, \citenamefont {Friedrich},\ and\ \citenamefont {Maziero}}]{farias2025short}%
  \BibitemOpen
  \bibfield  {author} {\bibinfo {author} {\bibfnamefont {T.}~\bibnamefont {De~Souza~Farias}}, \bibinfo {author} {\bibfnamefont {L.}~\bibnamefont {Friedrich}},\ and\ \bibinfo {author} {\bibfnamefont {J.}~\bibnamefont {Maziero}},\ }\href {https://arxiv.org/abs/2505.05158} {\bibinfo {title} {{A short review on qudit quantum machine learning}}} (\bibinfo {year} {2025}{\natexlab{a}})\BibitemShut {NoStop}%
\bibitem [{\citenamefont {Gokhale}\ \emph {et~al.}(2019)\citenamefont {Gokhale}, \citenamefont {Baker}, \citenamefont {Duckering}, \citenamefont {Brown}, \citenamefont {Brown},\ and\ \citenamefont {Chong}}]{gokhale2019asymptotic}%
  \BibitemOpen
  \bibfield  {author} {\bibinfo {author} {\bibfnamefont {P.}~\bibnamefont {Gokhale}}, \bibinfo {author} {\bibfnamefont {J.~M.}\ \bibnamefont {Baker}}, \bibinfo {author} {\bibfnamefont {C.}~\bibnamefont {Duckering}}, \bibinfo {author} {\bibfnamefont {N.~C.}\ \bibnamefont {Brown}}, \bibinfo {author} {\bibfnamefont {K.~R.}\ \bibnamefont {Brown}},\ and\ \bibinfo {author} {\bibfnamefont {F.~T.}\ \bibnamefont {Chong}},\ }\href {https://doi.org/10.1145/3307650.3322253} {\emph {\bibinfo {title} {{Asymptotic improvements to quantum circuits via qutrits}}}}\ (\bibinfo {year} {2019})\ pp.\ \bibinfo {pages} {554--566}\BibitemShut {NoStop}%
\bibitem [{\citenamefont {Pavlidis}\ and\ \citenamefont {Floratos}(2021)}]{pavlidis2021quantum}%
  \BibitemOpen
  \bibfield  {author} {\bibinfo {author} {\bibfnamefont {A.}~\bibnamefont {Pavlidis}}\ and\ \bibinfo {author} {\bibfnamefont {E.}~\bibnamefont {Floratos}},\ }\bibfield  {title} {\bibinfo {title} {{Quantum-Fourier-transform-based quantum arithmetic with qudits}},\ }\bibfield  {journal} {\bibinfo  {journal} {Physical review. A/Physical review, A}\ }\textbf {\bibinfo {volume} {103}},\ \href {https://doi.org/10.1103/physreva.103.032417} {10.1103/physreva.103.032417} (\bibinfo {year} {2021})\BibitemShut {NoStop}%
\bibitem [{\citenamefont {Gedik}\ \emph {et~al.}(2015)\citenamefont {Gedik}, \citenamefont {Silva}, \citenamefont {{\c{C}}akmak}, \citenamefont {Karpat}, \citenamefont {Vidoto}, \citenamefont {Soares-Pinto}, \citenamefont {deAzevedo},\ and\ \citenamefont {Fanchini}}]{gedik2015computational}%
  \BibitemOpen
  \bibfield  {author} {\bibinfo {author} {\bibfnamefont {Z.}~\bibnamefont {Gedik}}, \bibinfo {author} {\bibfnamefont {I.~A.}\ \bibnamefont {Silva}}, \bibinfo {author} {\bibfnamefont {B.}~\bibnamefont {{\c{C}}akmak}}, \bibinfo {author} {\bibfnamefont {G.}~\bibnamefont {Karpat}}, \bibinfo {author} {\bibfnamefont {E.~L.~G.}\ \bibnamefont {Vidoto}}, \bibinfo {author} {\bibfnamefont {D.~O.}\ \bibnamefont {Soares-Pinto}}, \bibinfo {author} {\bibfnamefont {E.~R.}\ \bibnamefont {deAzevedo}},\ and\ \bibinfo {author} {\bibfnamefont {F.~F.}\ \bibnamefont {Fanchini}},\ }\bibfield  {title} {\bibinfo {title} {{Computational speed-up with a single qudit}},\ }\href {https://doi.org/10.1038/srep14671} {\bibfield  {journal} {\bibinfo  {journal} {Scientific Reports}\ }\textbf {\bibinfo {volume} {5}},\ \bibinfo {pages} {14671} (\bibinfo {year} {2015})}\BibitemShut {NoStop}%
\bibitem [{\citenamefont {Deller}\ \emph {et~al.}(2023)\citenamefont {Deller}, \citenamefont {Schmitt}, \citenamefont {Lewenstein}, \citenamefont {Lenk}, \citenamefont {Federer}, \citenamefont {Jendrzejewski}, \citenamefont {Hauke},\ and\ \citenamefont {Kasper}}]{deller-2023}%
  \BibitemOpen
  \bibfield  {author} {\bibinfo {author} {\bibfnamefont {Y.}~\bibnamefont {Deller}}, \bibinfo {author} {\bibfnamefont {S.}~\bibnamefont {Schmitt}}, \bibinfo {author} {\bibfnamefont {M.}~\bibnamefont {Lewenstein}}, \bibinfo {author} {\bibfnamefont {S.}~\bibnamefont {Lenk}}, \bibinfo {author} {\bibfnamefont {M.}~\bibnamefont {Federer}}, \bibinfo {author} {\bibfnamefont {F.}~\bibnamefont {Jendrzejewski}}, \bibinfo {author} {\bibfnamefont {P.}~\bibnamefont {Hauke}},\ and\ \bibinfo {author} {\bibfnamefont {V.}~\bibnamefont {Kasper}},\ }\bibfield  {title} {\bibinfo {title} {{Quantum approximate optimization algorithm for qudit systems}},\ }\bibfield  {journal} {\bibinfo  {journal} {Physical review. A/Physical review, A}\ }\textbf {\bibinfo {volume} {107}},\ \href {https://doi.org/10.1103/physreva.107.062410} {10.1103/physreva.107.062410} (\bibinfo {year} {2023})\BibitemShut {NoStop}%
\bibitem [{\citenamefont {Roca-Jerat}\ \emph {et~al.}(2024)\citenamefont {Roca-Jerat}, \citenamefont {Rom{\'a}n-Roche},\ and\ \citenamefont {Zueco}}]{roca-jerat-2024}%
  \BibitemOpen
  \bibfield  {author} {\bibinfo {author} {\bibfnamefont {S.}~\bibnamefont {Roca-Jerat}}, \bibinfo {author} {\bibfnamefont {J.}~\bibnamefont {Rom{\'a}n-Roche}},\ and\ \bibinfo {author} {\bibfnamefont {D.}~\bibnamefont {Zueco}},\ }\bibfield  {title} {\bibinfo {title} {{Qudit machine learning}},\ }\href {https://doi.org/10.1088/2632-2153/ad360d} {\bibfield  {journal} {\bibinfo  {journal} {Machine Learning Science and Technology}\ }\textbf {\bibinfo {volume} {5}},\ \bibinfo {pages} {015057} (\bibinfo {year} {2024})}\BibitemShut {NoStop}%
\bibitem [{\citenamefont {Anand}\ \emph {et~al.}(2026)\citenamefont {Anand}, \citenamefont {Marshall}, \citenamefont {Saied}, \citenamefont {Rieffel},\ and\ \citenamefont {Morello}}]{anand-2026}%
  \BibitemOpen
  \bibfield  {author} {\bibinfo {author} {\bibfnamefont {N.}~\bibnamefont {Anand}}, \bibinfo {author} {\bibfnamefont {J.}~\bibnamefont {Marshall}}, \bibinfo {author} {\bibfnamefont {J.}~\bibnamefont {Saied}}, \bibinfo {author} {\bibfnamefont {E.}~\bibnamefont {Rieffel}},\ and\ \bibinfo {author} {\bibfnamefont {A.}~\bibnamefont {Morello}},\ }\href {https://arxiv.org/abs/2603.02659} {\bibinfo {title} {{Qudit Designs and Where to Find Them}}} (\bibinfo {year} {2026})\BibitemShut {NoStop}%
\bibitem [{\citenamefont {Wang}\ \emph {et~al.}(2020)\citenamefont {Wang}, \citenamefont {Hu}, \citenamefont {Sanders},\ and\ \citenamefont {Kais}}]{wang-2020}%
  \BibitemOpen
  \bibfield  {author} {\bibinfo {author} {\bibfnamefont {Y.}~\bibnamefont {Wang}}, \bibinfo {author} {\bibfnamefont {Z.}~\bibnamefont {Hu}}, \bibinfo {author} {\bibfnamefont {B.~C.}\ \bibnamefont {Sanders}},\ and\ \bibinfo {author} {\bibfnamefont {S.}~\bibnamefont {Kais}},\ }\bibfield  {title} {\bibinfo {title} {{Qudits and High-Dimensional Quantum Computing}},\ }\bibfield  {journal} {\bibinfo  {journal} {Frontiers in Physics}\ }\textbf {\bibinfo {volume} {8}},\ \href {https://doi.org/10.3389/fphy.2020.589504} {10.3389/fphy.2020.589504} (\bibinfo {year} {2020})\BibitemShut {NoStop}%
\bibitem [{\citenamefont {Ringbauer}\ \emph {et~al.}(2022)\citenamefont {Ringbauer}, \citenamefont {Meth}, \citenamefont {Postler}, \citenamefont {Stricker}, \citenamefont {Blatt}, \citenamefont {Schindler},\ and\ \citenamefont {Monz}}]{ringbauer2022universal}%
  \BibitemOpen
  \bibfield  {author} {\bibinfo {author} {\bibfnamefont {M.}~\bibnamefont {Ringbauer}}, \bibinfo {author} {\bibfnamefont {M.}~\bibnamefont {Meth}}, \bibinfo {author} {\bibfnamefont {L.}~\bibnamefont {Postler}}, \bibinfo {author} {\bibfnamefont {R.}~\bibnamefont {Stricker}}, \bibinfo {author} {\bibfnamefont {R.}~\bibnamefont {Blatt}}, \bibinfo {author} {\bibfnamefont {P.}~\bibnamefont {Schindler}},\ and\ \bibinfo {author} {\bibfnamefont {T.}~\bibnamefont {Monz}},\ }\bibfield  {title} {\bibinfo {title} {{A universal qudit quantum processor with trapped ions}},\ }\href {https://doi.org/10.1038/s41567-022-01658-0} {\bibfield  {journal} {\bibinfo  {journal} {Nature Physics}\ }\textbf {\bibinfo {volume} {18}},\ \bibinfo {pages} {1053} (\bibinfo {year} {2022})}\BibitemShut {NoStop}%
\bibitem [{\citenamefont {Blok}\ \emph {et~al.}(2021)\citenamefont {Blok}, \citenamefont {Ramasesh}, \citenamefont {Schuster}, \citenamefont {O'Brien}, \citenamefont {Kreikebaum}, \citenamefont {Dahlen}, \citenamefont {Morvan}, \citenamefont {Yoshida}, \citenamefont {Yao},\ and\ \citenamefont {Siddiqi}}]{blok2021quantum}%
  \BibitemOpen
  \bibfield  {author} {\bibinfo {author} {\bibfnamefont {M.~S.}\ \bibnamefont {Blok}}, \bibinfo {author} {\bibfnamefont {V.~V.}\ \bibnamefont {Ramasesh}}, \bibinfo {author} {\bibfnamefont {T.}~\bibnamefont {Schuster}}, \bibinfo {author} {\bibfnamefont {K.}~\bibnamefont {O'Brien}}, \bibinfo {author} {\bibfnamefont {J.~M.}\ \bibnamefont {Kreikebaum}}, \bibinfo {author} {\bibfnamefont {D.}~\bibnamefont {Dahlen}}, \bibinfo {author} {\bibfnamefont {A.}~\bibnamefont {Morvan}}, \bibinfo {author} {\bibfnamefont {B.}~\bibnamefont {Yoshida}}, \bibinfo {author} {\bibfnamefont {N.~Y.}\ \bibnamefont {Yao}},\ and\ \bibinfo {author} {\bibfnamefont {I.}~\bibnamefont {Siddiqi}},\ }\bibfield  {title} {\bibinfo {title} {{Quantum Information Scrambling on a Superconducting Qutrit Processor}},\ }\href {https://doi.org/10.1103/PhysRevX.11.021010} {\bibfield  {journal} {\bibinfo  {journal} {Physical Review X}\ }\textbf {\bibinfo {volume} {11}},\ \bibinfo {pages} {021010} (\bibinfo {year} {2021})}\BibitemShut {NoStop}%
\bibitem [{\citenamefont {Mandilara}\ \emph {et~al.}(2024)\citenamefont {Mandilara}, \citenamefont {Dellen}, \citenamefont {Jaekel}, \citenamefont {Valtinos},\ and\ \citenamefont {Syvridis}}]{mandilara2024classification}%
  \BibitemOpen
  \bibfield  {author} {\bibinfo {author} {\bibfnamefont {A.}~\bibnamefont {Mandilara}}, \bibinfo {author} {\bibfnamefont {B.}~\bibnamefont {Dellen}}, \bibinfo {author} {\bibfnamefont {U.}~\bibnamefont {Jaekel}}, \bibinfo {author} {\bibfnamefont {T.}~\bibnamefont {Valtinos}},\ and\ \bibinfo {author} {\bibfnamefont {D.}~\bibnamefont {Syvridis}},\ }\bibfield  {title} {\bibinfo {title} {{Classification of data with a qudit, a geometric approach}},\ }\bibfield  {journal} {\bibinfo  {journal} {Quantum Machine Intelligence}\ }\textbf {\bibinfo {volume} {6}},\ \href {https://doi.org/10.1007/s42484-024-00146-3} {10.1007/s42484-024-00146-3} (\bibinfo {year} {2024})\BibitemShut {NoStop}%
\bibitem [{\citenamefont {Wach}\ \emph {et~al.}(2023)\citenamefont {Wach}, \citenamefont {Rudolph}, \citenamefont {Jendrzejewski},\ and\ \citenamefont {Schmitt}}]{wach2023data}%
  \BibitemOpen
  \bibfield  {author} {\bibinfo {author} {\bibfnamefont {N.~L.}\ \bibnamefont {Wach}}, \bibinfo {author} {\bibfnamefont {M.~S.}\ \bibnamefont {Rudolph}}, \bibinfo {author} {\bibfnamefont {F.}~\bibnamefont {Jendrzejewski}},\ and\ \bibinfo {author} {\bibfnamefont {S.}~\bibnamefont {Schmitt}},\ }\bibfield  {title} {\bibinfo {title} {{Data re-uploading with a single qudit}},\ }\bibfield  {journal} {\bibinfo  {journal} {Quantum Machine Intelligence}\ }\textbf {\bibinfo {volume} {5}},\ \href {https://doi.org/10.1007/s42484-023-00125-0} {10.1007/s42484-023-00125-0} (\bibinfo {year} {2023})\BibitemShut {NoStop}%
\bibitem [{\citenamefont {Valtinos}\ \emph {et~al.}(2024)\citenamefont {Valtinos}, \citenamefont {Mandilara},\ and\ \citenamefont {Syvridis}}]{valtinos2024gell}%
  \BibitemOpen
  \bibfield  {author} {\bibinfo {author} {\bibfnamefont {T.}~\bibnamefont {Valtinos}}, \bibinfo {author} {\bibfnamefont {A.}~\bibnamefont {Mandilara}},\ and\ \bibinfo {author} {\bibfnamefont {D.}~\bibnamefont {Syvridis}},\ }\bibfield  {title} {\bibinfo {title} {{The Gell-Mann feature map of qutrits and its applications in classification tasks}},\ }in\ \href {https://doi.org/10.1117/12.3001127} {\emph {\bibinfo {booktitle} {Quantum Computing, Communication, and Simulation IV}}},\ Vol.\ \bibinfo {volume} {12911},\ \bibinfo {editor} {edited by\ \bibinfo {editor} {\bibfnamefont {P.~R.}\ \bibnamefont {Hemmer}}\ and\ \bibinfo {editor} {\bibfnamefont {A.~L.}\ \bibnamefont {Migdall}}},\ \bibinfo {organization} {International Society for Optics and Photonics}\ (\bibinfo  {publisher} {SPIE},\ \bibinfo {year} {2024})\ p.\ \bibinfo {pages} {129110O}\BibitemShut {NoStop}%
\bibitem [{\citenamefont {Acar}\ and\ \citenamefont {Y{\i}lmaz}(2025)}]{acar2025unlocking}%
  \BibitemOpen
  \bibfield  {author} {\bibinfo {author} {\bibfnamefont {E.}~\bibnamefont {Acar}}\ and\ \bibinfo {author} {\bibfnamefont {{\.{I}}.}~\bibnamefont {Y{\i}lmaz}},\ }\bibfield  {title} {\bibinfo {title} {{Unlocking the high dimensional' potential: Comparative analysis of qubits and qutrits in variational quantum neural networks}},\ }\href {https://doi.org/10.1016/j.neucom.2025.129404} {\bibfield  {journal} {\bibinfo  {journal} {Neurocomputing}\ }\textbf {\bibinfo {volume} {623}},\ \bibinfo {pages} {129404} (\bibinfo {year} {2025})}\BibitemShut {NoStop}%
\bibitem [{\citenamefont {Souza}\ and\ \citenamefont {Portugal}(2025)}]{souza-2025}%
  \BibitemOpen
  \bibfield  {author} {\bibinfo {author} {\bibfnamefont {L.~C.}\ \bibnamefont {Souza}}\ and\ \bibinfo {author} {\bibfnamefont {R.}~\bibnamefont {Portugal}},\ }\bibfield  {title} {\bibinfo {title} {{Single-qudit quantum neural networks for multiclass classification}},\ }\bibfield  {journal} {\bibinfo  {journal} {Quantum Information Processing}\ }\textbf {\bibinfo {volume} {24}},\ \href {https://doi.org/10.1007/s11128-025-04998-x} {10.1007/s11128-025-04998-x} (\bibinfo {year} {2025})\BibitemShut {NoStop}%
\bibitem [{\citenamefont {Useche}\ \emph {et~al.}(2021)\citenamefont {Useche}, \citenamefont {Giraldo-Carvajal}, \citenamefont {Zuluaga-Bucheli}, \citenamefont {Jaramillo-Villegas},\ and\ \citenamefont {Gonz{\'a}lez}}]{useche-2021}%
  \BibitemOpen
  \bibfield  {author} {\bibinfo {author} {\bibfnamefont {D.~H.}\ \bibnamefont {Useche}}, \bibinfo {author} {\bibfnamefont {A.}~\bibnamefont {Giraldo-Carvajal}}, \bibinfo {author} {\bibfnamefont {H.~M.}\ \bibnamefont {Zuluaga-Bucheli}}, \bibinfo {author} {\bibfnamefont {J.~A.}\ \bibnamefont {Jaramillo-Villegas}},\ and\ \bibinfo {author} {\bibfnamefont {F.~A.}\ \bibnamefont {Gonz{\'a}lez}},\ }\bibfield  {title} {\bibinfo {title} {{Quantum measurement classification with qudits}},\ }\bibfield  {journal} {\bibinfo  {journal} {Quantum Information Processing}\ }\textbf {\bibinfo {volume} {21}},\ \href {https://doi.org/10.1007/s11128-021-03363-y} {10.1007/s11128-021-03363-y} (\bibinfo {year} {2021})\BibitemShut {NoStop}%
\bibitem [{\citenamefont {Lai{\~n}o}\ \emph {et~al.}(2025)\citenamefont {Lai{\~n}o}, \citenamefont {Chobanova},\ and\ \citenamefont {Martínez}}]{laino-2025}%
  \BibitemOpen
  \bibfield  {author} {\bibinfo {author} {\bibfnamefont {M.~C.}\ \bibnamefont {Lai{\~n}o}}, \bibinfo {author} {\bibfnamefont {V.}~\bibnamefont {Chobanova}},\ and\ \bibinfo {author} {\bibfnamefont {M.~L.}\ \bibnamefont {Martínez}},\ }\href {https://arxiv.org/abs/2510.14001} {\bibinfo {title} {{Qutrits for physics at the LHC}}} (\bibinfo {year} {2025})\BibitemShut {NoStop}%
\bibitem [{\citenamefont {Havl{\'i}{\v{c}}ek}\ \emph {et~al.}(2019)\citenamefont {Havl{\'i}{\v{c}}ek}, \citenamefont {C{\'o}rcoles}, \citenamefont {Temme}, \citenamefont {Harrow}, \citenamefont {Kandala}, \citenamefont {Chow},\ and\ \citenamefont {Gambetta}}]{havlicek-2019}%
  \BibitemOpen
  \bibfield  {author} {\bibinfo {author} {\bibfnamefont {V.}~\bibnamefont {Havl{\'i}{\v{c}}ek}}, \bibinfo {author} {\bibfnamefont {A.~D.}\ \bibnamefont {C{\'o}rcoles}}, \bibinfo {author} {\bibfnamefont {K.}~\bibnamefont {Temme}}, \bibinfo {author} {\bibfnamefont {A.~W.}\ \bibnamefont {Harrow}}, \bibinfo {author} {\bibfnamefont {A.}~\bibnamefont {Kandala}}, \bibinfo {author} {\bibfnamefont {J.~M.}\ \bibnamefont {Chow}},\ and\ \bibinfo {author} {\bibfnamefont {J.~M.}\ \bibnamefont {Gambetta}},\ }\bibfield  {title} {\bibinfo {title} {{Supervised learning with quantum-enhanced feature spaces}},\ }\href {https://doi.org/10.1038/s41586-019-0980-2} {\bibfield  {journal} {\bibinfo  {journal} {Nature}\ }\textbf {\bibinfo {volume} {567}},\ \bibinfo {pages} {209} (\bibinfo {year} {2019})}\BibitemShut {NoStop}%
\bibitem [{\citenamefont {Schuld}\ and\ \citenamefont {Killoran}(2019)}]{schuld-2019-hilbert-spaces}%
  \BibitemOpen
  \bibfield  {author} {\bibinfo {author} {\bibfnamefont {M.}~\bibnamefont {Schuld}}\ and\ \bibinfo {author} {\bibfnamefont {N.}~\bibnamefont {Killoran}},\ }\bibfield  {title} {\bibinfo {title} {{Quantum Machine Learning in Feature Hilbert Spaces}},\ }\href {https://doi.org/10.1103/physrevlett.122.040504} {\bibfield  {journal} {\bibinfo  {journal} {Physical Review Letters}\ }\textbf {\bibinfo {volume} {122}},\ \bibinfo {pages} {040504} (\bibinfo {year} {2019})}\BibitemShut {NoStop}%
\bibitem [{\citenamefont {Bartkiewicz}\ \emph {et~al.}(2020)\citenamefont {Bartkiewicz}, \citenamefont {Gneiting}, \citenamefont {{\v{C}}ernoch}, \citenamefont {Jir{\'a}kov{\'a}}, \citenamefont {Lemr},\ and\ \citenamefont {Nori}}]{bartkiewicz-2020}%
  \BibitemOpen
  \bibfield  {author} {\bibinfo {author} {\bibfnamefont {K.}~\bibnamefont {Bartkiewicz}}, \bibinfo {author} {\bibfnamefont {C.}~\bibnamefont {Gneiting}}, \bibinfo {author} {\bibfnamefont {A.}~\bibnamefont {{\v{C}}ernoch}}, \bibinfo {author} {\bibfnamefont {K.}~\bibnamefont {Jir{\'a}kov{\'a}}}, \bibinfo {author} {\bibfnamefont {K.}~\bibnamefont {Lemr}},\ and\ \bibinfo {author} {\bibfnamefont {F.}~\bibnamefont {Nori}},\ }\bibfield  {title} {\bibinfo {title} {{Experimental kernel-based quantum machine learning in finite feature space}},\ }\href {https://doi.org/10.1038/s41598-020-68911-5} {\bibfield  {journal} {\bibinfo  {journal} {Scientific Reports}\ }\textbf {\bibinfo {volume} {10}},\ \bibinfo {pages} {12356} (\bibinfo {year} {2020})}\BibitemShut {NoStop}%
\bibitem [{\citenamefont {Huang}\ \emph {et~al.}(2021)\citenamefont {Huang}, \citenamefont {Broughton}, \citenamefont {Mohseni}, \citenamefont {Babbush}, \citenamefont {Boixo}, \citenamefont {Neven},\ and\ \citenamefont {McClean}}]{huang-2021}%
  \BibitemOpen
  \bibfield  {author} {\bibinfo {author} {\bibfnamefont {H.-Y.}\ \bibnamefont {Huang}}, \bibinfo {author} {\bibfnamefont {M.}~\bibnamefont {Broughton}}, \bibinfo {author} {\bibfnamefont {M.}~\bibnamefont {Mohseni}}, \bibinfo {author} {\bibfnamefont {R.}~\bibnamefont {Babbush}}, \bibinfo {author} {\bibfnamefont {S.}~\bibnamefont {Boixo}}, \bibinfo {author} {\bibfnamefont {H.}~\bibnamefont {Neven}},\ and\ \bibinfo {author} {\bibfnamefont {J.~R.}\ \bibnamefont {McClean}},\ }\bibfield  {title} {\bibinfo {title} {{Power of data in quantum machine learning}},\ }\href {https://doi.org/10.1038/s41467-021-22539-9} {\bibfield  {journal} {\bibinfo  {journal} {Nature Communications}\ }\textbf {\bibinfo {volume} {12}},\ \bibinfo {pages} {2631} (\bibinfo {year} {2021})}\BibitemShut {NoStop}%
\bibitem [{\citenamefont {Peters}\ \emph {et~al.}(2021)\citenamefont {Peters}, \citenamefont {Caldeira}, \citenamefont {Ho}, \citenamefont {Leichenauer}, \citenamefont {Mohseni}, \citenamefont {Neven}, \citenamefont {Spentzouris}, \citenamefont {Strain},\ and\ \citenamefont {Perdue}}]{peters-2021}%
  \BibitemOpen
  \bibfield  {author} {\bibinfo {author} {\bibfnamefont {E.}~\bibnamefont {Peters}}, \bibinfo {author} {\bibfnamefont {J.}~\bibnamefont {Caldeira}}, \bibinfo {author} {\bibfnamefont {A.}~\bibnamefont {Ho}}, \bibinfo {author} {\bibfnamefont {S.}~\bibnamefont {Leichenauer}}, \bibinfo {author} {\bibfnamefont {M.}~\bibnamefont {Mohseni}}, \bibinfo {author} {\bibfnamefont {H.}~\bibnamefont {Neven}}, \bibinfo {author} {\bibfnamefont {P.}~\bibnamefont {Spentzouris}}, \bibinfo {author} {\bibfnamefont {D.}~\bibnamefont {Strain}},\ and\ \bibinfo {author} {\bibfnamefont {G.~N.}\ \bibnamefont {Perdue}},\ }\bibfield  {title} {\bibinfo {title} {{Machine learning of high dimensional data on a noisy quantum processor}},\ }\bibfield  {journal} {\bibinfo  {journal} {npj Quantum Information}\ }\textbf {\bibinfo {volume} {7}},\ \href {https://doi.org/10.1038/s41534-021-00498-9} {10.1038/s41534-021-00498-9} (\bibinfo {year} {2021})\BibitemShut {NoStop}%
\bibitem [{\citenamefont {Kusumoto}\ \emph {et~al.}(2021)\citenamefont {Kusumoto}, \citenamefont {Mitarai}, \citenamefont {Fujii}, \citenamefont {Kitagawa},\ and\ \citenamefont {Negoro}}]{kusumoto-2021}%
  \BibitemOpen
  \bibfield  {author} {\bibinfo {author} {\bibfnamefont {T.}~\bibnamefont {Kusumoto}}, \bibinfo {author} {\bibfnamefont {K.}~\bibnamefont {Mitarai}}, \bibinfo {author} {\bibfnamefont {K.}~\bibnamefont {Fujii}}, \bibinfo {author} {\bibfnamefont {M.}~\bibnamefont {Kitagawa}},\ and\ \bibinfo {author} {\bibfnamefont {M.}~\bibnamefont {Negoro}},\ }\bibfield  {title} {\bibinfo {title} {{Experimental quantum kernel trick with nuclear spins in a solid}},\ }\href {https://doi.org/10.1038/s41534-021-00423-0} {\bibfield  {journal} {\bibinfo  {journal} {npj Quantum Information}\ }\textbf {\bibinfo {volume} {7}},\ \bibinfo {pages} {94} (\bibinfo {year} {2021})}\BibitemShut {NoStop}%
\bibitem [{\citenamefont {Schuld}(2021)}]{schuld-2021}%
  \BibitemOpen
  \bibfield  {author} {\bibinfo {author} {\bibfnamefont {M.}~\bibnamefont {Schuld}},\ }\href {https://arxiv.org/abs/2101.11020} {\bibinfo {title} {{Supervised quantum machine learning models are kernel methods}}} (\bibinfo {year} {2021})\BibitemShut {NoStop}%
\bibitem [{\citenamefont {Liu}\ \emph {et~al.}(2021{\natexlab{b}})\citenamefont {Liu}, \citenamefont {Arunachalam},\ and\ \citenamefont {Temme}}]{liu-2021}%
  \BibitemOpen
  \bibfield  {author} {\bibinfo {author} {\bibfnamefont {Y.}~\bibnamefont {Liu}}, \bibinfo {author} {\bibfnamefont {S.}~\bibnamefont {Arunachalam}},\ and\ \bibinfo {author} {\bibfnamefont {K.}~\bibnamefont {Temme}},\ }\bibfield  {title} {\bibinfo {title} {{A rigorous and robust quantum speed-up in supervised machine learning}},\ }\href {https://doi.org/10.1038/s41567-021-01287-z} {\bibfield  {journal} {\bibinfo  {journal} {Nature Physics}\ }\textbf {\bibinfo {volume} {17}},\ \bibinfo {pages} {1013} (\bibinfo {year} {2021}{\natexlab{b}})}\BibitemShut {NoStop}%
\bibitem [{\citenamefont {Kyriienko}\ and\ \citenamefont {Magnusson}(2022)}]{kyriienko-2022}%
  \BibitemOpen
  \bibfield  {author} {\bibinfo {author} {\bibfnamefont {O.}~\bibnamefont {Kyriienko}}\ and\ \bibinfo {author} {\bibfnamefont {E.~B.}\ \bibnamefont {Magnusson}},\ }\href {https://arxiv.org/abs/2208.01203} {\bibinfo {title} {{Unsupervised quantum machine learning for fraud detection}}} (\bibinfo {year} {2022})\BibitemShut {NoStop}%
\bibitem [{\citenamefont {Wu}\ \emph {et~al.}(2023)\citenamefont {Wu}, \citenamefont {Wu}, \citenamefont {Wang},\ and\ \citenamefont {Yuan}}]{wu-2023}%
  \BibitemOpen
  \bibfield  {author} {\bibinfo {author} {\bibfnamefont {Y.}~\bibnamefont {Wu}}, \bibinfo {author} {\bibfnamefont {B.}~\bibnamefont {Wu}}, \bibinfo {author} {\bibfnamefont {J.}~\bibnamefont {Wang}},\ and\ \bibinfo {author} {\bibfnamefont {X.}~\bibnamefont {Yuan}},\ }\bibfield  {title} {\bibinfo {title} {{Quantum Phase Recognition via Quantum Kernel Methods}},\ }\href {https://doi.org/10.22331/q-2023-04-17-981} {\bibfield  {journal} {\bibinfo  {journal} {Quantum}\ }\textbf {\bibinfo {volume} {7}},\ \bibinfo {pages} {981} (\bibinfo {year} {2023})}\BibitemShut {NoStop}%
\bibitem [{\citenamefont {Tomasi}\ \emph {et~al.}(2025)\citenamefont {Tomasi}, \citenamefont {Anthoine},\ and\ \citenamefont {Kadri}}]{tomasi-2025}%
  \BibitemOpen
  \bibfield  {author} {\bibinfo {author} {\bibfnamefont {J.}~\bibnamefont {Tomasi}}, \bibinfo {author} {\bibfnamefont {S.}~\bibnamefont {Anthoine}},\ and\ \bibinfo {author} {\bibfnamefont {H.}~\bibnamefont {Kadri}},\ }\href {https://arxiv.org/abs/2503.17020} {\bibinfo {title} {{Benign Overfitting with Quantum Kernels}}} (\bibinfo {year} {2025})\BibitemShut {NoStop}%
\bibitem [{\citenamefont {Rodriguez-Grasa}\ \emph {et~al.}(2025{\natexlab{a}})\citenamefont {Rodriguez-Grasa}, \citenamefont {Ban},\ and\ \citenamefont {Sanz}}]{nqk_pablo}%
  \BibitemOpen
  \bibfield  {author} {\bibinfo {author} {\bibfnamefont {P.}~\bibnamefont {Rodriguez-Grasa}}, \bibinfo {author} {\bibfnamefont {Y.}~\bibnamefont {Ban}},\ and\ \bibinfo {author} {\bibfnamefont {M.}~\bibnamefont {Sanz}},\ }\bibfield  {title} {\bibinfo {title} {{Neural quantum kernels: Training quantum kernels with quantum neural networks}},\ }\bibfield  {journal} {\bibinfo  {journal} {Physical Review Research}\ }\textbf {\bibinfo {volume} {7}},\ \href {https://doi.org/10.1103/xphb-x2g4} {10.1103/xphb-x2g4} (\bibinfo {year} {2025}{\natexlab{a}})\BibitemShut {NoStop}%
\bibitem [{\citenamefont {Rodriguez-Grasa}\ \emph {et~al.}(2025{\natexlab{b}})\citenamefont {Rodriguez-Grasa}, \citenamefont {Farzan-Rodriguez}, \citenamefont {Novelli}, \citenamefont {Ban},\ and\ \citenamefont {Sanz}}]{nqk_satellite}%
  \BibitemOpen
  \bibfield  {author} {\bibinfo {author} {\bibfnamefont {P.}~\bibnamefont {Rodriguez-Grasa}}, \bibinfo {author} {\bibfnamefont {R.}~\bibnamefont {Farzan-Rodriguez}}, \bibinfo {author} {\bibfnamefont {G.}~\bibnamefont {Novelli}}, \bibinfo {author} {\bibfnamefont {Y.}~\bibnamefont {Ban}},\ and\ \bibinfo {author} {\bibfnamefont {M.}~\bibnamefont {Sanz}},\ }\bibfield  {title} {\bibinfo {title} {{Satellite image classification with neural quantum kernels}},\ }\href {https://doi.org/10.1088/2632-2153/ada86c} {\bibfield  {journal} {\bibinfo  {journal} {Machine Learning Science and Technology}\ }\textbf {\bibinfo {volume} {6}},\ \bibinfo {pages} {015043} (\bibinfo {year} {2025}{\natexlab{b}})}\BibitemShut {NoStop}%
\bibitem [{\citenamefont {Rodriguez-Grasa}\ \emph {et~al.}(2025{\natexlab{c}})\citenamefont {Rodriguez-Grasa}, \citenamefont {Zhelnin}, \citenamefont {Arg{\"u}elles},\ and\ \citenamefont {Sanz}}]{nqk_neutrinos}%
  \BibitemOpen
  \bibfield  {author} {\bibinfo {author} {\bibfnamefont {P.}~\bibnamefont {Rodriguez-Grasa}}, \bibinfo {author} {\bibfnamefont {P.}~\bibnamefont {Zhelnin}}, \bibinfo {author} {\bibfnamefont {C.~A.}\ \bibnamefont {Arg{\"u}elles}},\ and\ \bibinfo {author} {\bibfnamefont {M.}~\bibnamefont {Sanz}},\ }\href {https://arxiv.org/abs/2506.16530} {\bibinfo {title} {{Neutrino Telescope Event Classification on Quantum Computers}}} (\bibinfo {year} {2025}{\natexlab{c}})\BibitemShut {NoStop}%
\bibitem [{\citenamefont {Sch{\"o}lkopf}\ and\ \citenamefont {Smola}(2002)}]{scholkopf2002learning}%
  \BibitemOpen
  \bibfield  {author} {\bibinfo {author} {\bibfnamefont {B.}~\bibnamefont {Sch{\"o}lkopf}}\ and\ \bibinfo {author} {\bibfnamefont {A.~J.}\ \bibnamefont {Smola}},\ }\href@noop {} {\emph {\bibinfo {title} {Learning with Kernels: Support Vector Machines, Regularization, Optimization, and Beyond}}},\ Adaptive Computation and Machine Learning\ (\bibinfo  {publisher} {MIT Press},\ \bibinfo {address} {Cambridge, MA},\ \bibinfo {year} {2002})\ p.\ \bibinfo {pages} {644}\BibitemShut {NoStop}%
\bibitem [{\citenamefont {Schuld}\ and\ \citenamefont {Petruccione}(2021)}]{schuld2021quantum}%
  \BibitemOpen
  \bibfield  {author} {\bibinfo {author} {\bibfnamefont {M.}~\bibnamefont {Schuld}}\ and\ \bibinfo {author} {\bibfnamefont {F.}~\bibnamefont {Petruccione}},\ }\bibinfo {title} {Quantum models as kernel methods},\ in\ \href {https://doi.org/10.1007/978-3-030-63023-2_6} {\emph {\bibinfo {booktitle} {Machine Learning with Quantum Computers}}}\ (\bibinfo  {publisher} {Springer International Publishing},\ \bibinfo {address} {Cham},\ \bibinfo {year} {2021})\ pp.\ \bibinfo {pages} {217--245}\BibitemShut {NoStop}%
\bibitem [{\citenamefont {Buhrman}\ \emph {et~al.}(2001)\citenamefont {Buhrman}, \citenamefont {Cleve}, \citenamefont {Watrous},\ and\ \citenamefont {De~Wolf}}]{buhrman-2001}%
  \BibitemOpen
  \bibfield  {author} {\bibinfo {author} {\bibfnamefont {H.}~\bibnamefont {Buhrman}}, \bibinfo {author} {\bibfnamefont {R.}~\bibnamefont {Cleve}}, \bibinfo {author} {\bibfnamefont {J.}~\bibnamefont {Watrous}},\ and\ \bibinfo {author} {\bibfnamefont {R.}~\bibnamefont {De~Wolf}},\ }\bibfield  {title} {\bibinfo {title} {{Quantum fingerprinting}},\ }\href {https://doi.org/10.1103/physrevlett.87.167902} {\bibfield  {journal} {\bibinfo  {journal} {Physical Review Letters}\ }\textbf {\bibinfo {volume} {87}},\ \bibinfo {pages} {167902} (\bibinfo {year} {2001})}\BibitemShut {NoStop}%
\bibitem [{\citenamefont {Fanizza}\ \emph {et~al.}(2020)\citenamefont {Fanizza}, \citenamefont {Rosati}, \citenamefont {Skotiniotis}, \citenamefont {Calsamiglia},\ and\ \citenamefont {Giovannetti}}]{fanizza-2020}%
  \BibitemOpen
  \bibfield  {author} {\bibinfo {author} {\bibfnamefont {M.}~\bibnamefont {Fanizza}}, \bibinfo {author} {\bibfnamefont {M.}~\bibnamefont {Rosati}}, \bibinfo {author} {\bibfnamefont {M.}~\bibnamefont {Skotiniotis}}, \bibinfo {author} {\bibfnamefont {J.}~\bibnamefont {Calsamiglia}},\ and\ \bibinfo {author} {\bibfnamefont {V.}~\bibnamefont {Giovannetti}},\ }\bibfield  {title} {\bibinfo {title} {{Beyond the Swap Test: Optimal Estimation of Quantum State Overlap}},\ }\href {https://doi.org/10.1103/physrevlett.124.060503} {\bibfield  {journal} {\bibinfo  {journal} {Physical Review Letters}\ }\textbf {\bibinfo {volume} {124}},\ \bibinfo {pages} {060503} (\bibinfo {year} {2020})}\BibitemShut {NoStop}%
\bibitem [{\citenamefont {Cincio}\ \emph {et~al.}(2018)\citenamefont {Cincio}, \citenamefont {Suba{\c{s}}{\i}}, \citenamefont {Sornborger},\ and\ \citenamefont {Coles}}]{cincio-2018}%
  \BibitemOpen
  \bibfield  {author} {\bibinfo {author} {\bibfnamefont {L.}~\bibnamefont {Cincio}}, \bibinfo {author} {\bibfnamefont {Y.}~\bibnamefont {Suba{\c{s}}{\i}}}, \bibinfo {author} {\bibfnamefont {A.~T.}\ \bibnamefont {Sornborger}},\ and\ \bibinfo {author} {\bibfnamefont {P.~J.}\ \bibnamefont {Coles}},\ }\bibfield  {title} {\bibinfo {title} {{Learning the quantum algorithm for state overlap}},\ }\href {https://doi.org/10.1088/1367-2630/aae94a} {\bibfield  {journal} {\bibinfo  {journal} {New Journal of Physics}\ }\textbf {\bibinfo {volume} {20}},\ \bibinfo {pages} {113022} (\bibinfo {year} {2018})}\BibitemShut {NoStop}%
\bibitem [{\citenamefont {Gil-Fuster}\ \emph {et~al.}(2024)\citenamefont {Gil-Fuster}, \citenamefont {Eisert},\ and\ \citenamefont {Dunjko}}]{gil2024expressivity}%
  \BibitemOpen
  \bibfield  {author} {\bibinfo {author} {\bibfnamefont {E.}~\bibnamefont {Gil-Fuster}}, \bibinfo {author} {\bibfnamefont {J.}~\bibnamefont {Eisert}},\ and\ \bibinfo {author} {\bibfnamefont {V.}~\bibnamefont {Dunjko}},\ }\bibfield  {title} {\bibinfo {title} {On the expressivity of embedding quantum kernels},\ }\href {https://doi.org/10.1088/2632-2153/ad2f51} {\bibfield  {journal} {\bibinfo  {journal} {Machine Learning: Science and Technology}\ }\textbf {\bibinfo {volume} {5}},\ \bibinfo {pages} {025003} (\bibinfo {year} {2024})}\BibitemShut {NoStop}%
\bibitem [{\citenamefont {Lloyd}\ \emph {et~al.}(2020)\citenamefont {Lloyd}, \citenamefont {Schuld}, \citenamefont {Ijaz}, \citenamefont {Izaac},\ and\ \citenamefont {Killoran}}]{lloyd-2020}%
  \BibitemOpen
  \bibfield  {author} {\bibinfo {author} {\bibfnamefont {S.}~\bibnamefont {Lloyd}}, \bibinfo {author} {\bibfnamefont {M.}~\bibnamefont {Schuld}}, \bibinfo {author} {\bibfnamefont {A.}~\bibnamefont {Ijaz}}, \bibinfo {author} {\bibfnamefont {J.}~\bibnamefont {Izaac}},\ and\ \bibinfo {author} {\bibfnamefont {N.}~\bibnamefont {Killoran}},\ }\href {https://arxiv.org/abs/2001.03622} {\bibinfo {title} {{Quantum embeddings for machine learning}}} (\bibinfo {year} {2020})\BibitemShut {NoStop}%
\bibitem [{\citenamefont {Blank}\ \emph {et~al.}(2020)\citenamefont {Blank}, \citenamefont {Park}, \citenamefont {Rhee},\ and\ \citenamefont {Petruccione}}]{blank-2020}%
  \BibitemOpen
  \bibfield  {author} {\bibinfo {author} {\bibfnamefont {C.}~\bibnamefont {Blank}}, \bibinfo {author} {\bibfnamefont {D.~K.}\ \bibnamefont {Park}}, \bibinfo {author} {\bibfnamefont {J.-K.~K.}\ \bibnamefont {Rhee}},\ and\ \bibinfo {author} {\bibfnamefont {F.}~\bibnamefont {Petruccione}},\ }\bibfield  {title} {\bibinfo {title} {{Quantum classifier with tailored quantum kernel}},\ }\href {https://doi.org/10.1038/s41534-020-0272-6} {\bibfield  {journal} {\bibinfo  {journal} {npj Quantum Information}\ }\textbf {\bibinfo {volume} {6}},\ \bibinfo {pages} {57} (\bibinfo {year} {2020})}\BibitemShut {NoStop}%
\bibitem [{\citenamefont {K\"{u}bler}\ \emph {et~al.}(2021)\citenamefont {K\"{u}bler}, \citenamefont {Buchholz},\ and\ \citenamefont {Sch\"{o}lkopf}}]{kubler-2021}%
  \BibitemOpen
  \bibfield  {author} {\bibinfo {author} {\bibfnamefont {J.~M.}\ \bibnamefont {K\"{u}bler}}, \bibinfo {author} {\bibfnamefont {S.}~\bibnamefont {Buchholz}},\ and\ \bibinfo {author} {\bibfnamefont {B.}~\bibnamefont {Sch\"{o}lkopf}},\ }\bibfield  {title} {\bibinfo {title} {The inductive bias of quantum kernels},\ }in\ \href@noop {} {\emph {\bibinfo {booktitle} {Proceedings of the 35th International Conference on Neural Information Processing Systems}}},\ \bibinfo {series and number} {NIPS '21}\ (\bibinfo  {publisher} {Curran Associates Inc.},\ \bibinfo {address} {Red Hook, NY, USA},\ \bibinfo {year} {2021})\BibitemShut {NoStop}%
\bibitem [{\citenamefont {Salmenper{\"a}}\ \emph {et~al.}(2024)\citenamefont {Salmenper{\"a}}, \citenamefont {Kuhtarskis}, \citenamefont {De~Griend Arianne~Meijer},\ and\ \citenamefont {Nurminen}}]{salmenpera-2024}%
  \BibitemOpen
  \bibfield  {author} {\bibinfo {author} {\bibfnamefont {I.}~\bibnamefont {Salmenper{\"a}}}, \bibinfo {author} {\bibfnamefont {I.}~\bibnamefont {Kuhtarskis}}, \bibinfo {author} {\bibfnamefont {V.}~\bibnamefont {De~Griend Arianne~Meijer}},\ and\ \bibinfo {author} {\bibfnamefont {J.~K.}\ \bibnamefont {Nurminen}},\ }\bibfield  {title} {\bibinfo {title} {{The Impact of Feature Embedding Placement in the Ansatz of a Quantum Kernel in QSVMs}},\ }\bibfield  {journal} {\bibinfo  {journal} {arXiv (Cornell University)}\ }\href {https://doi.org/10.48550/arxiv.2409.13147} {10.48550/arxiv.2409.13147} (\bibinfo {year} {2024})\BibitemShut {NoStop}%
\bibitem [{\citenamefont {Shirai}\ \emph {et~al.}(2024)\citenamefont {Shirai}, \citenamefont {Kubo}, \citenamefont {Mitarai},\ and\ \citenamefont {Fujii}}]{shirai-2024}%
  \BibitemOpen
  \bibfield  {author} {\bibinfo {author} {\bibfnamefont {N.}~\bibnamefont {Shirai}}, \bibinfo {author} {\bibfnamefont {K.}~\bibnamefont {Kubo}}, \bibinfo {author} {\bibfnamefont {K.}~\bibnamefont {Mitarai}},\ and\ \bibinfo {author} {\bibfnamefont {K.}~\bibnamefont {Fujii}},\ }\bibfield  {title} {\bibinfo {title} {{Quantum tangent kernel}},\ }\bibfield  {journal} {\bibinfo  {journal} {Physical Review Research}\ }\textbf {\bibinfo {volume} {6}},\ \href {https://doi.org/10.1103/physrevresearch.6.033179} {10.1103/physrevresearch.6.033179} (\bibinfo {year} {2024})\BibitemShut {NoStop}%
\bibitem [{\citenamefont {Vedaie}\ \emph {et~al.}(2020)\citenamefont {Vedaie}, \citenamefont {Noori}, \citenamefont {Oberoi}, \citenamefont {Sanders},\ and\ \citenamefont {Zahedinejad}}]{vedaie-2020}%
  \BibitemOpen
  \bibfield  {author} {\bibinfo {author} {\bibfnamefont {S.~S.}\ \bibnamefont {Vedaie}}, \bibinfo {author} {\bibfnamefont {M.}~\bibnamefont {Noori}}, \bibinfo {author} {\bibfnamefont {J.~S.}\ \bibnamefont {Oberoi}}, \bibinfo {author} {\bibfnamefont {B.~C.}\ \bibnamefont {Sanders}},\ and\ \bibinfo {author} {\bibfnamefont {E.}~\bibnamefont {Zahedinejad}},\ }\href {https://arxiv.org/abs/2011.09694} {\bibinfo {title} {{Quantum Multiple Kernel Learning}}} (\bibinfo {year} {2020})\BibitemShut {NoStop}%
\bibitem [{\citenamefont {Hubregtsen}\ \emph {et~al.}(2022)\citenamefont {Hubregtsen}, \citenamefont {Wierichs}, \citenamefont {Gil-Fuster}, \citenamefont {Derks}, \citenamefont {Faehrmann},\ and\ \citenamefont {Meyer}}]{hubregtsen-2022}%
  \BibitemOpen
  \bibfield  {author} {\bibinfo {author} {\bibfnamefont {T.}~\bibnamefont {Hubregtsen}}, \bibinfo {author} {\bibfnamefont {D.}~\bibnamefont {Wierichs}}, \bibinfo {author} {\bibfnamefont {E.}~\bibnamefont {Gil-Fuster}}, \bibinfo {author} {\bibfnamefont {P.-J. H.~S.}\ \bibnamefont {Derks}}, \bibinfo {author} {\bibfnamefont {P.~K.}\ \bibnamefont {Faehrmann}},\ and\ \bibinfo {author} {\bibfnamefont {J.~J.}\ \bibnamefont {Meyer}},\ }\bibfield  {title} {\bibinfo {title} {{Training quantum embedding kernels on near-term quantum computers}},\ }\bibfield  {journal} {\bibinfo  {journal} {Physical review. A/Physical review, A}\ }\textbf {\bibinfo {volume} {106}},\ \href {https://doi.org/10.1103/physreva.106.042431} {10.1103/physreva.106.042431} (\bibinfo {year} {2022})\BibitemShut {NoStop}%
\bibitem [{\citenamefont {Ghukasyan}\ \emph {et~al.}(2023)\citenamefont {Ghukasyan}, \citenamefont {Baker}, \citenamefont {Goktas}, \citenamefont {Carrasquilla},\ and\ \citenamefont {Radha}}]{ghukasyan-2023}%
  \BibitemOpen
  \bibfield  {author} {\bibinfo {author} {\bibfnamefont {A.}~\bibnamefont {Ghukasyan}}, \bibinfo {author} {\bibfnamefont {J.~S.}\ \bibnamefont {Baker}}, \bibinfo {author} {\bibfnamefont {O.}~\bibnamefont {Goktas}}, \bibinfo {author} {\bibfnamefont {J.}~\bibnamefont {Carrasquilla}},\ and\ \bibinfo {author} {\bibfnamefont {S.~K.}\ \bibnamefont {Radha}},\ }\href {https://arxiv.org/abs/2305.17707} {\bibinfo {title} {{Quantum-Classical Multiple kernel learning}}} (\bibinfo {year} {2023})\BibitemShut {NoStop}%
\bibitem [{\citenamefont {Thanasilp}\ \emph {et~al.}(2023)\citenamefont {Thanasilp}, \citenamefont {Wang}, \citenamefont {Nghiem}, \citenamefont {Coles},\ and\ \citenamefont {Cerezo}}]{thanasilp2023subtleties}%
  \BibitemOpen
  \bibfield  {author} {\bibinfo {author} {\bibfnamefont {S.}~\bibnamefont {Thanasilp}}, \bibinfo {author} {\bibfnamefont {S.}~\bibnamefont {Wang}}, \bibinfo {author} {\bibfnamefont {N.~A.}\ \bibnamefont {Nghiem}}, \bibinfo {author} {\bibfnamefont {P.}~\bibnamefont {Coles}},\ and\ \bibinfo {author} {\bibfnamefont {M.}~\bibnamefont {Cerezo}},\ }\bibfield  {title} {\bibinfo {title} {Subtleties in the trainability of quantum machine learning models},\ }\href@noop {} {\bibfield  {journal} {\bibinfo  {journal} {Quantum Machine Intelligence}\ }\textbf {\bibinfo {volume} {5}},\ \bibinfo {pages} {21} (\bibinfo {year} {2023})}\BibitemShut {NoStop}%
\bibitem [{\citenamefont {Shaydulin}\ and\ \citenamefont {Wild}(2022)}]{shaydulin-2022}%
  \BibitemOpen
  \bibfield  {author} {\bibinfo {author} {\bibfnamefont {R.}~\bibnamefont {Shaydulin}}\ and\ \bibinfo {author} {\bibfnamefont {S.~M.}\ \bibnamefont {Wild}},\ }\bibfield  {title} {\bibinfo {title} {{Importance of kernel bandwidth in quantum machine learning}},\ }\bibfield  {journal} {\bibinfo  {journal} {Physical review. A/Physical review, A}\ }\textbf {\bibinfo {volume} {106}},\ \href {https://doi.org/10.1103/physreva.106.042407} {10.1103/physreva.106.042407} (\bibinfo {year} {2022})\BibitemShut {NoStop}%
\bibitem [{\citenamefont {Kairon}\ \emph {et~al.}(2025)\citenamefont {Kairon}, \citenamefont {J{\"a}ger},\ and\ \citenamefont {Krems}}]{kairon2025equivalenceexponentialconcentrationquantum}%
  \BibitemOpen
  \bibfield  {author} {\bibinfo {author} {\bibfnamefont {P.}~\bibnamefont {Kairon}}, \bibinfo {author} {\bibfnamefont {J.}~\bibnamefont {J{\"a}ger}},\ and\ \bibinfo {author} {\bibfnamefont {R.~V.}\ \bibnamefont {Krems}},\ }\href {https://arxiv.org/abs/2501.07433} {\bibinfo {title} {Equivalence between exponential concentration in quantum machine learning kernels and barren plateaus in variational algorithms}} (\bibinfo {year} {2025}),\ \Eprint {https://arxiv.org/abs/2501.07433} {arXiv:2501.07433 [quant-ph]} \BibitemShut {NoStop}%
\bibitem [{\citenamefont {Egger}\ \emph {et~al.}(2021)\citenamefont {Egger}, \citenamefont {Mare{\v{c}}ek},\ and\ \citenamefont {Woerner}}]{egger-2021}%
  \BibitemOpen
  \bibfield  {author} {\bibinfo {author} {\bibfnamefont {D.~J.}\ \bibnamefont {Egger}}, \bibinfo {author} {\bibfnamefont {J.}~\bibnamefont {Mare{\v{c}}ek}},\ and\ \bibinfo {author} {\bibfnamefont {S.}~\bibnamefont {Woerner}},\ }\bibfield  {title} {\bibinfo {title} {{Warm-starting quantum optimization}},\ }\href {https://doi.org/10.22331/q-2021-06-17-479} {\bibfield  {journal} {\bibinfo  {journal} {Quantum}\ }\textbf {\bibinfo {volume} {5}},\ \bibinfo {pages} {479} (\bibinfo {year} {2021})}\BibitemShut {NoStop}%
\bibitem [{\citenamefont {Mhiri}\ \emph {et~al.}(2025)\citenamefont {Mhiri}, \citenamefont {Puig}, \citenamefont {Lerch}, \citenamefont {Rudolph}, \citenamefont {Chotibut}, \citenamefont {Thanasilp},\ and\ \citenamefont {Holmes}}]{mhiri-2025}%
  \BibitemOpen
  \bibfield  {author} {\bibinfo {author} {\bibfnamefont {H.}~\bibnamefont {Mhiri}}, \bibinfo {author} {\bibfnamefont {R.}~\bibnamefont {Puig}}, \bibinfo {author} {\bibfnamefont {S.}~\bibnamefont {Lerch}}, \bibinfo {author} {\bibfnamefont {M.~S.}\ \bibnamefont {Rudolph}}, \bibinfo {author} {\bibfnamefont {T.}~\bibnamefont {Chotibut}}, \bibinfo {author} {\bibfnamefont {S.}~\bibnamefont {Thanasilp}},\ and\ \bibinfo {author} {\bibfnamefont {Z.}~\bibnamefont {Holmes}},\ }\href {https://arxiv.org/abs/2502.07889} {\bibinfo {title} {{A unifying account of warm start guarantees for patches of quantum landscapes}}} (\bibinfo {year} {2025})\BibitemShut {NoStop}%
\bibitem [{\citenamefont {Puig}\ \emph {et~al.}(2025)\citenamefont {Puig}, \citenamefont {Drudis}, \citenamefont {Thanasilp},\ and\ \citenamefont {Holmes}}]{puig-2025}%
  \BibitemOpen
  \bibfield  {author} {\bibinfo {author} {\bibfnamefont {R.}~\bibnamefont {Puig}}, \bibinfo {author} {\bibfnamefont {M.}~\bibnamefont {Drudis}}, \bibinfo {author} {\bibfnamefont {S.}~\bibnamefont {Thanasilp}},\ and\ \bibinfo {author} {\bibfnamefont {Z.}~\bibnamefont {Holmes}},\ }\bibfield  {title} {\bibinfo {title} {{Variational Quantum Simulation: A Case Study for Understanding Warm Starts}},\ }\bibfield  {journal} {\bibinfo  {journal} {PRX Quantum}\ }\textbf {\bibinfo {volume} {6}},\ \href {https://doi.org/10.1103/prxquantum.6.010317} {10.1103/prxquantum.6.010317} (\bibinfo {year} {2025})\BibitemShut {NoStop}%
\bibitem [{\citenamefont {Balantekin}\ and\ \citenamefont {Suliga}(2024)}]{balantekin-2024}%
  \BibitemOpen
  \bibfield  {author} {\bibinfo {author} {\bibfnamefont {A.~B.}\ \bibnamefont {Balantekin}}\ and\ \bibinfo {author} {\bibfnamefont {A.~M.}\ \bibnamefont {Suliga}},\ }\bibfield  {title} {\bibinfo {title} {{On the properties of qudits}},\ }\bibfield  {journal} {\bibinfo  {journal} {The European Physical Journal A}\ }\textbf {\bibinfo {volume} {60}},\ \href {https://doi.org/10.1140/epja/s10050-024-01347-x} {10.1140/epja/s10050-024-01347-x} (\bibinfo {year} {2024})\BibitemShut {NoStop}%
\bibitem [{\citenamefont {Nikolaeva}\ \emph {et~al.}(2024)\citenamefont {Nikolaeva}, \citenamefont {Kiktenko},\ and\ \citenamefont {Fedorov}}]{nikolaeva-2024}%
  \BibitemOpen
  \bibfield  {author} {\bibinfo {author} {\bibfnamefont {A.~S.}\ \bibnamefont {Nikolaeva}}, \bibinfo {author} {\bibfnamefont {E.~O.}\ \bibnamefont {Kiktenko}},\ and\ \bibinfo {author} {\bibfnamefont {A.~K.}\ \bibnamefont {Fedorov}},\ }\bibfield  {title} {\bibinfo {title} {{Efficient realization of quantum algorithms with qudits}},\ }\bibfield  {journal} {\bibinfo  {journal} {EPJ Quantum Technology}\ }\textbf {\bibinfo {volume} {11}},\ \href {https://doi.org/10.1140/epjqt/s40507-024-00250-0} {10.1140/epjqt/s40507-024-00250-0} (\bibinfo {year} {2024})\BibitemShut {NoStop}%
\bibitem [{\citenamefont {Tilma}\ and\ \citenamefont {Sudarshan}(2002)}]{tilma-2002}%
  \BibitemOpen
  \bibfield  {author} {\bibinfo {author} {\bibfnamefont {T.}~\bibnamefont {Tilma}}\ and\ \bibinfo {author} {\bibfnamefont {E.~C.~G.}\ \bibnamefont {Sudarshan}},\ }\bibfield  {title} {\bibinfo {title} {{Generalized Euler angle parametrization forSU(N)}},\ }\href {https://doi.org/10.1088/0305-4470/35/48/316} {\bibfield  {journal} {\bibinfo  {journal} {Journal of Physics A Mathematical and General}\ }\textbf {\bibinfo {volume} {35}},\ \bibinfo {pages} {10467} (\bibinfo {year} {2002})}\BibitemShut {NoStop}%
\bibitem [{\citenamefont {Vitanov}(2012)}]{vitanov-2012}%
  \BibitemOpen
  \bibfield  {author} {\bibinfo {author} {\bibfnamefont {N.~V.}\ \bibnamefont {Vitanov}},\ }\bibfield  {title} {\bibinfo {title} {{Synthesis of arbitrary SU(3) transformations of atomic qutrits}},\ }\bibfield  {journal} {\bibinfo  {journal} {Physical Review A}\ }\textbf {\bibinfo {volume} {85}},\ \href {https://doi.org/10.1103/physreva.85.032331} {10.1103/physreva.85.032331} (\bibinfo {year} {2012})\BibitemShut {NoStop}%
\bibitem [{\citenamefont {Pudda}\ \emph {et~al.}(2024)\citenamefont {Pudda}, \citenamefont {Chizzini},\ and\ \citenamefont {Crippa}}]{pudda-2024}%
  \BibitemOpen
  \bibfield  {author} {\bibinfo {author} {\bibfnamefont {F.}~\bibnamefont {Pudda}}, \bibinfo {author} {\bibfnamefont {M.}~\bibnamefont {Chizzini}},\ and\ \bibinfo {author} {\bibfnamefont {L.}~\bibnamefont {Crippa}},\ }\href {https://arxiv.org/abs/2410.05122} {\bibinfo {title} {{Generalised Quantum Gates for Qudits and their Application in Quantum Fourier Transform}}} (\bibinfo {year} {2024})\BibitemShut {NoStop}%
\bibitem [{\citenamefont {De~Souza~Farias}\ \emph {et~al.}(2025{\natexlab{b}})\citenamefont {De~Souza~Farias}, \citenamefont {Friedrich},\ and\ \citenamefont {Maziero}}]{de-souza-farias-2025}%
  \BibitemOpen
  \bibfield  {author} {\bibinfo {author} {\bibfnamefont {T.}~\bibnamefont {De~Souza~Farias}}, \bibinfo {author} {\bibfnamefont {L.}~\bibnamefont {Friedrich}},\ and\ \bibinfo {author} {\bibfnamefont {J.}~\bibnamefont {Maziero}},\ }\bibfield  {title} {\bibinfo {title} {{QuForge: A library for qudits simulation}},\ }\href {https://doi.org/10.1016/j.cpc.2025.109687} {\bibfield  {journal} {\bibinfo  {journal} {Computer Physics Communications}\ }\textbf {\bibinfo {volume} {314}},\ \bibinfo {pages} {109687} (\bibinfo {year} {2025}{\natexlab{b}})}\BibitemShut {NoStop}%
\bibitem [{\citenamefont {Xiao}\ \emph {et~al.}(2017)\citenamefont {Xiao}, \citenamefont {Rasul},\ and\ \citenamefont {Vollgraf}}]{fmnist}%
  \BibitemOpen
  \bibfield  {author} {\bibinfo {author} {\bibfnamefont {H.}~\bibnamefont {Xiao}}, \bibinfo {author} {\bibfnamefont {K.}~\bibnamefont {Rasul}},\ and\ \bibinfo {author} {\bibfnamefont {R.}~\bibnamefont {Vollgraf}},\ }\bibfield  {title} {\bibinfo {title} {Fashion-mnist: a novel image dataset for benchmarking machine learning algorithms},\ }\href {http://arxiv.org/abs/1708.07747} {\bibfield  {journal} {\bibinfo  {journal} {CoRR}\ }\textbf {\bibinfo {volume} {abs/1708.07747}} (\bibinfo {year} {2017})},\ \Eprint {https://arxiv.org/abs/1708.07747} {1708.07747} \BibitemShut {NoStop}%
\bibitem [{\citenamefont {{Zalando Research}}(2017)}]{fmnist_github}%
  \BibitemOpen
  \bibfield  {author} {\bibinfo {author} {\bibnamefont {{Zalando Research}}},\ }\href {https://github.com/zalandoresearch/fashion-mnist} {\bibinfo {title} {Fashion-{MNIST}}},\ \bibinfo {howpublished} {GitHub repository} (\bibinfo {year} {2017}),\ \bibinfo {note} {accessed: 2026-04-20}\BibitemShut {NoStop}%
\bibitem [{\citenamefont {Wang}\ \emph {et~al.}(2021)\citenamefont {Wang}, \citenamefont {Fontana}, \citenamefont {Cerezo}, \citenamefont {Sharma}, \citenamefont {Sone}, \citenamefont {Cincio},\ and\ \citenamefont {Coles}}]{wang2021noise}%
  \BibitemOpen
  \bibfield  {author} {\bibinfo {author} {\bibfnamefont {S.}~\bibnamefont {Wang}}, \bibinfo {author} {\bibfnamefont {E.}~\bibnamefont {Fontana}}, \bibinfo {author} {\bibfnamefont {M.}~\bibnamefont {Cerezo}}, \bibinfo {author} {\bibfnamefont {K.}~\bibnamefont {Sharma}}, \bibinfo {author} {\bibfnamefont {A.}~\bibnamefont {Sone}}, \bibinfo {author} {\bibfnamefont {L.}~\bibnamefont {Cincio}},\ and\ \bibinfo {author} {\bibfnamefont {P.~J.}\ \bibnamefont {Coles}},\ }\bibfield  {title} {\bibinfo {title} {Noise-induced barren plateaus in variational quantum algorithms},\ }\href@noop {} {\bibfield  {journal} {\bibinfo  {journal} {Nature communications}\ }\textbf {\bibinfo {volume} {12}},\ \bibinfo {pages} {6961} (\bibinfo {year} {2021})}\BibitemShut {NoStop}%
\bibitem [{\citenamefont {Sonawane}\ \emph {et~al.}(2024)\citenamefont {Sonawane}, \citenamefont {Dhayalkar}, \citenamefont {Waje}, \citenamefont {Markhelkar}, \citenamefont {Wattamwar},\ and\ \citenamefont {Shrawne}}]{human_activity_recognition_using_smartphones_240}%
  \BibitemOpen
  \bibfield  {author} {\bibinfo {author} {\bibfnamefont {M.}~\bibnamefont {Sonawane}}, \bibinfo {author} {\bibfnamefont {S.~R.}\ \bibnamefont {Dhayalkar}}, \bibinfo {author} {\bibfnamefont {S.}~\bibnamefont {Waje}}, \bibinfo {author} {\bibfnamefont {S.}~\bibnamefont {Markhelkar}}, \bibinfo {author} {\bibfnamefont {A.}~\bibnamefont {Wattamwar}},\ and\ \bibinfo {author} {\bibfnamefont {S.~C.}\ \bibnamefont {Shrawne}},\ }\href {https://arxiv.org/abs/2404.02869} {\bibinfo {title} {{Human Activity Recognition using Smartphones}}} (\bibinfo {year} {2024})\BibitemShut {NoStop}%
\bibitem [{\citenamefont {Bock}(2004)}]{magic_gamma_telescope_159}%
  \BibitemOpen
  \bibfield  {author} {\bibinfo {author} {\bibfnamefont {R.}~\bibnamefont {Bock}},\ }\href@noop {} {\bibinfo {title} {{MAGIC Gamma Telescope}}},\ \bibinfo {howpublished} {UCI Machine Learning Repository} (\bibinfo {year} {2004}),\ \bibinfo {note} {{DOI}: https://doi.org/10.24432/C52C8B}\BibitemShut {NoStop}%
\bibitem [{\citenamefont {Blackard}(1998)}]{covertype_31}%
  \BibitemOpen
  \bibfield  {author} {\bibinfo {author} {\bibfnamefont {J.}~\bibnamefont {Blackard}},\ }\href@noop {} {\bibinfo {title} {{Covertype}}},\ \bibinfo {howpublished} {UCI Machine Learning Repository} (\bibinfo {year} {1998}),\ \bibinfo {note} {{DOI}: https://doi.org/10.24432/C50K5N}\BibitemShut {NoStop}%
\end{thebibliography}%

\newpage
\clearpage
\onecolumngrid
\appendix
\raggedbottom

\section{Datasets descriptions}
\label{sec:appendix_datasets_description}

\subsubsection*{Fashion-MNIST}

Fashion-MNIST is a 10-class benchmark dataset designed as a drop-in replacement for MNIST. It contains $28\times 28$ grayscale images of $70{,}000$ fashion products, with $7{,}000$ images per class and a standard split of $60{,}000$ training images and $10{,}000$ test images \cite{fmnist}.

The dataset is derived from product images from Zalando. The authors construct Fashion-MNIST from front-view thumbnail images and apply a standardized conversion pipeline to match the MNIST format, including cropping/trimming, resizing to $28\times 28$, sharpening, intensity negation, and conversion to 8-bit grayscale. The final dataset uses the following class labels: 0 (T-shirt/top), 1 (Trouser), 2 (Pullover), 3 (Dress), 4 (Coat), 5 (Sandal), 6 (Shirt), 7 (Sneaker), 8 (Bag), and 9 (Ankle boot).

In our experiments, we form binary and three-class classification tasks by selecting subsets of these labels, namely $(5,7)$ for binary (Sandal vs.\ Sneaker) and $(5,7,9)$ for three-class (Sandal vs.\ Sneaker vs.\ Boot).

\subsubsection*{HAR (Human Activity Recognition Using Smartphones)}

The Human Activity Recognition Using Smartphones (HAR) dataset is a benchmark for activity classification from wearable sensing data. It was collected from 30 volunteers (ages 19--48) performing six activities: \textsc{Walking}, \textsc{Walking Upstairs}, \textsc{Walking Downstairs}, \textsc{Sitting}, \textsc{Standing}, and \textsc{Laying}, while carrying a smartphone (Samsung Galaxy S II) on the waist \cite{human_activity_recognition_using_smartphones_240}. Tri-axial linear acceleration and tri-axial angular velocity were recorded from the embedded accelerometer and gyroscope at 50\,Hz, and the sessions were video-recorded to enable manual labeling. For each window, the dataset provides a 561-dimensional feature vector computed from time- and frequency-domain statistics, together with the corresponding activity label and subject identifier.

In our experiments, we restrict HAR to a binary task \textsc{Sitting} vs.\ \textsc{Laying} (labels 4 vs.\ 6) and a three-class task \textsc{Sitting} vs.\ \textsc{Standing} vs.\ \textsc{Laying} (labels 4/5/6).

\subsubsection*{MAGIC Gamma Telescope}

The MAGIC Gamma Telescope dataset is a tabular binary classification benchmark from the UCI Machine Learning Repository \cite{magic_gamma_telescope_159}. The data were computer-generated to simulate the detection of high-energy gamma particles using a ground-based atmospheric Cherenkov gamma telescope. In such instruments, gamma rays produce electromagnetic showers in the atmosphere, generating Cherenkov radiation that is recorded by photomultiplier tubes arranged in a camera. The recorded shower images can be statistically discriminated between primary gamma (g) events and hadronic showers initiated by cosmic rays (background).

The dataset provides a set of engineered features derived from the recorded shower images, together with a binary class label (gamma vs.\ hadron).

\subsubsection*{Covertype}

The Covertype dataset is a large-scale tabular benchmark from the UCI Machine Learning Repository \cite{covertype_31}. It contains $581{,}012$ observations, each corresponding to a $30\times 30$\,m cell in four wilderness areas of the Roosevelt National Forest (Colorado, USA). The prediction target is the forest cover type, with seven classes (labels 1--7) representing dominant vegetation categories such as Spruce/Fir, Lodgepole Pine, and Ponderosa Pine.

Each instance is described by cartographic variables derived from USGS and USFS data. The dataset includes 10 quantitative attributes (e.g., elevation, slope, distances to hydrology/roads/fire points, and hillshade indices) and two groups of one-hot encoded categorical variables: 4 binary wilderness-area indicators and 40 binary soil-type indicators, for a total of 54 input columns.

In our experiments, we restrict Covertype to a binary task using cover types 1 vs.\ 2 and a three-class task using cover types 1 vs.\ 2 vs.\ 3.

\FloatBarrier

\section{Additional results on HAR, MAGIC Gamma Telescope, and Covertype}
\label{sec:appendix_datasets_results}

\begin{figure*}[h]
    \centering
    \begin{minipage}[t]{0.32\textwidth}
    \centering
    \subfloat[]{\includegraphics[width=\linewidth]{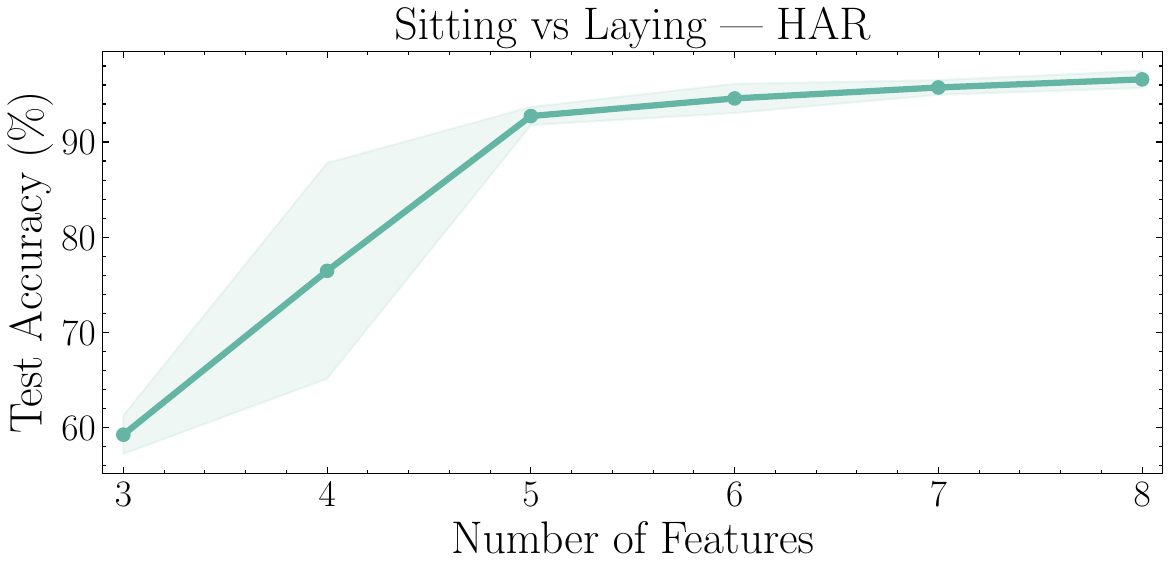}\label{fig:app_a}}
    \end{minipage}\hfill
    \begin{minipage}[t]{0.32\textwidth}
    \centering
    \subfloat[]{\includegraphics[width=\linewidth]{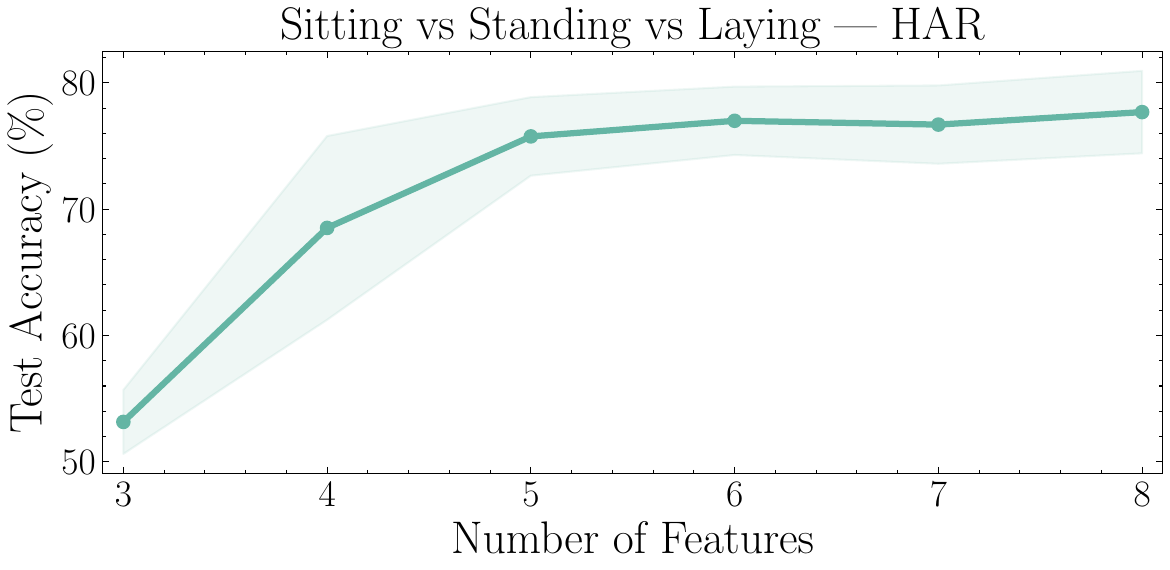}\label{fig:app_b}}
    \end{minipage}\hfill
    \begin{minipage}[t]{0.32\textwidth}
    \centering
    \subfloat[]{\includegraphics[width=\linewidth]{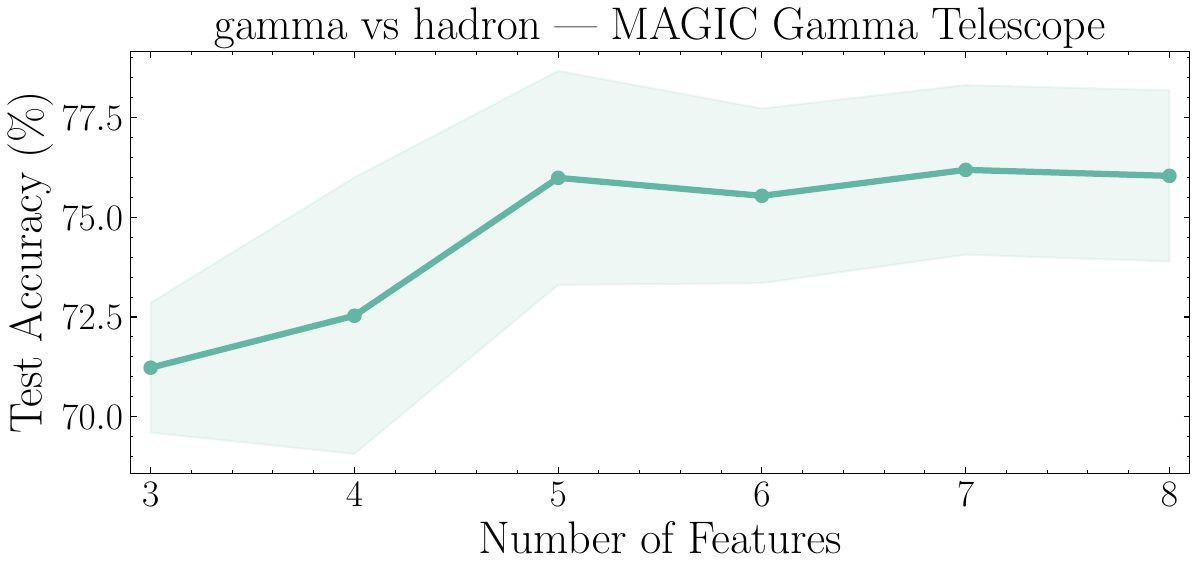}\label{fig:app_c}}
    \end{minipage}
    
    \vspace{2mm}
    \makebox[\textwidth][c]{%
        \begin{minipage}[t]{0.32\textwidth}
        \centering
        \subfloat[]{\includegraphics[width=\linewidth]{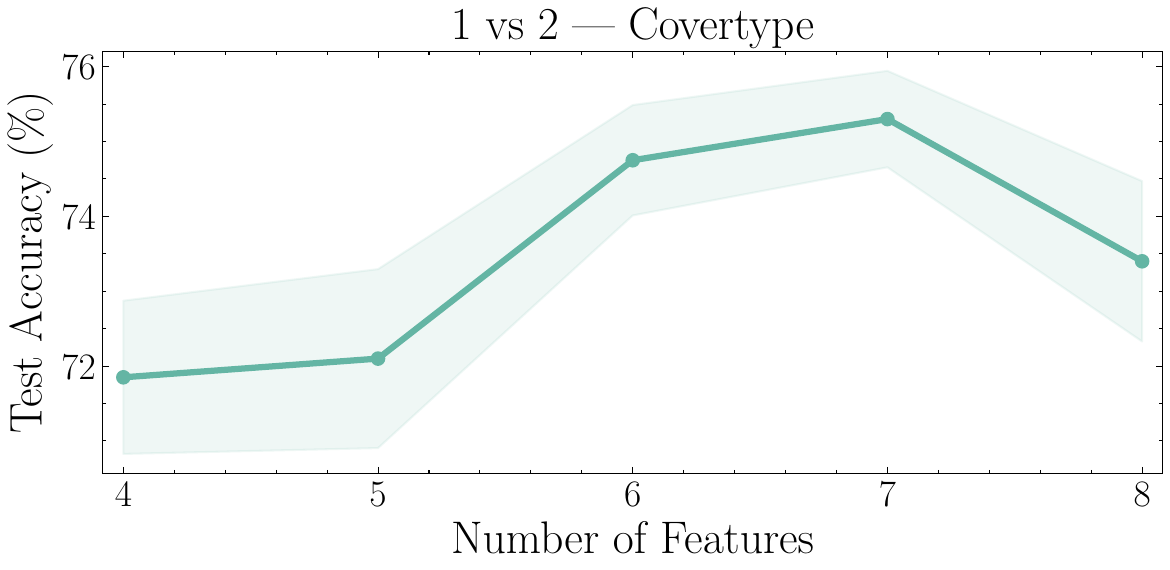}\label{fig:app_d}}
        \end{minipage}
        \hspace{0.04\textwidth}
        \begin{minipage}[t]{0.32\textwidth}
        \centering
        \subfloat[]{\includegraphics[width=\linewidth]{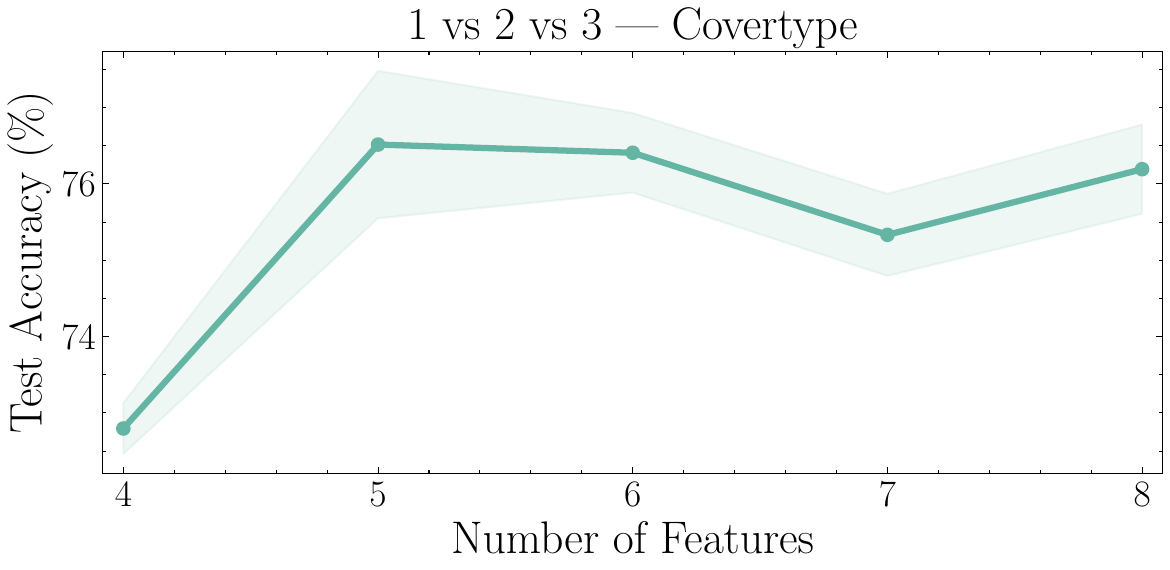}\label{fig:app_e}}
        \end{minipage}
    }
    \caption{Test accuracy of the $1$-qutrit QNN as a function of the number of encoded features on additional benchmarks (mean $\pm$ standard error over 5 stratified folds). Panels correspond to: (a) HAR (Sitting vs.\ Laying), (b) HAR (Sitting vs.\ Standing vs.\ Laying), (c) MAGIC Gamma Telescope (gamma vs.\ hadron), (d) Covertype (1 vs.\ 2), and (e) Covertype (1 vs.\ 2 vs.\ 3). While increasing the feature budget often improves performance, the trend is not strictly monotonic across tasks; in particular, for Covertype accuracy peaks at $p=7$ in the binary task and at $p=5$ in the three-class task, decreasing thereafter.}
    \label{fig:app_1qnn_features}
\end{figure*}

\begin{figure*}[h]
    \centering
    
    \begin{minipage}[t]{0.32\textwidth}
    \centering
    \subfloat[]{\includegraphics[width=\linewidth]{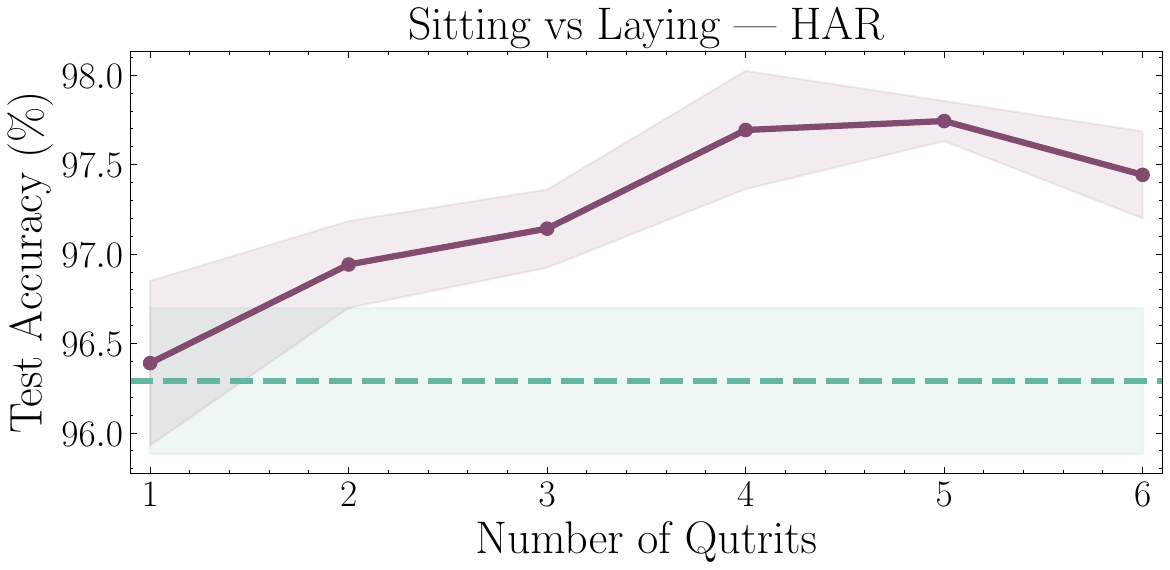}}
    \end{minipage}\hfill
    \begin{minipage}[t]{0.32\textwidth}
    \centering
    \subfloat[]{\includegraphics[width=\linewidth]{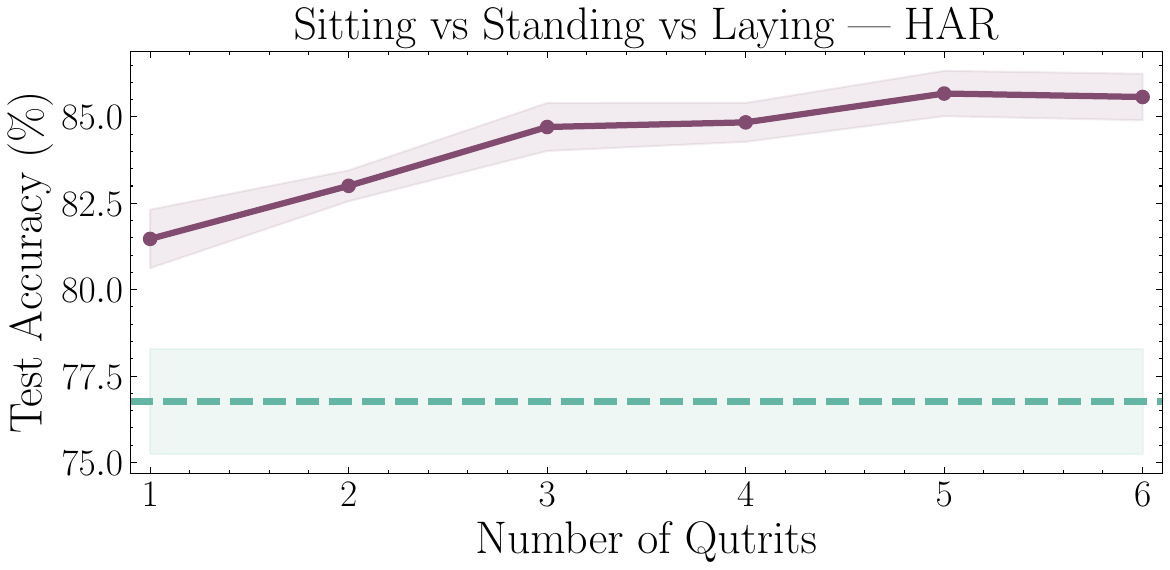}}
    \end{minipage}\hfill
    \begin{minipage}[t]{0.32\textwidth}
    \centering
    \subfloat[]{\includegraphics[width=\linewidth]{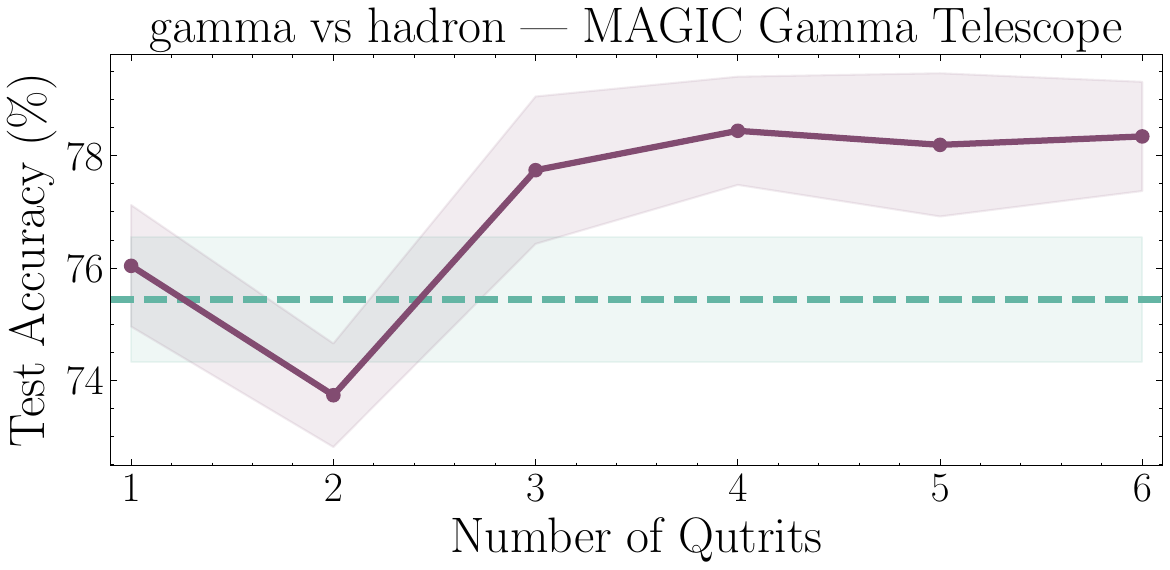}}
    \end{minipage}
    
    \vspace{2mm}
    
    \makebox[\textwidth][c]{%
        \begin{minipage}[t]{0.32\textwidth}
        \centering
        \subfloat[]{\includegraphics[width=\linewidth]{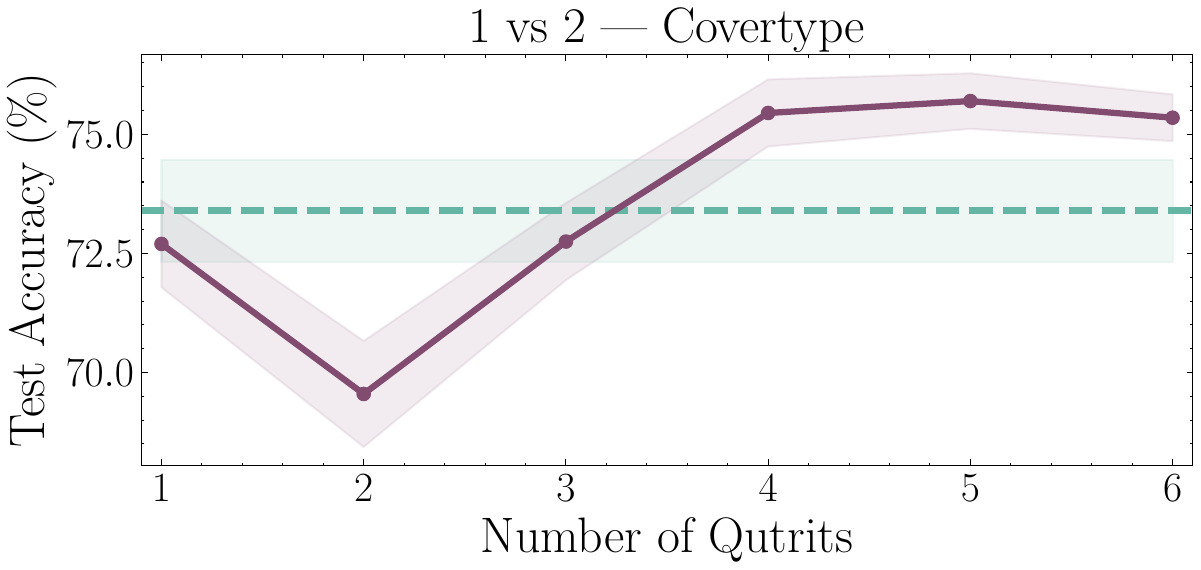}}
        \end{minipage}
        \hspace{0.04\textwidth}
        \begin{minipage}[t]{0.32\textwidth}
        \centering
        \subfloat[]{\includegraphics[width=\linewidth]{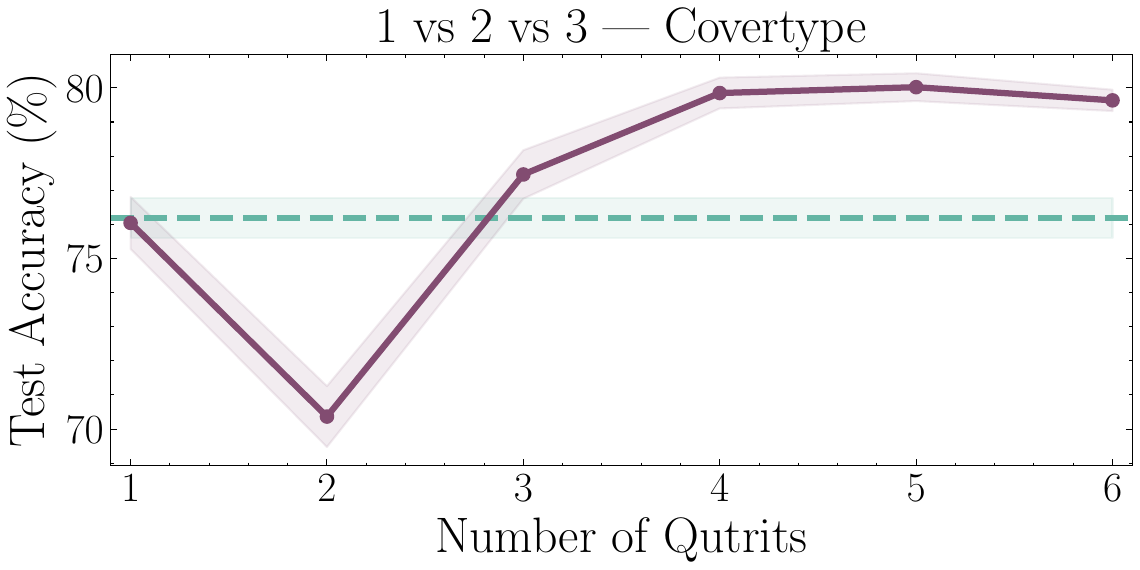}}
        \end{minipage}
    }
    
    \vspace{2mm}
    
    \makebox[\textwidth][c]{%
        \begin{minipage}[t]{0.32\textwidth}
        \centering
        \includegraphics[width=\linewidth]{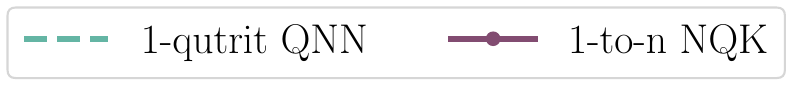}
        \end{minipage}
    }
    
    \caption{\justifying Test accuracy of the $1$-to-$n$ NQK as a function of the number of qutrits with the feature budget fixed to $p=8$ (mean $\pm$ standard error over 5 stratified folds). Panels correspond to: (a) HAR (Sitting vs.\ Laying), (b) HAR (Sitting vs.\ Standing vs.\ Laying), (c) MAGIC Gamma Telescope (gamma vs.\ hadron), (d) Covertype (1 vs.\ 2), and (e) Covertype (1 vs.\ 2 vs.\ 3).}
    \label{fig:appendix_1ton_qutrits}
\end{figure*}

\begin{figure*}[h]
    \centering
    
    \begin{minipage}[t]{0.32\textwidth}
    \centering
    \subfloat[]{\includegraphics[width=\linewidth]{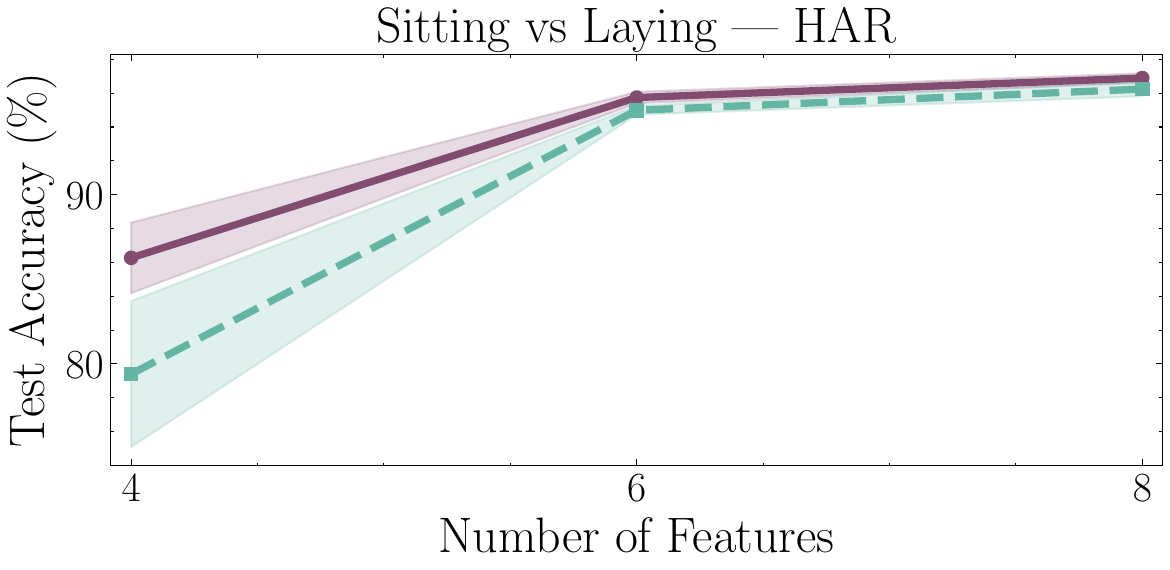}}
    \end{minipage}\hfill
    \begin{minipage}[t]{0.32\textwidth}
    \centering
    \subfloat[]{\includegraphics[width=\linewidth]{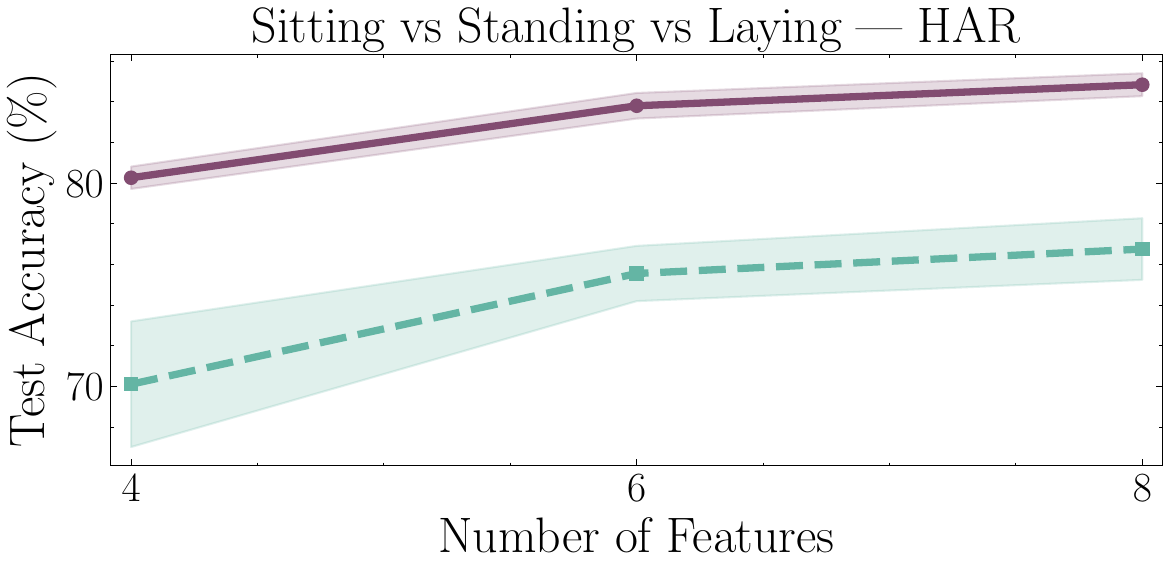}}
    \end{minipage}\hfill
    \begin{minipage}[t]{0.32\textwidth}
    \centering
    \subfloat[]{\includegraphics[width=\linewidth]{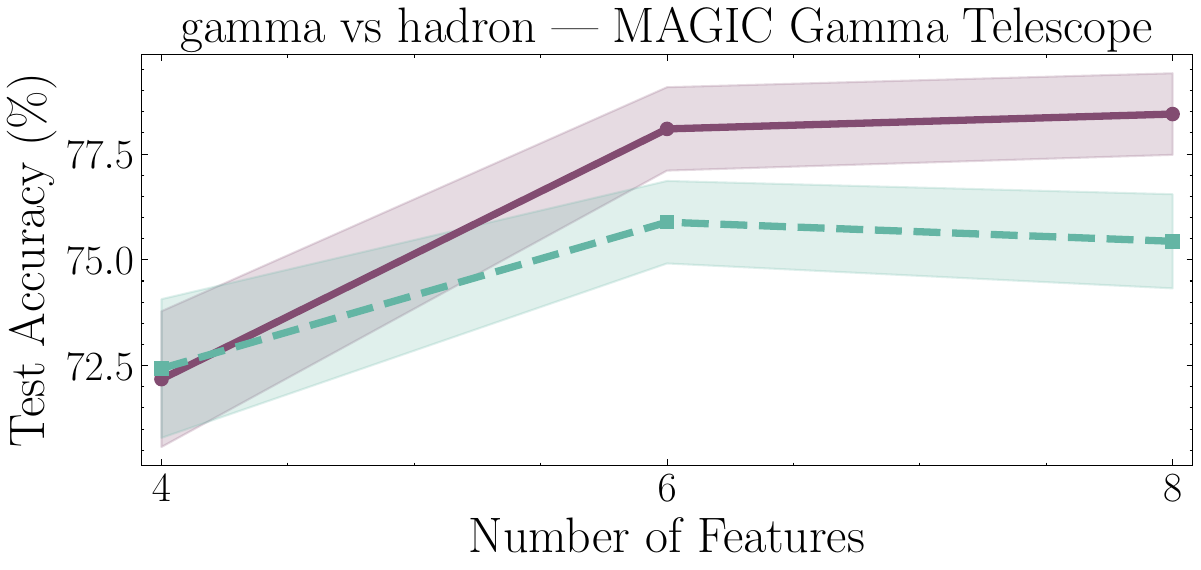}}
    \end{minipage}
    
    \vspace{2mm}
    
    \makebox[\textwidth][c]{%
        \begin{minipage}[t]{0.32\textwidth}
        \centering
        \subfloat[]{\includegraphics[width=\linewidth]{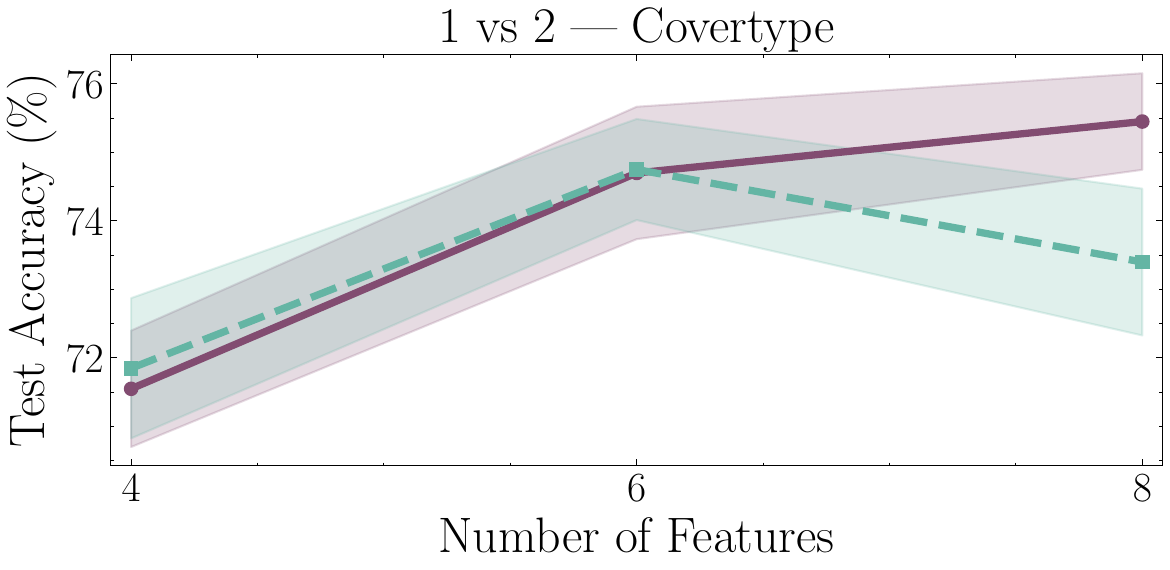}}
        \end{minipage}
        \hspace{0.04\textwidth}
        \begin{minipage}[t]{0.32\textwidth}
        \centering
        \subfloat[]{\includegraphics[width=\linewidth]{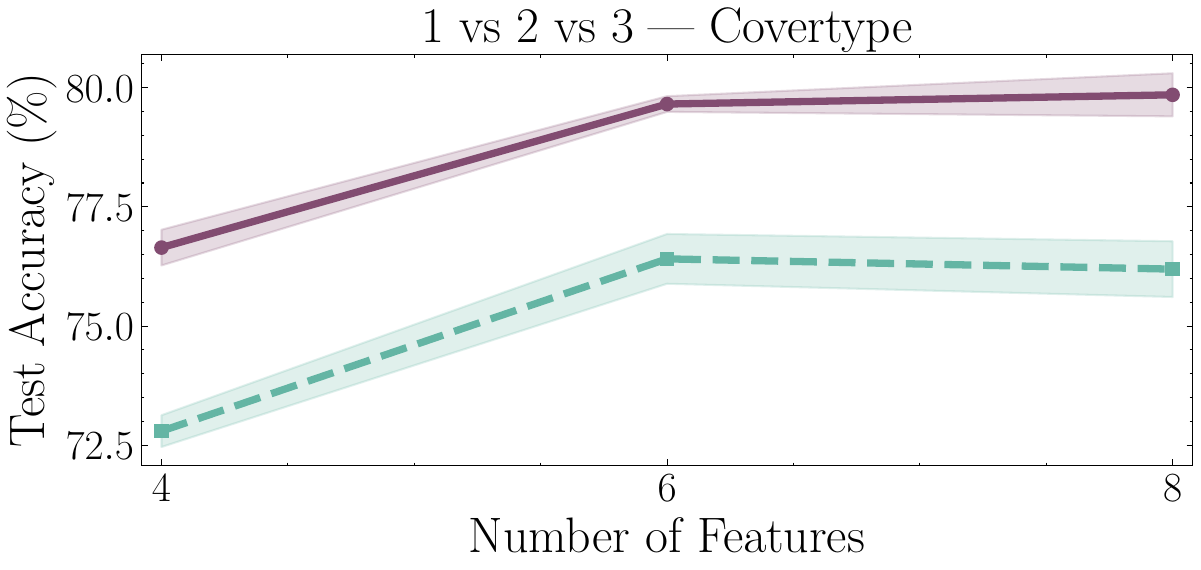}}
        \end{minipage}
    }
    
    \vspace{2mm}
    
    \makebox[\textwidth][c]{%
        \begin{minipage}[t]{0.32\textwidth}
        \centering
        \includegraphics[width=\linewidth]{plots/1-to-n_legend.pdf}
        \end{minipage}
    }
    \caption{\justifying Test accuracy of the $1$-to-$n$ NQK as a function of the number of encoded features at fixed system size $n=4$ (mean $\pm$ standard error over 5 stratified folds). Panels correspond to: (a) HAR (Sitting vs.\ Laying), (b) HAR (Sitting vs.\ Standing vs.\ Laying), (c) MAGIC Gamma Telescope (gamma vs.\ hadron), (d) Covertype (1 vs.\ 2), and (e) Covertype (1 vs.\ 2 vs.\ 3). Compared to the corresponding $1$-qutrit QNN baseline, feature scaling in the kernel setting is more regular, with accuracy increasing consistently across the considered feature budgets.}
    \label{fig:appendix_1ton_features}
\end{figure*}

\begin{figure*}[h]
    \centering
    
    \begin{minipage}[t]{0.32\textwidth}
    \centering
    \subfloat[]{\includegraphics[width=\linewidth]{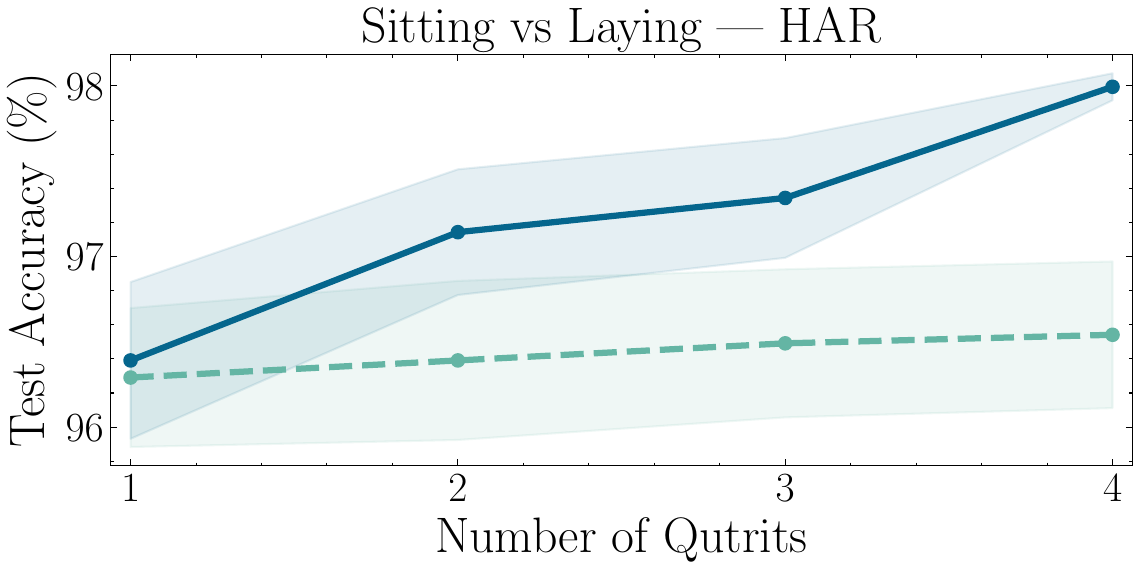}}
    \end{minipage}\hfill
    \begin{minipage}[t]{0.32\textwidth}
    \centering
    \subfloat[]{\includegraphics[width=\linewidth]{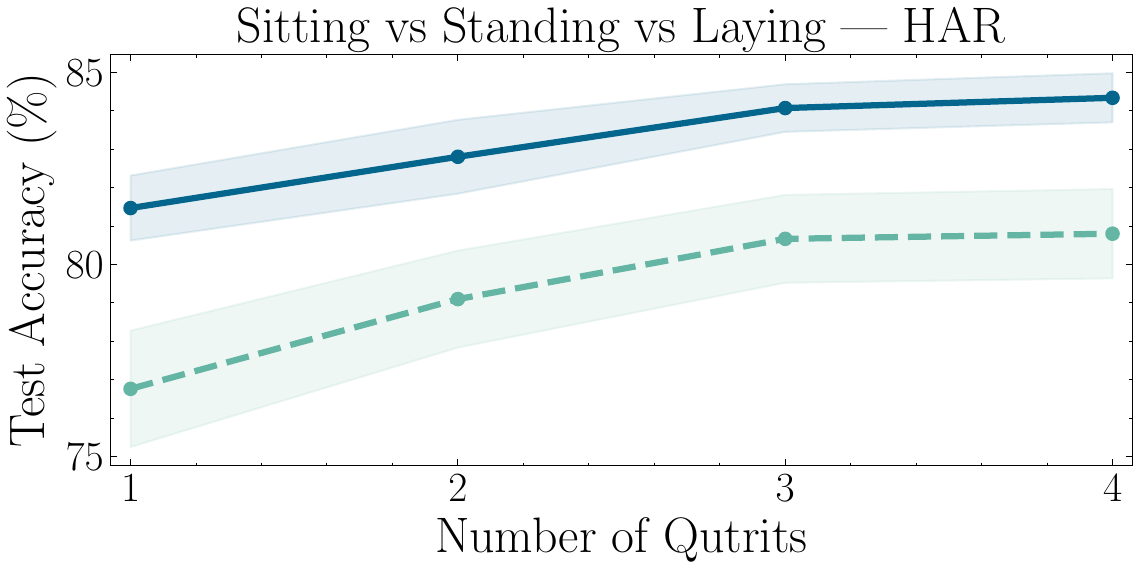}}
    \end{minipage}\hfill
    \begin{minipage}[t]{0.32\textwidth}
    \centering
    \subfloat[]{\includegraphics[width=\linewidth]{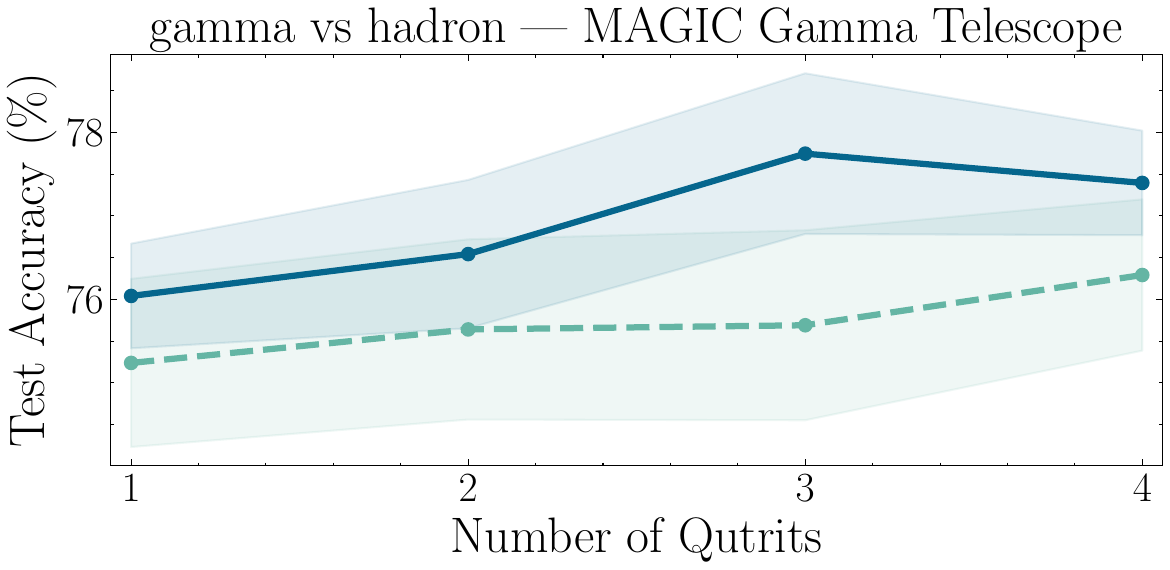}}
    \end{minipage}
    
    \vspace{2mm}
    
    \makebox[\textwidth][c]{%
        \begin{minipage}[t]{0.32\textwidth}
        \centering
        \subfloat[]{\includegraphics[width=\linewidth]{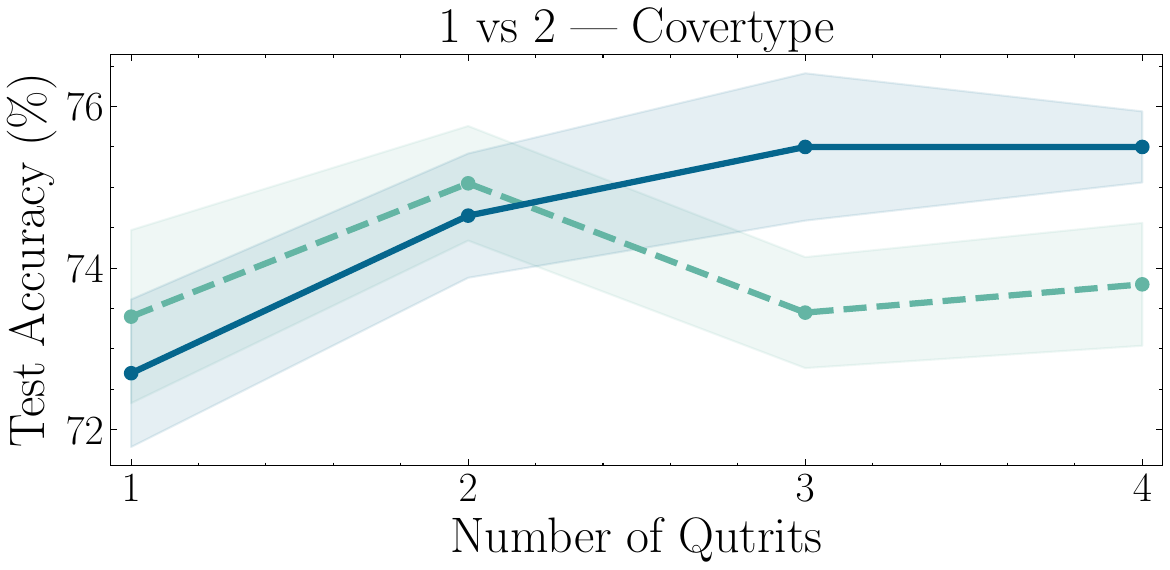}}
        \end{minipage}
        \hspace{0.04\textwidth}
        \begin{minipage}[t]{0.32\textwidth}
        \centering
        \subfloat[]{\includegraphics[width=\linewidth]{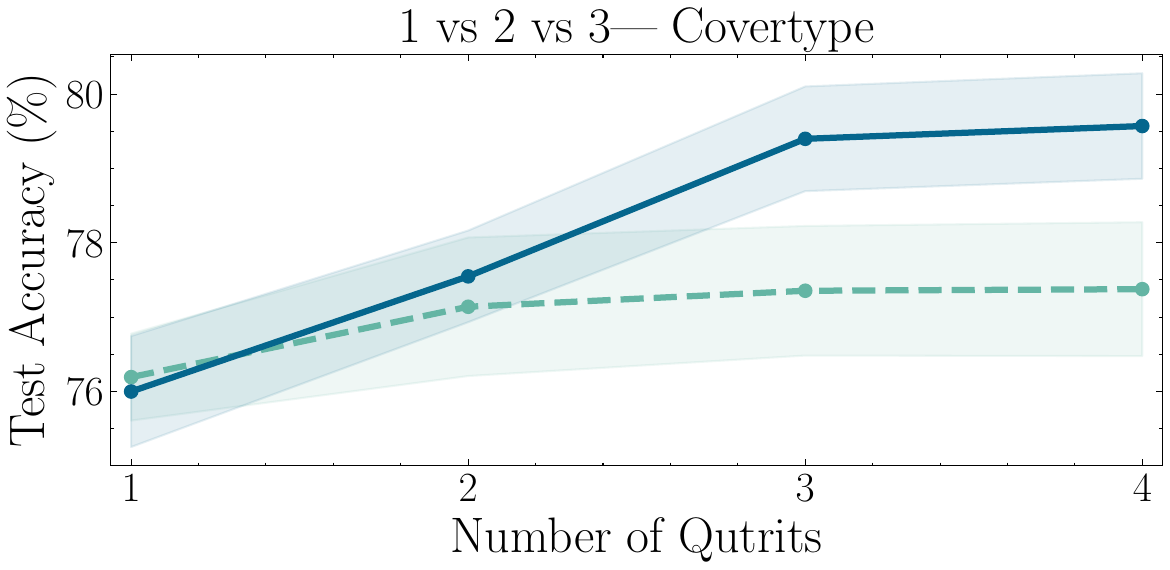}}
        \end{minipage}
    }
    
    \vspace{2mm}
    
    \makebox[\textwidth][c]{%
        \begin{minipage}[t]{0.28\textwidth}
        \centering
        \includegraphics[width=\linewidth]{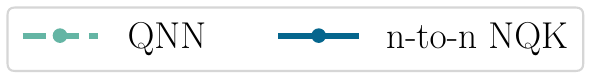}
        \end{minipage}
    }
    \caption{\justifying Test accuracy of the $n$-to-$n$ NQK as a function of the number of qutrits at fixed feature budget ($p=8$ for HAR and MAGIC; $p=6$ for Covertype, panels (d)--(e)) (mean $\pm$ standard error over 5 stratified folds). Panels correspond to: (a) HAR (Sitting vs.\ Laying), (b) HAR (Sitting vs.\ Standing vs.\ Laying), (c) MAGIC Gamma Telescope (gamma vs.\ hadron), (d) Covertype (1 vs.\ 2), and (e) Covertype (1 vs.\ 2 vs.\ 3). Overall, the $n$-to-$n$ kernel outperforms the matched $n$-qutrit QNN baseline in most tasks, except in the binary Covertype where accuracy falls below the baseline in $n\in[1, 2]$.}
    \label{fig:appendix_nton_qutrits}
\end{figure*}

\begin{figure*}[h]
    \centering
    
    \begin{minipage}[t]{0.32\textwidth}
    \centering
    \subfloat[]{\includegraphics[width=\linewidth]{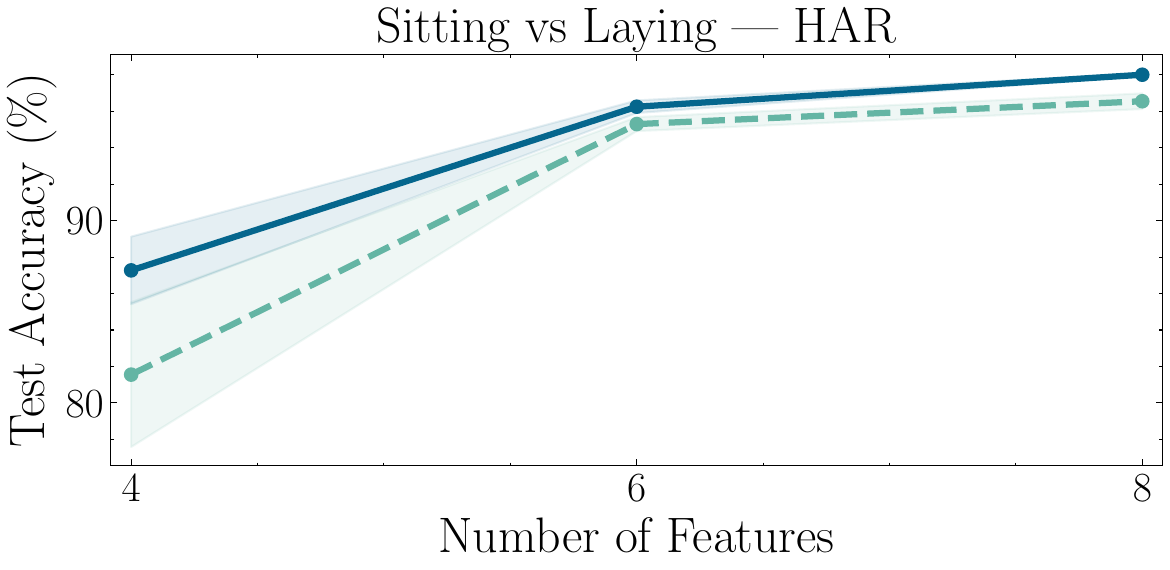}}
    \end{minipage}\hfill
    \begin{minipage}[t]{0.32\textwidth}
    \centering
    \subfloat[]{\includegraphics[width=\linewidth]{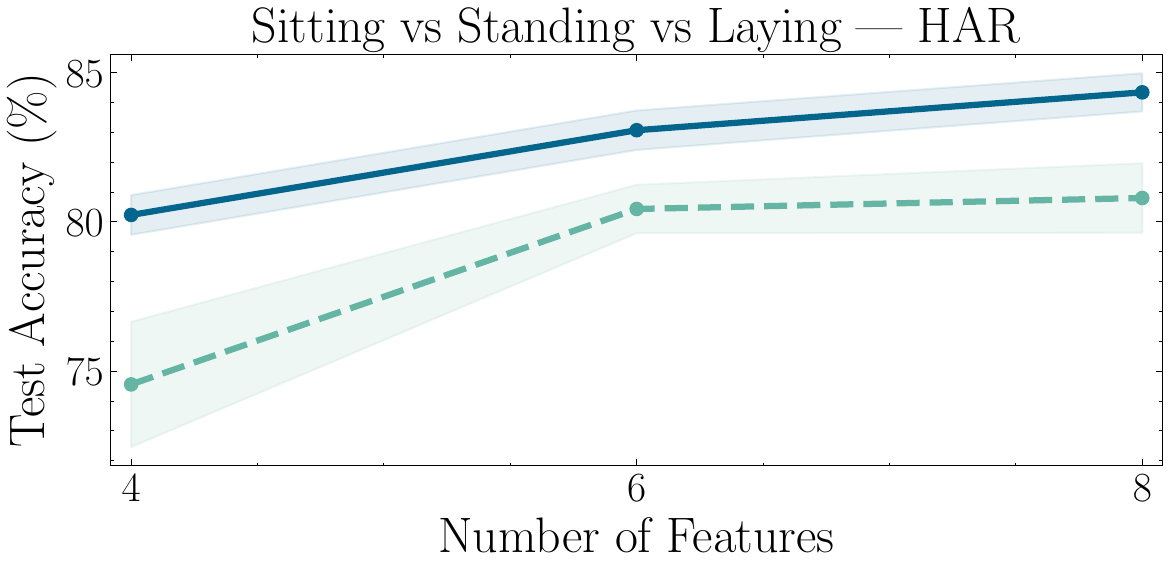}}
    \end{minipage}\hfill
    \begin{minipage}[t]{0.32\textwidth}
    \centering
    \subfloat[]{\includegraphics[width=\linewidth]{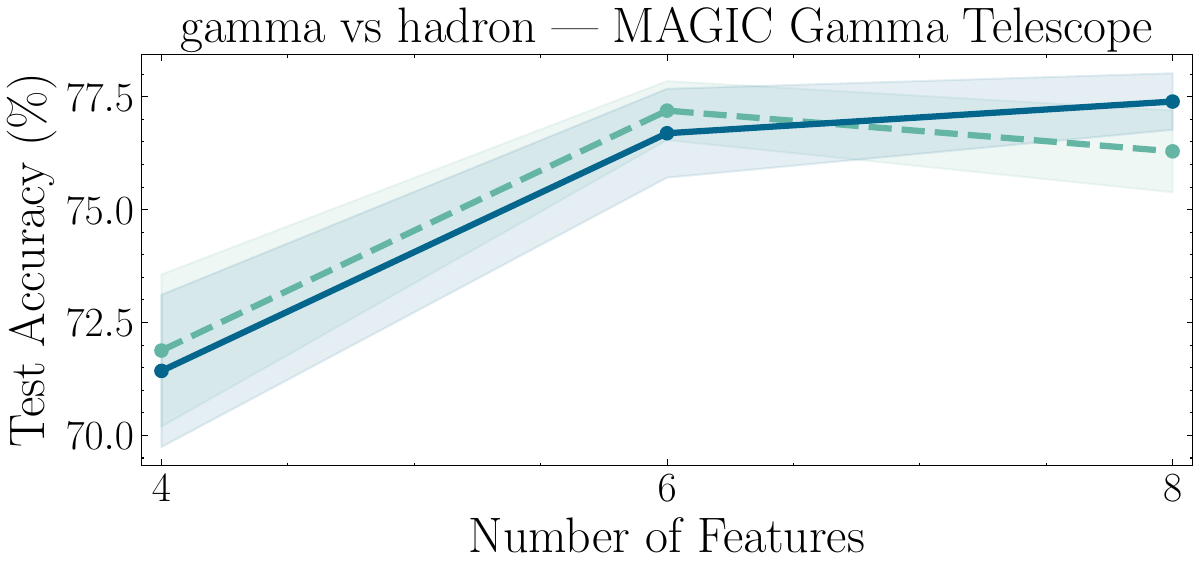}}
    \end{minipage}
    
    \vspace{2mm}
    
    \makebox[\textwidth][c]{%
        \begin{minipage}[t]{0.32\textwidth}
        \centering
        \subfloat[]{\includegraphics[width=\linewidth]{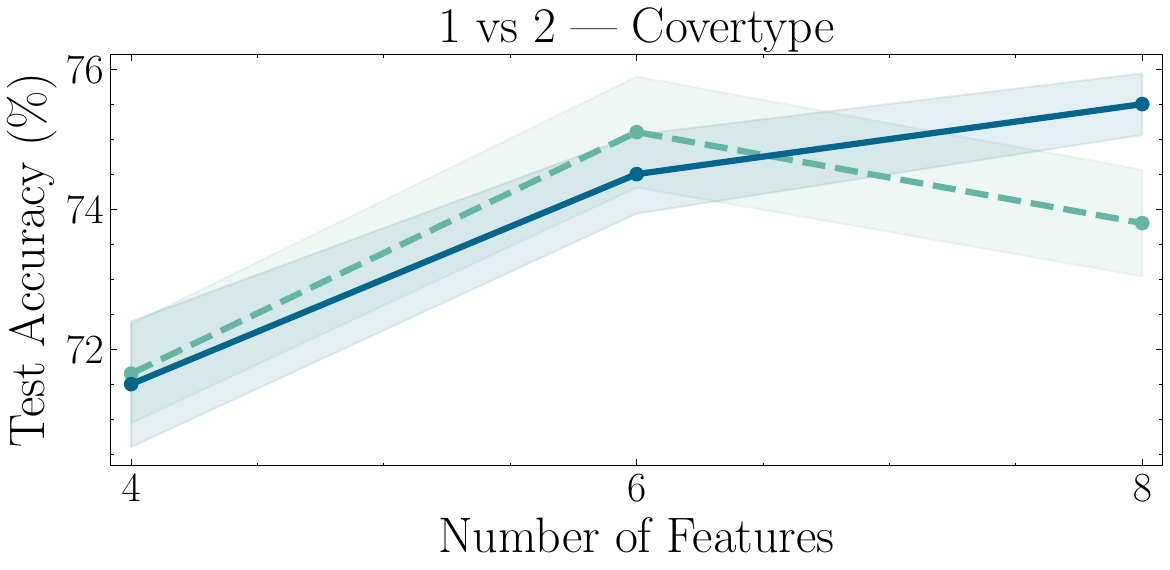}}
        \end{minipage}
        \hspace{0.04\textwidth}
        \begin{minipage}[t]{0.32\textwidth}
        \centering
        \subfloat[]{\includegraphics[width=\linewidth]{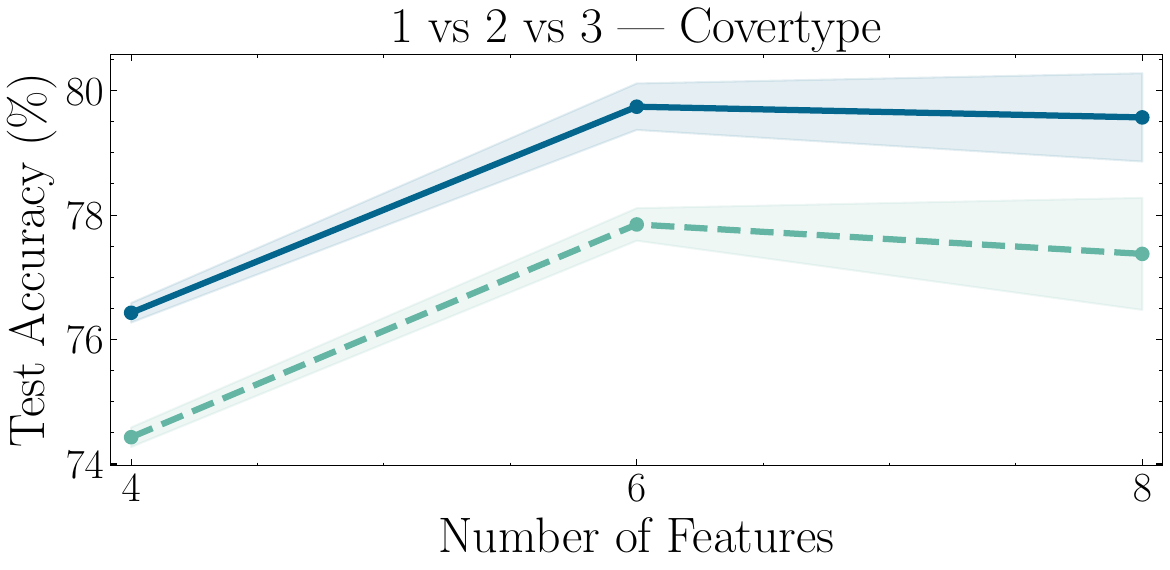}}
        \end{minipage}
    }
    
    \vspace{2mm}
    
    \makebox[\textwidth][c]{%
        \begin{minipage}[t]{0.28\textwidth}
        \centering
        \includegraphics[width=\linewidth]{plots/n-to-n_legend.pdf}
        \end{minipage}
    }
    \caption{\justifying Test accuracy of the $n$-to-$n$ NQK as a function of the number of encoded features at fixed system size $n=4$ (mean $\pm$ standard error over 5 stratified folds). Panels correspond to: (a) HAR (Sitting vs.\ Laying), (b) HAR (Sitting vs.\ Standing vs.\ Laying), (c) MAGIC Gamma Telescope (gamma vs.\ hadron), (d) Covertype (1 vs.\ 2), and (e) Covertype (1 vs.\ 2 vs.\ 3).}
    \label{fig:appendix_nton_features}
\end{figure*}

\begin{figure*}[h]
    \centering
    
    \begin{minipage}[t]{0.32\textwidth}
    \centering
    \subfloat[]{\includegraphics[width=\linewidth]{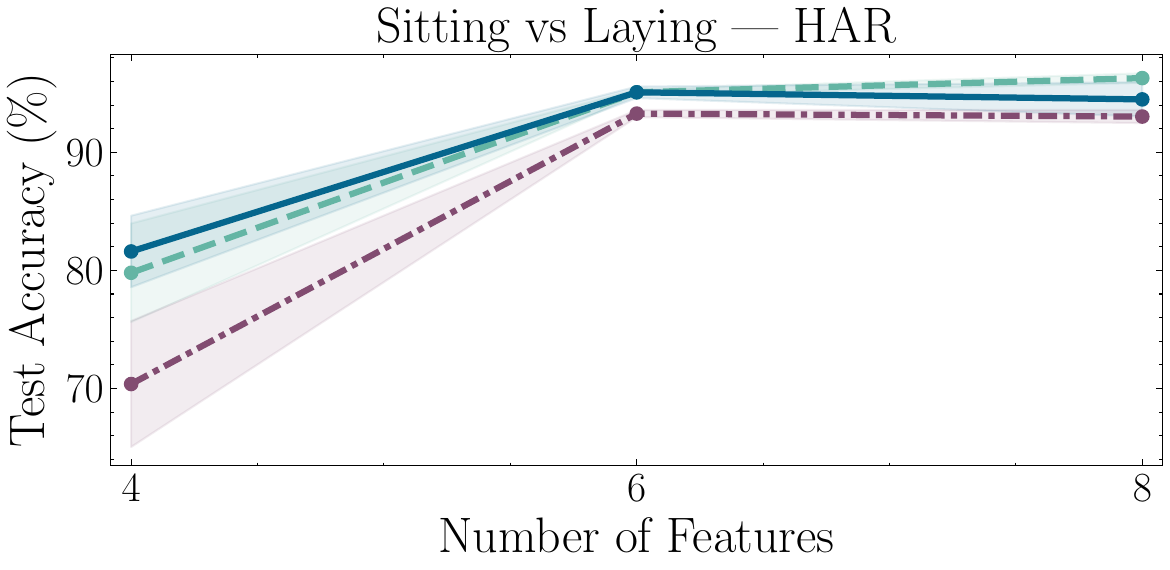}}
    \end{minipage}\hfill
    \begin{minipage}[t]{0.32\textwidth}
    \centering
    \subfloat[]{\includegraphics[width=\linewidth]{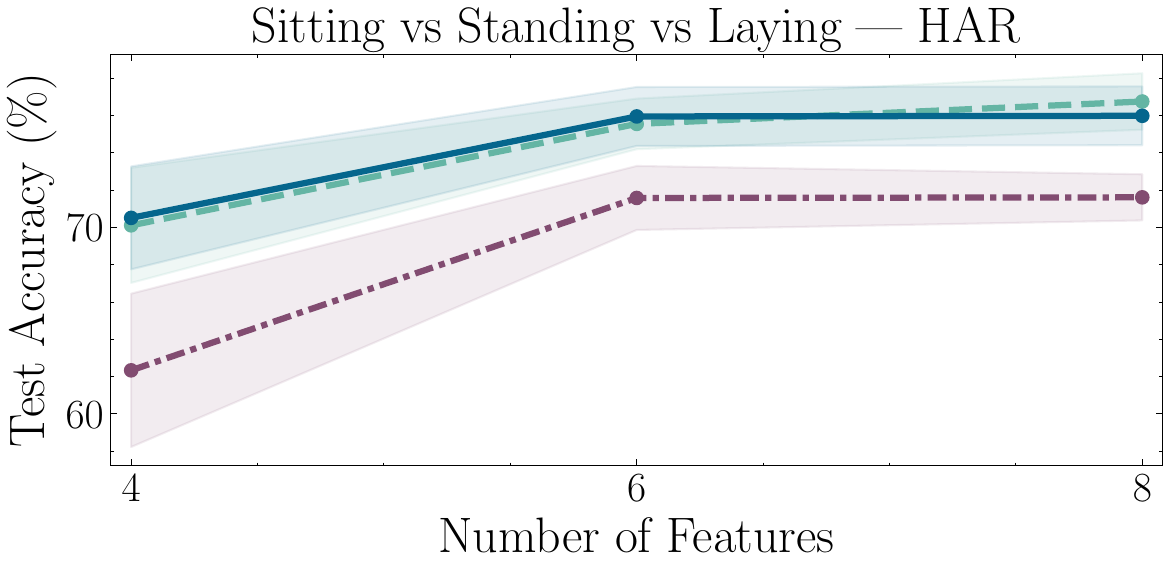}}
    \end{minipage}\hfill
    \begin{minipage}[t]{0.32\textwidth}
    \centering
    \subfloat[]{\includegraphics[width=\linewidth]{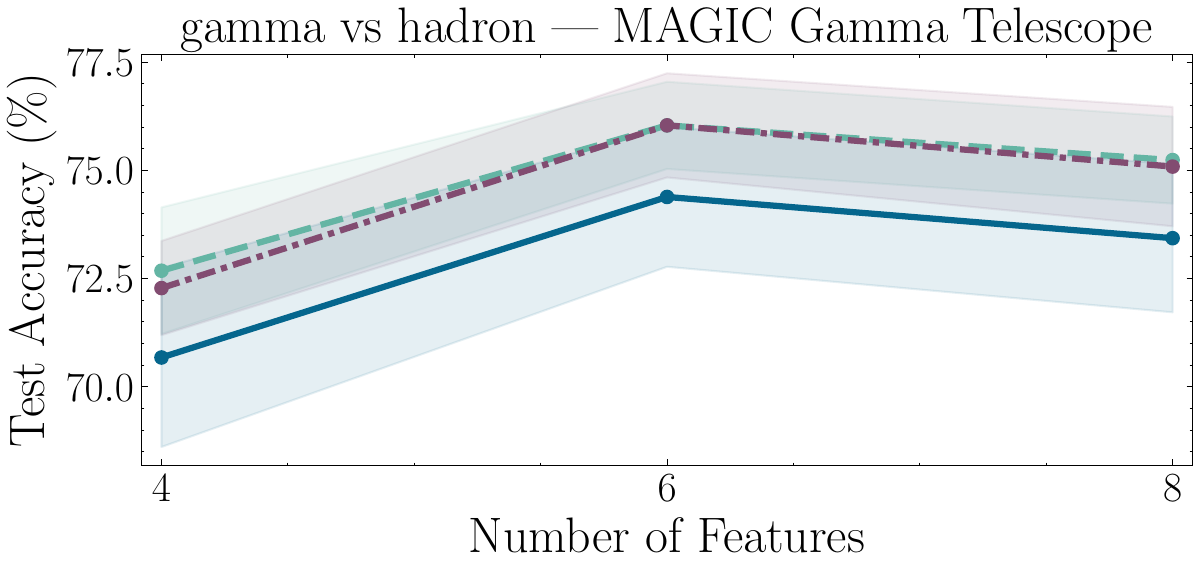}}
    \end{minipage}
    
    \vspace{2mm}
    
    \makebox[\textwidth][c]{%
        \begin{minipage}[t]{0.32\textwidth}
        \centering
        \subfloat[]{\includegraphics[width=\linewidth]{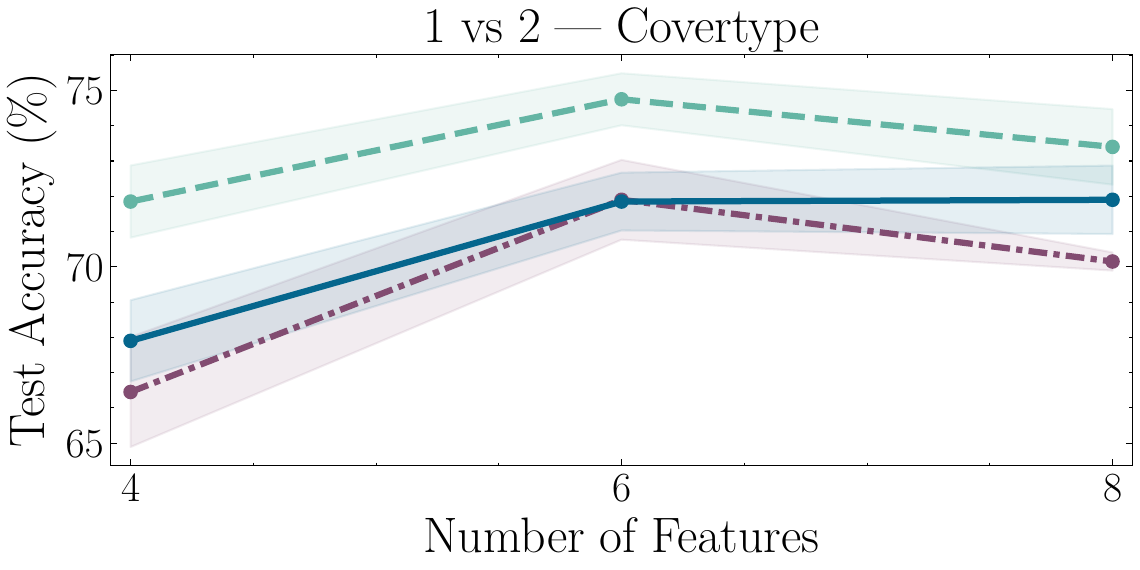}}
        \end{minipage}
        \hspace{0.04\textwidth}
        \begin{minipage}[t]{0.32\textwidth}
        \centering
        \subfloat[]{\includegraphics[width=\linewidth]{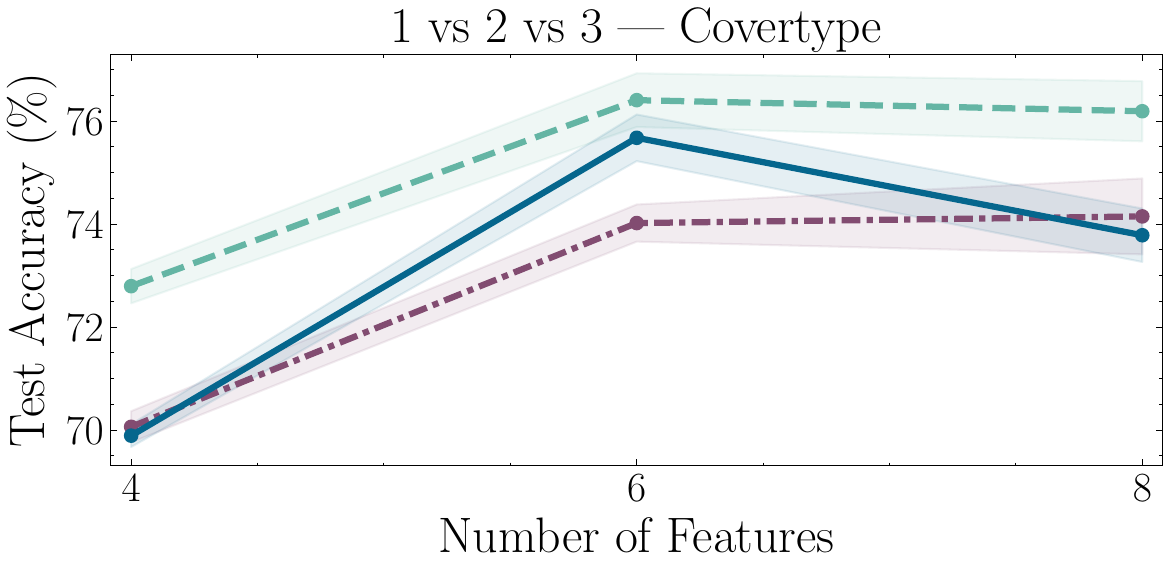}}
        \end{minipage}
    }
    
    \vspace{2mm}
    
    \makebox[\textwidth][c]{%
        \begin{minipage}[t]{0.35\textwidth}
        \centering
        \includegraphics[width=\linewidth]{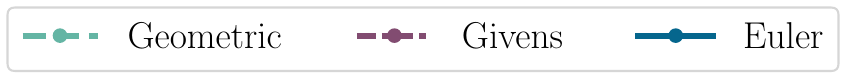}
        \end{minipage}
    }
    \caption{\justifying Test accuracy of the $1$-qutrit QNN as a function of the number of encoded features on additional benchmarks, comparing three $\mathrm{SU}(3)$ parameterizations (\textit{Geometric}, \textit{Euler-angles}, and \textit{Givens-rotation}; mean $\pm$ standard error over 5 stratified folds). Panels correspond to: (a) HAR (Sitting vs.\ Laying), (b) HAR (Sitting vs.\ Standing vs.\ Laying), (c) MAGIC Gamma Telescope (gamma vs.\ hadron), (d) Covertype (1 vs.\ 2), and (e) Covertype (1 vs.\ 2 vs.\ 3). Overall, the \textit{Geometric} parameterization tends to achieve the highest accuracy across the considered settings.}
    \label{fig:appendix_paramet_qnn_features}
\end{figure*}

\begin{figure*}[h]
    \centering
    
    \begin{minipage}[t]{0.32\textwidth}
    \centering
    \subfloat[]{\includegraphics[width=\linewidth]{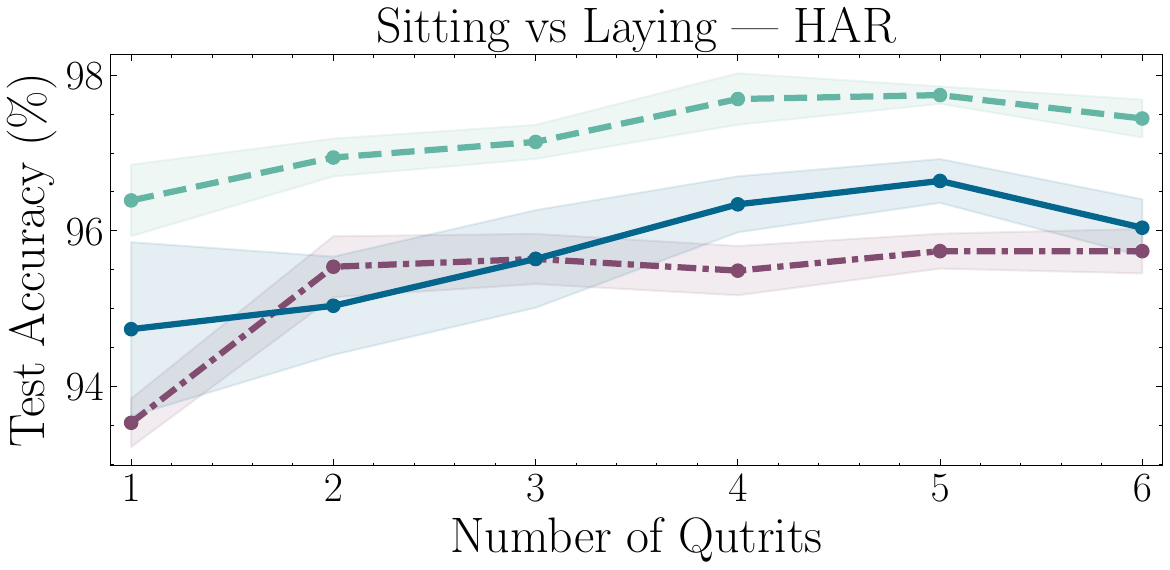}}
    \end{minipage}\hfill
    \begin{minipage}[t]{0.32\textwidth}
    \centering
    \subfloat[]{\includegraphics[width=\linewidth]{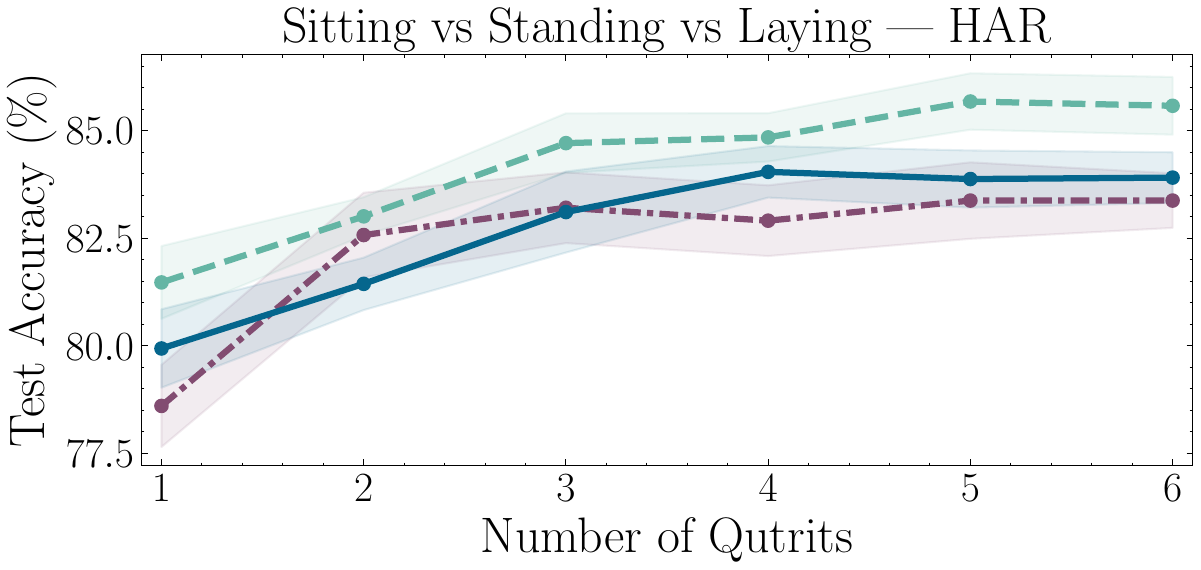}}
    \end{minipage}\hfill
    \begin{minipage}[t]{0.32\textwidth}
    \centering
    \subfloat[]{\includegraphics[width=\linewidth]{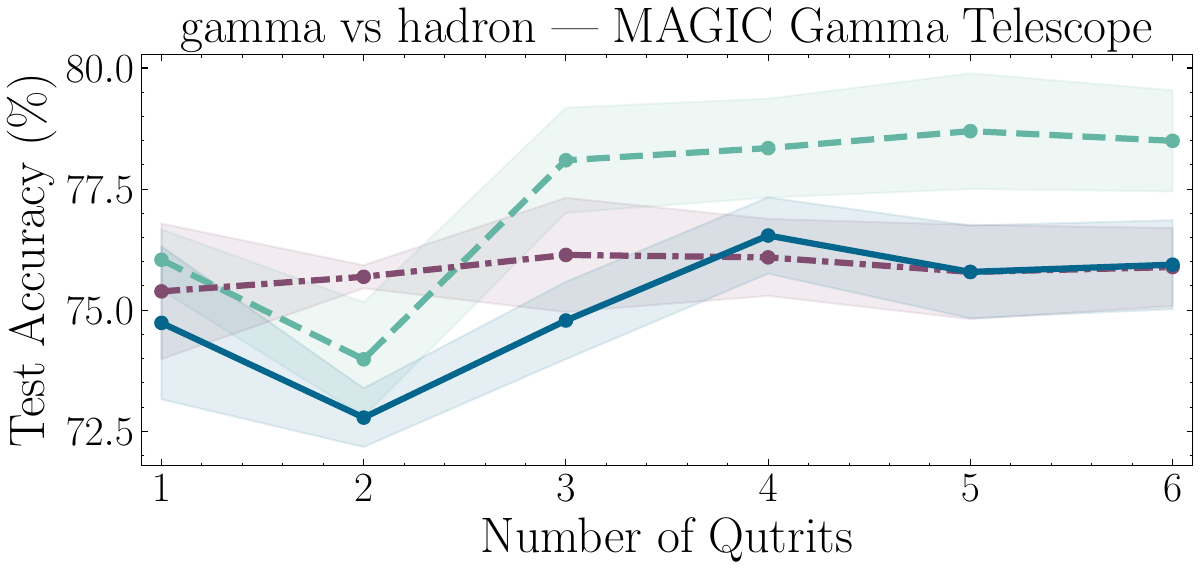}}
    \end{minipage}
    
    \vspace{2mm}
    
    \makebox[\textwidth][c]{%
        \begin{minipage}[t]{0.32\textwidth}
        \centering
        \subfloat[]{\includegraphics[width=\linewidth]{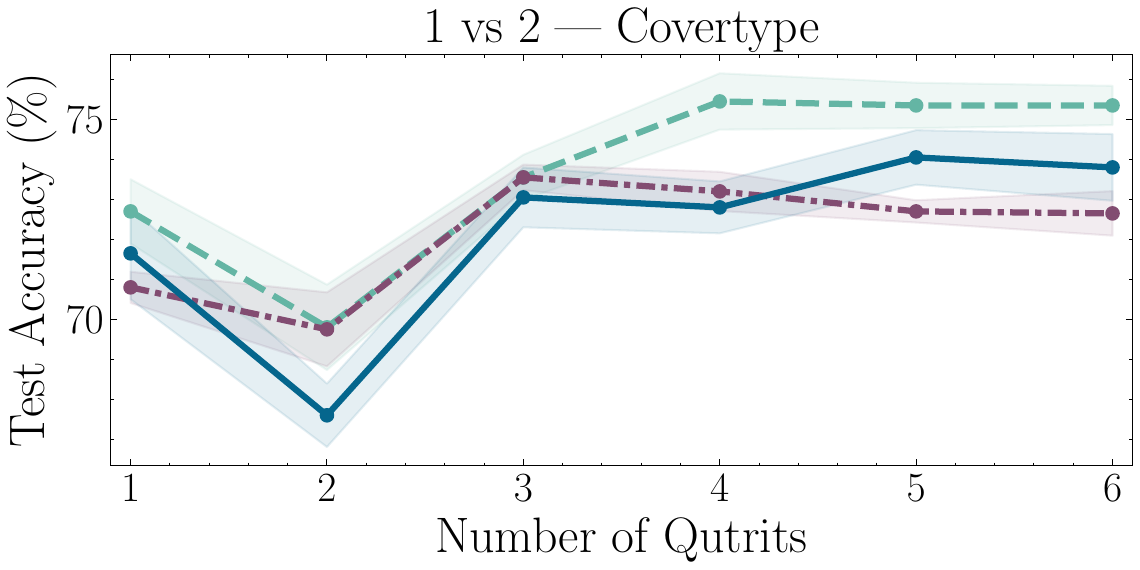}}
        \end{minipage}
        \hspace{0.04\textwidth}
        \begin{minipage}[t]{0.32\textwidth}
        \centering
        \subfloat[]{\includegraphics[width=\linewidth]{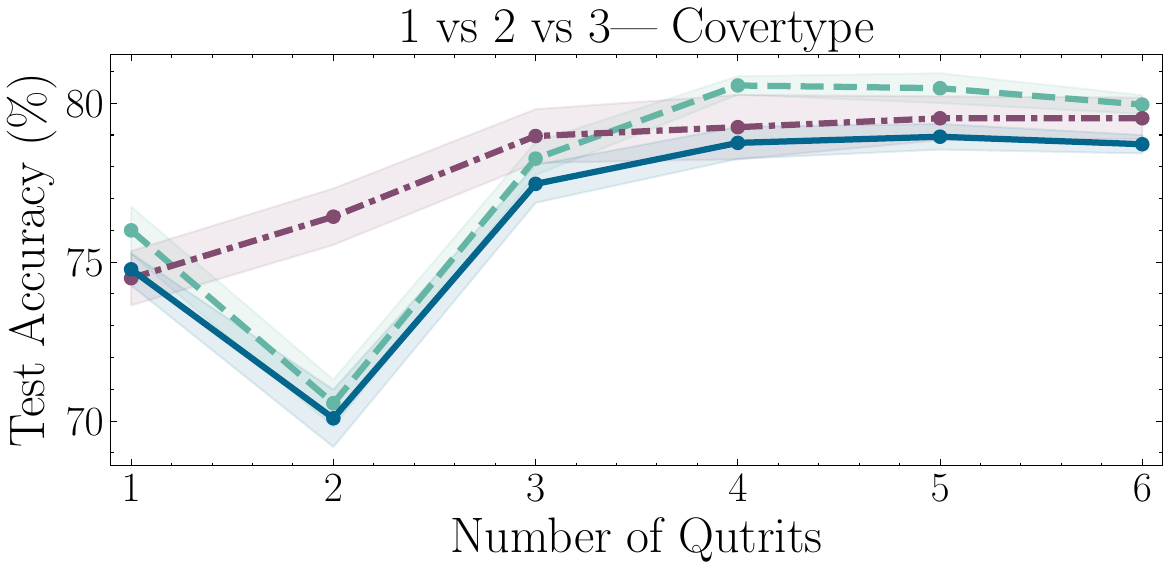}}
        \end{minipage}
    }
    
    \vspace{2mm}
    
    \makebox[\textwidth][c]{%
        \begin{minipage}[t]{0.35\textwidth}
        \centering
        \includegraphics[width=\linewidth]{plots/paramet_legend.pdf}
        \end{minipage}
    }
    \caption{\justifying Effect of the $\mathrm{SU}(3)$ parameterization (\textit{Geometric}, \textit{Euler-angles}, and \textit{Givens-rotation}) on the $1$-to-$n$ NQK test accuracy as a function of the number of qutrits (mean $\pm$ standard error over 5 stratified folds; feature budget fixed as in the main text). Panels correspond to: (a) HAR (Sitting vs.\ Laying), (b) HAR (Sitting vs.\ Standing vs.\ Laying), (c) MAGIC Gamma Telescope (gamma vs.\ hadron), (d) Covertype (1 vs.\ 2), and (e) Covertype (1 vs.\ 2 vs.\ 3). Across tasks, accuracy typically stabilizes around $n\simeq 4$, consistent with the Fashion-MNIST trends.}
    \label{fig:appendix_paramet_qnn_qutrits}
\end{figure*}

\begin{figure*}[h]
    \centering
    
    \begin{minipage}[t]{0.32\textwidth}
    \centering
    \subfloat[]
    {\includegraphics[width=\linewidth]{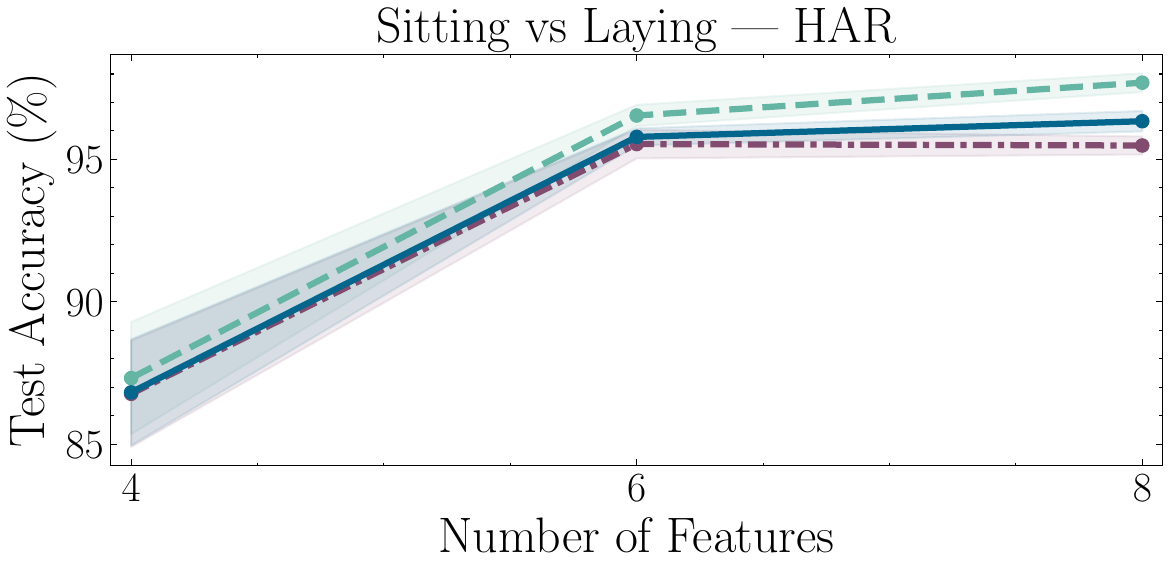}}
    \end{minipage}\hfill
    \begin{minipage}[t]{0.32\textwidth}
    \centering
    \subfloat[]
    {\includegraphics[width=\linewidth]{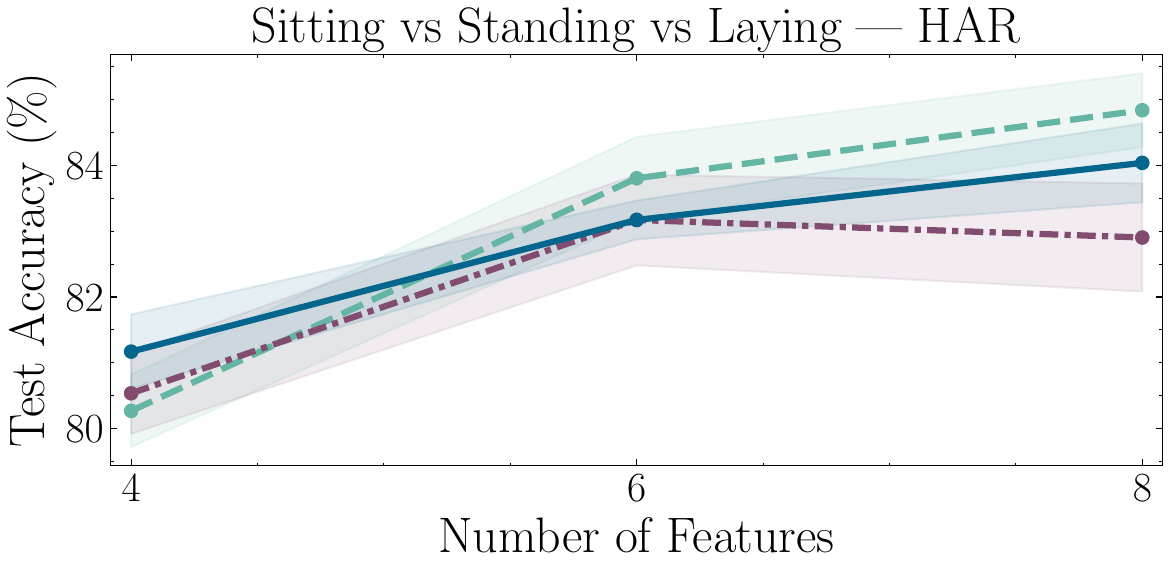}}
    \end{minipage}\hfill
    \begin{minipage}[t]{0.32\textwidth}
    \centering
    \subfloat[]
    {\includegraphics[width=\linewidth]{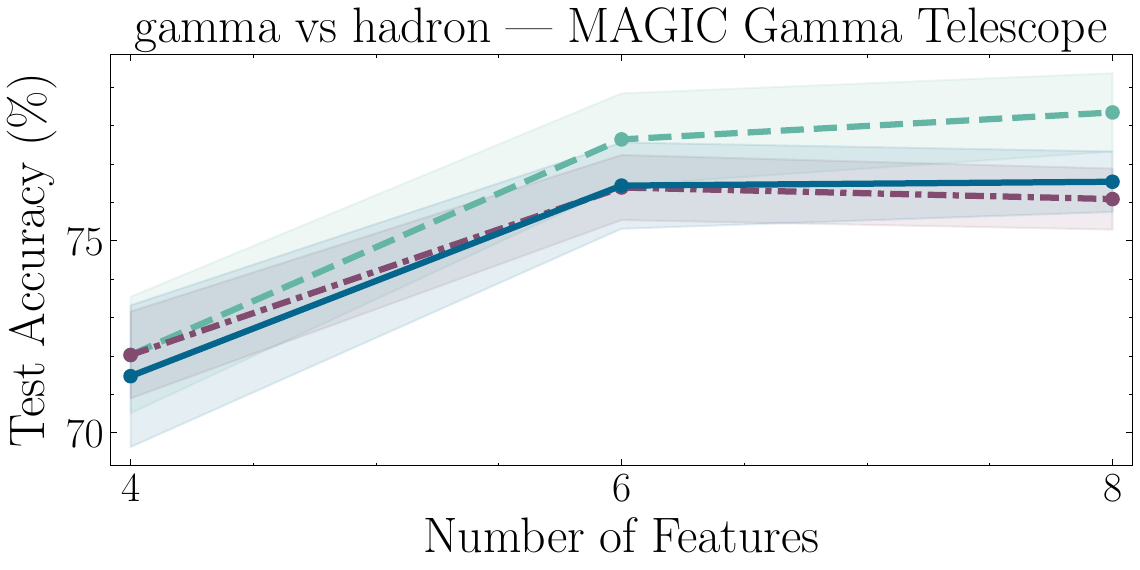}}
    \end{minipage}
    
    \vspace{2mm}
    
    \makebox[\textwidth][c]{%
        \begin{minipage}[t]{0.32\textwidth}
        \centering
        \subfloat[]
        {\includegraphics[width=\linewidth]{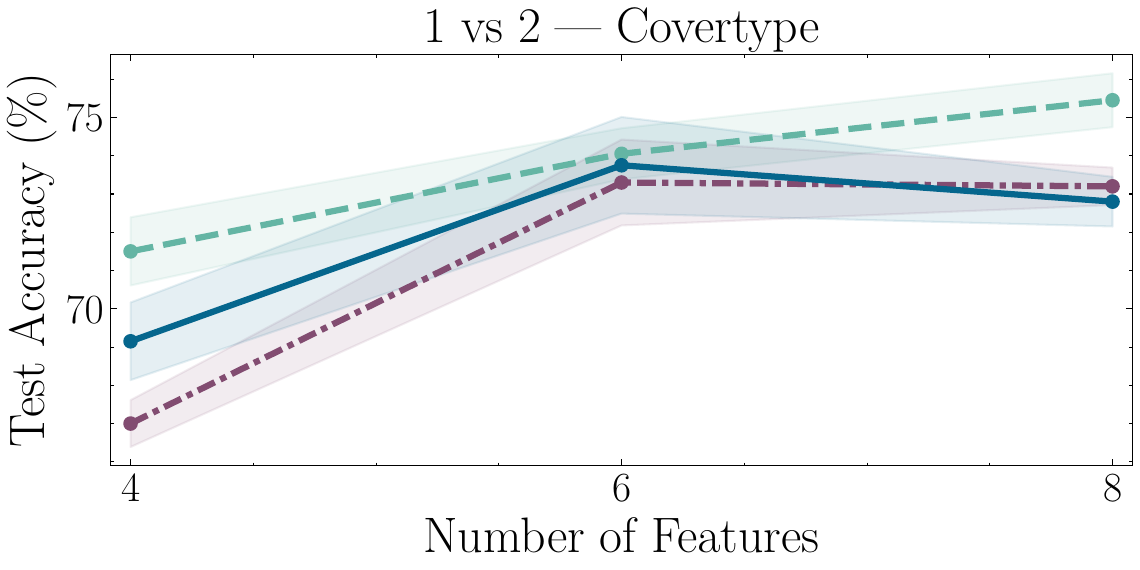}}
        \end{minipage}
        \hspace{0.04\textwidth}
        \begin{minipage}[t]{0.32\textwidth}
        \centering
        \subfloat[]{\includegraphics[width=\linewidth]{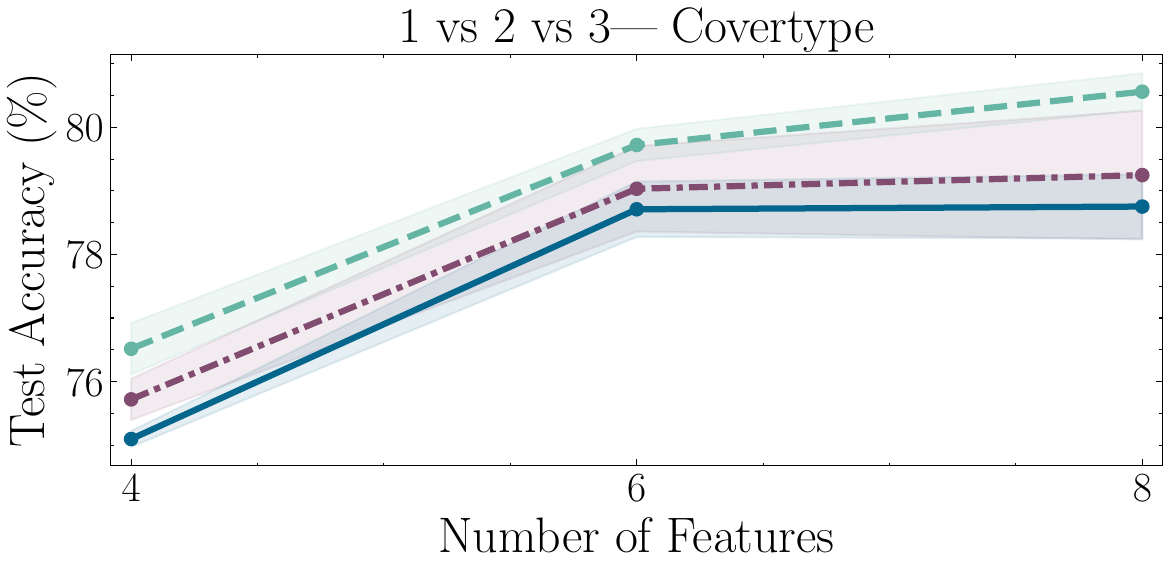}}
        \end{minipage}
    }
    
    \vspace{2mm}
    
    \makebox[\textwidth][c]{%
        \begin{minipage}[t]{0.35\textwidth}
        \centering
        \includegraphics[width=\linewidth]{plots/paramet_legend.pdf}
        \end{minipage}
    }
    \caption{\justifying Test accuracy of the $1$-to-$n$ NQK at fixed system size $n=4$ as a function of the number of encoded features on additional benchmarks, comparing three $\mathrm{SU}(3)$ parameterizations (\textit{Geometric}, \textit{Euler-angles}, and \textit{Givens-rotation}; mean $\pm$ standard error over 5 stratified folds). Panels correspond to: (a) HAR (Sitting vs.\ Laying), (b) HAR (Sitting vs.\ Standing vs.\ Laying), (c) MAGIC Gamma Telescope (gamma vs.\ hadron), (d) Covertype (1 vs.\ 2), and (e) Covertype (1 vs.\ 2 vs.\ 3). Consistent with the Fashion-MNIST results, feature scaling in the kernel setting is typically more regular than for the corresponding $1$-qutrit QNN.}
    \label{fig:appendix_paramet_1ton_features}
\end{figure*}

In this appendix we report results for the remaining benchmarks (HAR, MAGIC Gamma Telescope, and Covertype). The qualitative trends observed on Fashion-MNIST in the main text are broadly reproduced: (i) feature scaling can improve accuracy but is not necessarily monotonic for the $1$-qutrit QNN (Fig. \ref{fig:app_1qnn_features}); (ii) performance generally improves with system size for both the $1$-to-$n$ and $n$-to-$n$ constructions (Figs. \ref{fig:appendix_1ton_qutrits} and \ref{fig:appendix_nton_qutrits}); (iii) at fixed system size ($n=4$), feature scaling in the $1$-to-$n$ kernel setting is typically more regular than for the corresponding $1$-qutrit QNN (Fig. \ref{fig:appendix_1ton_features}), and similarly for the $n$-to-$n$ construction (Fig \ref{fig:appendix_nton_features}); (iv) the choice of $\mathrm{SU}(3)$ parameterization can materially affect both performance and scaling trends, as shown by the parameterization comparisons for the $1$-qutrit QNN and the $1$-to-$n$ construction (Figs. \ref{fig:appendix_paramet_qnn_features}, \ref{fig:appendix_paramet_qnn_qutrits}, and \ref{fig:appendix_paramet_1ton_features}); and (v) the $1$-to-$4$ and $4$-to-$4$ qutrit NQK's achieves performance comparable to a classical RBF-SVM baseline across the additional benchmarks (Table \ref{tab:appendix_nqk_rbf_svm_comparison}).
\vspace{1em}
\FloatBarrier
\begin{table}[t]
    \centering
    \caption{\justifying
    Comparison between the $1$-to-$4$ and $4$-to-$4$ qutrit NQK's and a classical RBF-SVM baseline on the additional benchmarks (test accuracy in \%, mean $\pm$ standard error over 5 folds).}
    \label{tab:appendix_nqk_rbf_svm_comparison}
    \small
    \setlength{\tabcolsep}{6pt}
    \renewcommand{\arraystretch}{1.2}

    \begin{tabular}{lccc}
    \Xhline{1.2pt}
    \textbf{Task} & \textbf{$1$-to-$4$ NQK} & \textbf{$4$-to-$4$ NQK} & \textbf{RBF-SVM} \\
    \Xhline{1.2pt}
    binary - HAR binary  & 97.69$\pm$00.33 & 97.99$\pm$0.08 & 98.25$\pm$0.56 \\
    \Xhline{0.2pt}
    3-class - HAR & 84.84$\pm$0.56 & 84.34$\pm$0.64 & 85.78$\pm$1.31 \\
    \Xhline{0.2pt}
    MAGIC Gamma Telescope & 78.45$\pm$0.96 & 77.39$\pm$0.63 & 78.70$\pm$1.45 \\
    \Xhline{0.2pt}
    binary - Covertype & 75.45$\pm$0.70 & 75.50$\pm$0.44 & 76.45$\pm$2.62 \\
    \Xhline{0.2pt}
      3-class - Covertype & 80.56$\pm$0.29 & 79.57$\pm$0.71 & 78.00$\pm$1.83 \\
    \Xhline{1.2pt}
    \end{tabular}
\end{table}

\FloatBarrier

\section{Sensitivity to the cost function}
\label{sec:appendix_cost_functions}

For qutrit-based binary classification we predict the label from the sign of the spin-$1$ expectation value $\langle S_z\rangle$, which provides a natural two-way partition of the local state space. For three-class classification, however, the choice of training objective is less canonical. In the main text we use cross-entropy (Eq. \ref{eq:qutrit_cross_entropy}) over computational-basis probabilities, but alternative objectives are possible, e.g., maximizing fidelity to class states or defining decision regions via an $S_z$-based partition.

\paragraph{Fidelity-based cost function.}
Let $|\psi_j\rangle$ denote the output state of the readout qutrit for datapoint $j$, and let $|\phi_{y_j}\rangle\in\{|0\rangle,|1\rangle,|2\rangle\}$ be the computational-basis state associated with label $y_j$. We consider the fidelity loss
\begin{equation}
    f_{\mathrm{cost}}^{\mathrm{Fid}}(\theta)
    \;=\;
    \frac{1}{M}\sum_{j=1}^{M}
    \left(
    1 - |\langle \phi_{y_j} \mid \psi_j \rangle|^2
    \right)^2,
    \label{eq:cost_fidelity}
\end{equation}
which encourages the circuit output to align with the corresponding label state.

\paragraph{$S_z$-based regression with three-region thresholding.}
We also consider an objective based on the expectation value of the spin-1 operator
$S_z=\mathrm{diag}(1,0,-1)$, trained via mean-squared error against class-dependent target values in $[-1,1]$. Concretely, we map labels $y_j\in\{0,1,2\}$ to targets $t(y_j)\in\{+1,0,-1\}$ and minimize
\begin{equation}
    f_{\mathrm{cost}}^{S_z}(\theta)
    \;=\;
    \frac{1}{M}\sum_{j=1}^{M}
    \left(
    \langle S_z \rangle_{\bm{x}_j} - t(y_j)
    \right)^2.
    \label{eq:cost_sz_3class}
\end{equation}
At inference time, we discretize $\langle S_z\rangle_{\bm{x}}$ into three regions,
\begin{equation}
    \hat{y}(\bm{x})
    \;=\;
    \begin{cases}
     0, & \text{if } \langle S_z \rangle_{\bm{x}} \ge \frac{1}{3},\\[2pt]
     1, & \text{if } \frac{1}{3} > \langle S_z \rangle_{\bm{x}}  \ge -\frac{1}{3},\\[2pt]
     2, & \text{if } \langle S_z \rangle_{\bm{x}} < -\frac{1}{3}.
    \end{cases}
    \label{eq:sz_threshold_3class}
\end{equation}
The thresholds $\pm 1/3$ are motivated by an equal-area partition of the unit sphere into three latitudinal bands: since the surface area between polar angles depends linearly on $z=\cos\theta$, dividing the sphere into three equal-area bands yields cuts at $z=\pm 1/3$. We emphasize that this provides a convenient heuristic based on the one-dimensional projection $\langle S_z\rangle$, rather than a canonical partition of the full qutrit state space.

\begin{figure*}[h]
    \centering
    
    \begin{minipage}[t]{0.32\textwidth}
        \centering
        \subfloat[]{\includegraphics[width=\linewidth]{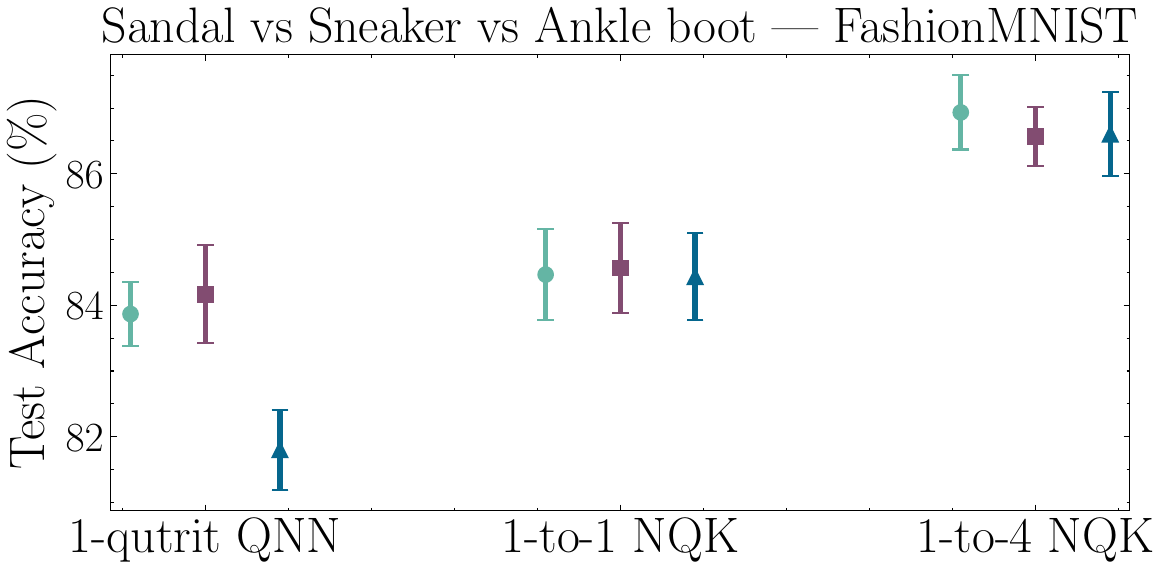}}
    \end{minipage}\hfill
    \begin{minipage}[t]{0.32\textwidth}
        \centering
        \subfloat[]{\includegraphics[width=\linewidth]{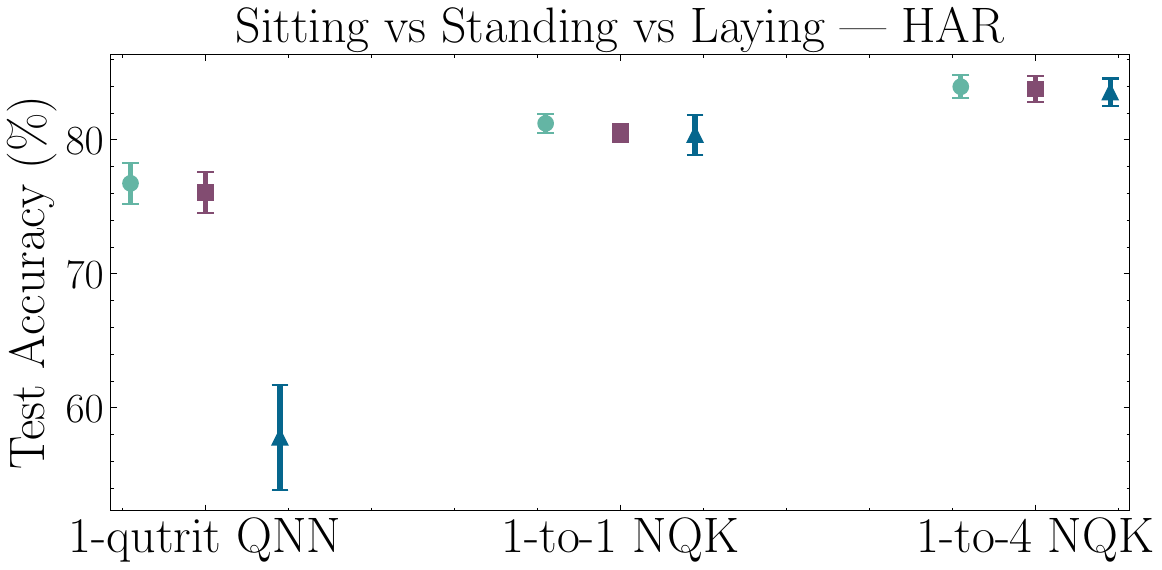}}
    \end{minipage}\hfill
    \begin{minipage}[t]{0.32\textwidth}
        \centering
        \subfloat[]{\includegraphics[width=\linewidth]{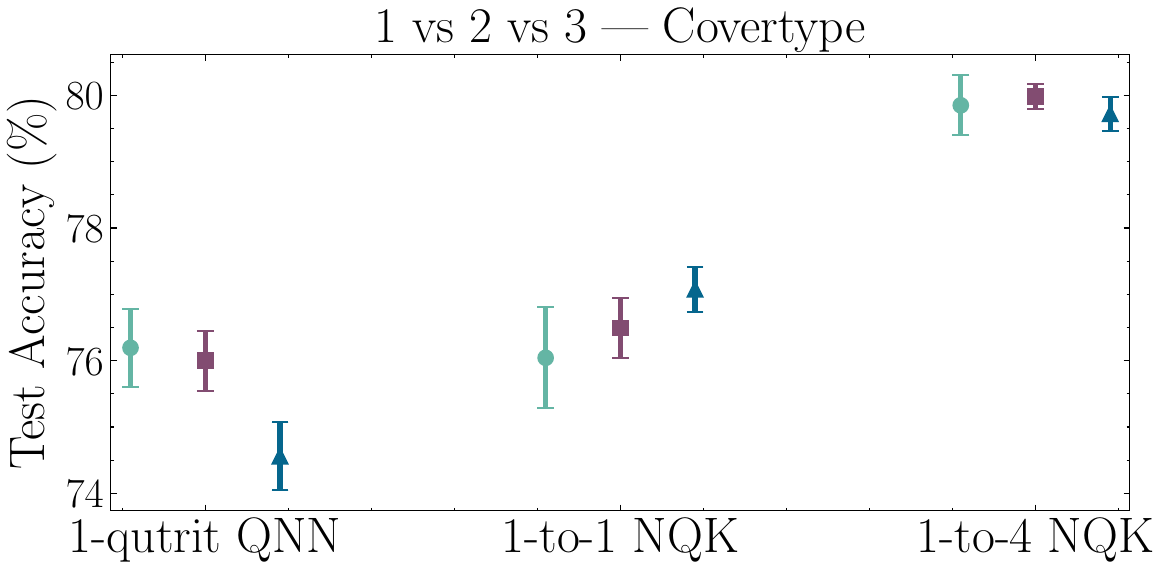}}
    \end{minipage}
    
    \vspace{2mm}

    \makebox[\textwidth][c]{%
        \begin{minipage}[t]{0.35\textwidth}
            \centering
            \includegraphics[width=\linewidth]{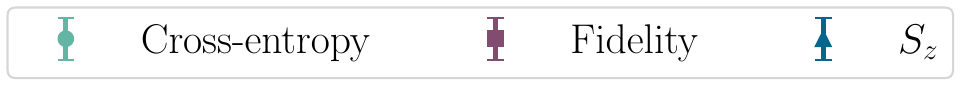}
        \end{minipage}
    }
    \caption{\justifying Effect of the training objective for three-class qutrit classification. Test accuracy for the $1$-qutrit QNN and the corresponding $1$-to-$n$ NQK constructed from it (shown for $n=1$ and $n=4$), comparing three objectives: cross-entropy, a fidelity-based loss, and an $S_z$-based objective (mean $\pm$ standard error over 5 stratified folds). Panels correspond to: (a) Fashion-MNIST (Sandal vs.\ Sneaker vs.\ Ankle boot), (b) HAR (Sitting vs.\ Standing vs.\ Laying), and (c) Covertype (1 vs.\ 2 vs.\ 3). Cross-entropy and fidelity yield similar performance for the $1$-qutrit QNN, whereas the $S_z$-based objective performs worse; however, the corresponding kernel models become much closer in performance, converging to nearly the same accuracy at $n=4$, suggesting increased robustness of the NQK construction to the specific pretraining objective.}
    \label{fig:appendix_cost_functions}
\end{figure*}

To assess how sensitive our conclusions are to this choice, we train the $1$-qutrit QNN using three loss functions: cross-entropy, a fidelity-based loss, and an $S_z$-based objective. As shown in Fig. \ref{fig:appendix_cost_functions}, cross-entropy and fidelity yield similar test accuracy for the $1$-qutrit QNN, whereas the $S_z$-based objective performs substantially worse. When the trained $1$-qutrit circuit is reused to construct the $1$-to-$n$ kernel, these differences are markedly reduced: the resulting NQKs achieve comparable accuracy and, for $n=4$, converge to nearly the same performance. This supports the view that the kernel construction provides a robust final predictor, reducing sensitivity to the particular supervised objective used during QNN pretraining.

\section{Training diagnostics for $\mathrm{SU}(3)$ parameterizations}
\label{sec:appendix_param_diagnostics}

\begin{figure*}[h]
    \centering
    
    \begin{minipage}[t]{0.35\textwidth}
    \centering
    \subfloat[]{\includegraphics[width=\linewidth]{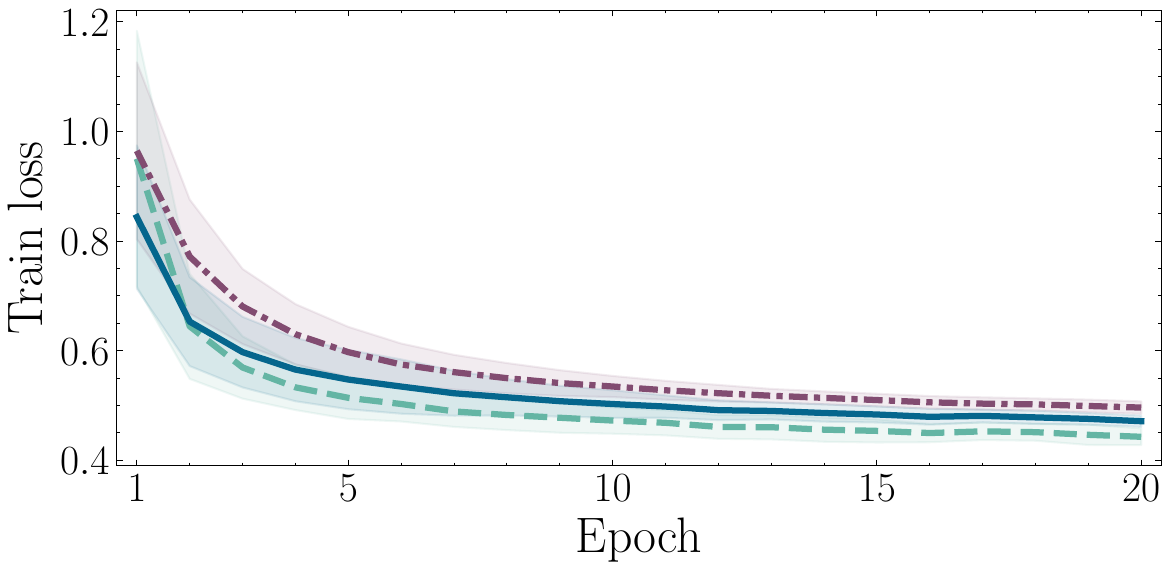}}
    \end{minipage}
    \hspace{1cm}
    \begin{minipage}[t]{0.35\textwidth}
    \centering
    \subfloat[]{\includegraphics[width=\linewidth]{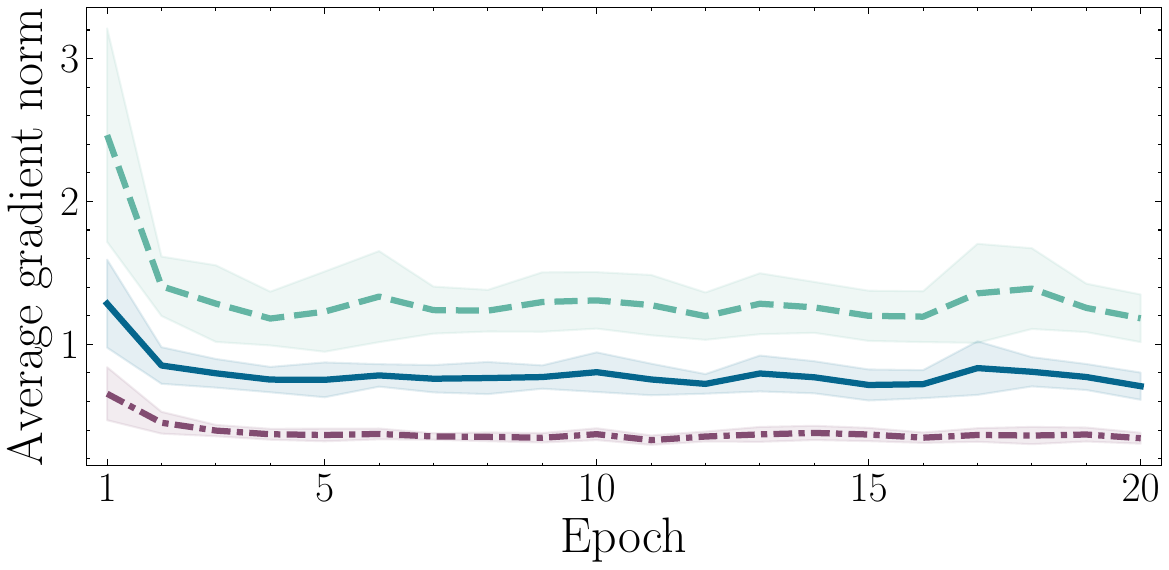}}
    \end{minipage}
    
    \vspace{2mm}
    
    \makebox[\textwidth][c]{%
        \begin{minipage}[t]{0.35\textwidth}
        \centering
        \includegraphics[width=\linewidth]{plots/paramet_legend.pdf}
        \end{minipage}
    }
    \caption{\justifying
    Training diagnostics for the $1$-qutrit QNN under the three $\mathrm{SU}(3)$ parameterizations (\textit{Geometric}, \textit{Euler-angles}, and \textit{Givens-rotation}) using identical data splits and optimizer settings. Panel (a) shows the training loss versus epoch and panel (b) shows the average gradient norm versus epoch (mean $\pm$ standard deviation over random initializations). The \textit{Geometric} parameterization typically reaches lower loss and maintains larger gradients than the \textit{Givens-rotations} and \textit{Euler-angles} parameterizations, consistent with the performance trends reported in Sec. \ref{sec:results_parameterization}.}
    \label{fig:param_diagnostics}
\end{figure*}

Figure \ref{fig:param_diagnostics} reports training loss and average gradient norms across epochs (mean $\pm$ standard deviation over random initializations) for the three $\mathrm{SU}(3)$ parameterizations under identical data and optimizer settings. The \textit{Geometric} parameterization reaches lower loss and exhibits larger gradients than \textit{Givens-rotations}, consistent with the performance trends reported in Sec. \ref{sec:results_parameterization}.

\FloatBarrier

\end{document}